%% file: main.tex
\documentclass[acmsmall,authorversion,nonacm]{acmart}
\usepackage{amsmath}

\usepackage{amssymb}

\usepackage{amsfonts}

\usepackage{mathtools}

\usepackage{marvosym}
\usepackage{wrapfig}

\usepackage{acro}[=v2]
\usepackage{geometry}
\usepackage{enumitem}
\usepackage{tabularx}
\usepackage{booktabs}
\usepackage{subcaption}
\usepackage[utf8]{inputenc}
\usepackage[english]{babel}
\usepackage{csquotes}
\usepackage{algorithm}
\usepackage{algorithmic}
\usepackage{wasysym}
\usepackage{multirow}
\usepackage{xspace}
\usepackage{pdflscape}
\usepackage{multicol}
\usepackage{arydshln}
\usepackage{adjustbox}
\usepackage{subcaption}
\usepackage{makecell}

\usepackage{array}
\newcolumntype{P}[1]{>{\hspace{0pt}}p{#1}}
\usepackage{xcolor}
\definecolor{pythonblue}{HTML}{1f77b4}

\newcommand{\revision}[1]{{#1}}

\PassOptionsToPackage{hyphens}{url}

\newcommand{\invt}{\revision{\mathcal{I}}}
\newcommand{\const}{\revision{\mathcal{K}}}
\newcommand{\coin}[1]{\texttt{#1}}
\newcommand{\tokenaddress}[2]{\href{https://etherscan.io/token/#2}{\texttt{#1}}}

\newcommand{\ie}{{i.e., }}
\newcommand{\etal}{{et al.}}

\AtBeginDocument{%
  \providecommand\BibTeX{{%
    \normalfont B\kern-0.5em{\scshape i\kern-0.25em b}\kern-0.8em\TeX}}}

\setcopyright{acmcopyright}
\copyrightyear{2022}
\acmYear{2022}
\acmDOI{XXXXXXX.XXXXXXX}

\acmJournal{CSUR}
\acmVolume{37}
\acmNumber{1}
\acmArticle{}
\acmMonth{1}

\makeatletter
\let\disable\@secondoftwo
\addto\extrasenglish{%
  \renewcommand{\sectionautorefname}{\revision{\S}\@gobble}%
}	
\addto\extrasenglish{%
  \renewcommand{\subsectionautorefname}{\revision{\S}\@gobble}%
}
\addto\extrasenglish{%
  \renewcommand{\subsubsectionautorefname}{\revision{\S}\@gobble}%
}
\addto\extrasenglish{%
  \renewcommand{\paragraphautorefname}{\revision{\S}\@gobble}%
}
\makeatother

\addto\extrasenglish{%
}
\addto\extrasenglish{%
  
  }

\input{exhibits/acronyms}

\begin{document}

\title{\acs{sok}: Decentralized Exchanges (DEX) with Automated Market Maker (\acs{amm}) Protocols}

\author{Jiahua Xu}
\affiliation{%
  \institution{UCL Centre for Blockchain Technologies}
  \city{London}
  \country{UK}}
\email{jiahua.xu@ucl.ac.uk}

\author{Krzysztof Paruch}
\affiliation{%
  \institution{Vienna University of Economics and Business}
  \country{Austria}}
\email{krzysztof.paruch@wu.ac.at}

\author{Simon Cousaert}
\affiliation{%
  \institution{The Block}
  \city{London}
  \country{UK}}
\email{scousaert@theblock.co}

\author{Yebo Feng}
\authornote{Corresponding Author: Yebo Feng.}
\affiliation{%
  \institution{University of Oregon}
  \city{Eugene}
  \country{USA}}
\email{yebof@uoregon.edu}

\input{sections/abstract}

\maketitle

\input{sections/intro}

\input{sections/preliminaries}

\input{sections/statespace}

\input{exhibits/comparison_table}

\input{sections/amm_properties}

\input{sections/comparison}

\input{sections/otherAMMprotocols}

\subsubsection*{\bf Summary}
\input{sections/discussion}

\input{sections/additionalAMMfeatures}

\input{sections/implementations}
\input{sections/layer2}

\input{sections/securityprivacy}

\input{sections/future}

\input{sections/relatedwork}

\input{sections/conclusion}

\appendix

\input{sections/comparison_app}

\input{exhibits/landscape_table}
\input{sections/attack_app}
\section{Overview tables for attacks and related work}
\begin{landscape}
    \input{exhibits/attacks_amm}

    \input{exhibits/literature_review}
\end{landscape}
\newpage

\input{sections/glossary}

\printacronyms

\input{sections/acknowledgements}

\bibliographystyle{ACM-Reference-Format}
\bibliography{references,references2}

\end{document}

%% file: exhibits/acronyms.tex
\DeclareAcronym{sok}{
  short = SoK,
  long  = systematization of knowledge
}
\DeclareAcronym{defi}{
  short = DeFi,
  long  = decentralized finance
}
\DeclareAcronym{amm}{
  short = AMM,
  long = automated market maker
}
\DeclareAcronym{cfmm}{
  short = CFMM,
  long = constant function market maker
}
\DeclareAcronym{dex}{
  short = DEX,
  long = decentralized exchange,
  short-plural-form = DEXs,
  long-plural-form = decentralized exchanges
}
\DeclareAcronym{lp}{
  short = LP,
  long = liquidity provider,
  short-plural-form = LPs,
  long-plural-form = liquidity providers
}
\DeclareAcronym{pmm}{
  short = PMM,
  long = proactive market maker
}
\DeclareAcronym{dao}{
  short = DAO,
  long = Decentralized Autonomous Organization
}
\DeclareAcronym{nft}{
  short = NFT,
  long = non-fungible token
}
\DeclareAcronym{lmsr}{
  short = LMSR,
  long = logarithmic market scoring rule
}
\DeclareAcronym{ido}{
  short = IDO,
  long = initial DEX offering
}
\DeclareAcronym{ieo}{
  short = IEO,
  long = initial exchange offering
}
\DeclareAcronym{ddos}{
  short = DDoS,
  long = distributed denial-of-service
}
\DeclareAcronym{dns}{
  short = DNS,
  long = domain name server
}
\DeclareAcronym{bgp}{
  short = BGP,
  long = border gateway protocol
}
\DeclareAcronym{mev}{
  short = MEV,
  long = miner extractable value
}
\DeclareAcronym{bdos}{
  short = BDoS,
  long = blockchain denial-of-service
}
\DeclareAcronym{zkp}{
  short = ZKP,
  long = zero-knowledge proof
}
\DeclareAcronym{mpc}{
  short = MPC,
  long = multiparty computation
}
\DeclareAcronym{l7ddos}{
  short = L7 DDoS,
  long = application-layer distributed denial-of-service
}
\DeclareAcronym{dlt}{
  short = DLT,
  long = distributed ledger technology
}
\DeclareAcronym{bsc}{
  short = BSC,
  long = Binance Smart Chain
}
\DeclareAcronym{dapp}{
  short = dApp,
  long = decentralized application
}
\DeclareAcronym{pbs}{
  short = PBS,
  long = proposer/builder separation
}
\DeclareAcronym{nizk}{
  short = NIZK,
  long = non-interactive zero-knowledge
}
\DeclareAcronym{lbp}{
  short = LBP,
  long = Liquidity Bootstrapping Pool
}
\DeclareAcronym{cmm}{
  short = CMM,
  long = Custom Market Maker
}
\DeclareAcronym{evm}{
  short = EVM,
  long = Ethereum Virtual Machine
}

%% file: sections/abstract.tex
\begin{abstract}
  As an integral part of the \ac{defi} ecosystem, \acfp{dex} with \acf{amm} protocols have gained massive traction with the recently revived interest in blockchain and \ac{dlt} in general.
  Instead of matching the buy and sell sides, \acp{amm} employ a peer-to-pool method and determine asset price algorithmically through a so-called conservation function.
  To facilitate the improvement and development of \ac{amm}-based \acp{dex}, we create the first systematization of knowledge in this area.
  We first establish a general \ac{amm} framework describing the economics and formalizing the system's state-space representation. We then employ our framework to systematically compare the top \ac{amm} protocols' mechanics, illustrating their conservation functions, as well as slippage and divergence loss functions. We further discuss security and privacy concerns, how they are enabled by  \ac{amm}-based \revision{\acp{dex}'} inherent properties, and explore mitigating solutions. Finally, we conduct a comprehensive literature review on related work covering both \ac{defi} and conventional market microstructure.
\end{abstract}

\ccsdesc[500]{Information systems~Digital cash}
\ccsdesc[300]{Information systems~Secure online transactions}
\ccsdesc[500]{Information systems~Electronic funds transfer}
\ccsdesc[500]{General and reference~Surveys and overviews}
\ccsdesc[500]{Social and professional topics~Economic impact}
\ccsdesc[300]{Security and privacy}

\keywords{decentralized finance, decentralized exchange, automated market maker, blockchain, Ethereum}

%% file: sections/intro.tex
\section{Introduction}
\label{sec:intro}


With the revived interest in blockchain and cryptocurrency among both the general populace and institutional actors, the past year has witnessed a surge in crypto trading activity and accelerated development in the \acf{defi} space.
Among all the prominent \ac{defi} applications, \revision{\acfp{dex} with \acf{amm} protocols} are in the ascendancy, with an aggregate value locked exceeding \$100 billion at the time of writing~\cite{2021DefiPulse}.

\revision{An \ac{amm}-based \ac{dex} bears} attractive features such as decentralization, automation and continuous liquidity.
With traditional order-book-based exchanges, the market price of an asset is determined by the last matched buy and sell orders, ultimately driven by the supply and demand of the asset. In contrast, on an \ac{amm}-based \ac{dex}, a liquidity pool acts as a single counterparty for each transaction, with a so-called conservation function that prices assets algorithmically by only allowing the price to move along predefined trajectories.
\acp{amm} implement a peer-to-pool method, where \acp{lp} contribute assets to liquidity pools while individual users exchange assets with a pool or pools containing the input and the output assets. Users obtain immediate liquidity without having to find an exchange counterparty, whereas \acp{lp} profit from asset supply with exchange fees from users.
Furthermore, by using a conservation function for price setting, \acp{amm} render moot the necessity of maintaining the state of an order book, which would be costly on a distributed ledger.

\acp{amm} benefit both \acp{lp} and exchange users with accessible liquidity provision and exchange, especially for illiquid assets.
Despite the apparent advantages, \acp{amm} are often characterized by high slippage and divergence loss, two implicit economic risks imposed on the funds of exchange users and \acp{lp} respectively. Moreover, \ac{amm}-based \ac{dex} are associated with myriads of security and privacy issues.
Throughout the last three years, new protocols have been introduced to the market one after another with incremental improvements, attempting to tackle different issues which had been identified as weak spots in a previous version.
On top of that, new use cases are being addressed, and new applications of \ac{amm} iterations are proposed to the market.
While innovative in certain aspects, the various \ac{amm} protocols generally consist of the same set of composed mechanisms to allow for multiple functionalities of the system. As such, these systems are structurally similar, and their main differences lie in parameter choices and/or mechanism adaptations.


As one of the earliest \ac{defi} application categories, \ac{amm}-based \acp{dex} also constitute the most fundamental and crucial building blocks of the \ac{defi} ecosystem. Their importance prompts the emergence of studies covering \revision{various aspects of \ac{amm}-based \acp{dex}:} economics \cite{Bartoletti2021}, security \cite{Zhou2021A2MM,Yuksel2021,Zhou2021High-Frequency}, privacy \cite{Baum2021,Bowe2020}, etc.
While there has arisen a large number of surveys on general \ac{defi} \cite{werner2020sokDeFi,Schar2021Defi} as well as various \ac{defi} applications \cite{Cousaert2021,Bartoletti2020sokLendingPools}, there is a dearth of literature that systematically studies and critically examines \acp{dex} with \ac{amm} protocols in a comprehensive manner.
To the best of our knowledge,
this paper \revision{fills this gap} as the first \ac{sok} on \ac{amm}-based \ac{dex} with examples of deployed protocols.
We contribute to the body of literature mainly by:
\begin{enumerate}
  \item generalizing mechanisms and economics of \ac{amm}-based \acp{dex} with a formalized state space modeling framework \revision{and a summary of key common properties};
  \item illustrating instantiations of our framework by comparing major \ac{amm}-based \acp{dex} with mathematical derivation and parameterized visualization on their conservation function, slippage and divergence loss functions;
  \item positioning \ac{amm}-based \acp{dex} within the broader taxonomy of \ac{defi}, and examining their relationships and interactions with other \ac{defi} protocols;
  \item establishing a taxonomy of security and privacy issues concerning \ac{amm}-based \acp{dex}, and exploring mitigation solutions;
  \item conducting a state-of-the-art literature review summarizing current research priorities as well as existing output in \ac{amm}-based \acp{dex} and related fields, and identifying potential directions for future research.
\end{enumerate}



The rest of the paper is structured as follows:
in \autoref{sec:prel}
we lay out fundamental concepts and components of \acp{amm};
in \autoref{sec:formal}
we formalize \ac{amm} mechanisms with a state-space representation;
in \autoref{sec:comp}
we compare main protocols in terms of conservation function, exchange rates, slippage and impermanent loss;
in \autoref{sec:security} we
address issues related to security and privacy of \ac{amm}-based \ac{dex};
in \autoref{sec:future} we indicate several avenues of future \ac{amm} research;
in \autoref{sec:related} we present related work;
in \autoref{sec:conc} we conclude.

%% file: sections/preliminaries.tex
\section{AMM Preliminaries}
\label{sec:prel}

This section presents \acp{amm}-based \acp{dex}' main components, including different actors and assets, as well as their generalized mechanism and economics.

\subsection{Actors}

\subsubsection{\Acf{lp}}
\label{sec:lp}

A liquidity pool can be deployed through a smart contract with some initial supply of crypto assets by the first \ac{lp}. Other \acp{lp} can subsequently increase the pool's reserve by adding more of the type of assets that are contained in the pool. In turn, they receive pool shares proportionate to their liquidity contribution as a fraction of the entire pool \cite{Evans2020}. \acp{lp} earn transaction fees paid by exchange users.
While sometimes subject to a withdrawal penalty, \acp{lp} can freely remove funds from the pool \cite{martinelli2019whitepaper} by surrendering a corresponding amount of pool shares~\cite{Evans2020}.



\subsubsection{Exchange user (Trader)}
\label{sec:trader}

A trader submits an exchange order to a liquidity pool by specifying the input and output asset and either an input asset or output asset quantity; the smart contract automatically calculates the exchange rate based on the conservation function as well as the transaction fee and executes the exchange order accordingly.


\revision{Arbitrageurs compare asset prices across different markets to execute trades whenever closing price gaps can extract profits~\cite{makarov2020trading}.}
\Acp{amm} such as DODO (see \autoref{sec:dodointro}) leverage users' arbitrage behavior through their protocol design.

\subsubsection{Protocol foundation}
\label{sec:protocolfoundation}

A protocol foundation consists of protocol founders, designers, and developers responsible for architecting and improving the protocol. The development activities are often funded directly or indirectly through accrued earnings such that the foundation members are financially incentivized to build a user-friendly protocol that can attract high trading volume.

\subsection{Assets}
\label{sec:assets}

Several distinct types of assets are used in \ac{amm} protocols for operations and governance; one asset may assume multiple roles.

\subsubsection{Risk assets}
Characterized by illiquidity, risk assets are the primary type of assets \ac{amm}-based \acp{dex} are designed for.
Like centralized exchanges, an \ac{amm}-based \ac{dex} can facilitate an \ac{ieo} to launch a new token through liquidity pool creation, a capital raising activity termed \enquote{\ac{ido}} that is particularly suitable for illiquid assets.
To be eligible for an \ac{ido}, a risk asset sometimes needs to be whitelisted, and must be compatible with the protocol's technical requirements (e.g. ERC20~\cite{2021EthereumStandard} for most \acp{amm} on Ethereum).

\subsubsection{Base assets} Some protocols require a trading pair always to consist of a risk asset and a designated base asset. In the case of Bancor, every risk asset is paired with \coin{BNT}, the protocol's native token
\cite{Bancor2020}.
Uniswap V1 requires every pool to be initiated with a risk asset paired with \coin{ETH}.
Many protocols, such as Balancer and Curve, can connect two or more risk assets directly in liquidity pools without a designated base asset.

\subsubsection{Pool shares} Also known as \enquote{liquidity shares} and \enquote{\ac{lp} shares}, pool shares represent ownership in the portfolio of assets within a pool, and are distributed to \acp{lp}. Shares accrue trading fees proportionally and can be redeemed at any time to withdraw funds from the pool.

\subsubsection{Protocol tokens}
\label{sec:protocoltoken}
Protocol tokens are used to represent voting rights on protocol governance matters and are thus also termed \enquote{governance tokens} (see \autoref{sec:govright}). Protocol tokens are typically valuable assets \cite 
{Xu2022DeFiFlow} that are tradeable outside of the \ac{amm} and can incentivize participation when e.g. rewarded to \acp{lp} proportionate to their liquidity supply.
\acp{amm} compete with each other to attract funds and trading volume.
To bootstrap an \ac{amm} in the early phase with incentivized early pool establishment and trading, a feature called liquidity mining can be installed where the native protocol's tokens are minted and issued to \acp{lp} and/or exchange users.

\subsection{Fundamental \ac{amm} dynamics} \label{sec:fundAMMdynamics}

\subsubsection{Invariant properties}
The functionality of an \ac{amm} depends upon a \textit{conservation function} which encodes a desired invariant property of the system. As an intuitive example, Uniswap's constant product function determines trading dynamics between assets in the pool as it always conserves the product of value-weighted quantities of both assets in the protocol---each trade has to be made in a way such that the value removed in one asset equals the value added in the other asset. This weight-preserving characteristic is one desired invariant property supported by the design of Uniswap.


\subsubsection{Mechanisms}
An \ac{amm} typically involves two types of interaction mechanism: asset swapping of assets and liquidity provision/withdrawal.
Interaction mechanisms have to be specified in a way such that desired invariant properties are upheld; therefore the class of admissible mechanisms is restricted to the ones which respect the defined conservation function, if one is specified, or conserve the defined properties otherwise.

\subsection{Fundamental \ac{amm} economics}

\subsubsection{Rewards}

\ac{amm} protocols often run several reward schemes, including liquidity reward, staking reward, governance rights and security reward distributed to different actors to encourage participation and contribution.

\paragraph{Liquidity reward}

\Acp{lp} are rewarded for supplying assets to a liquidity pool, as they have to bear the opportunity costs associated with funds being locked in the pool. \Acp{lp} receive their share of trading fees paid by exchange users.

\paragraph{Staking reward}

On top of the liquidity reward in the form of transaction income, \acp{lp} are offered the possibility to stake pool shares or other tokens as part of an initial incentive program from a certain token protocol. The ultimate goal of the individual token protocols (see e.g. \coin{GIV} \cite{CryptoLocally2020} and \coin{TRIPS} \cite{DeGiglio2021}) is to further encourage token holding, while simultaneously facilitating token liquidity on exchanges and product usage. These staking rewards are given by protocols other than the \ac{amm}.

\paragraph{Governance right}
\label{sec:govright}
An \ac{amm} may encourage liquidity provision and/or swapping by rewarding participants governance rights in the form of protocol tokens (see \autoref{sec:protocoltoken}).
Currently, governance issues such as protocol treasury management~\cite{Demosthenes.eth2021} are proposed and discussed mostly on off-chain governance portals such as snapshot (\url{snapshot.org}), Tally (\url{tally.xyz}) and Boardroom (\url{boardroom.io}), where protocol tokens are used as ballots (see \autoref{sec:protocoltoken}) to vote on proposals.

\paragraph{Security reward}
Just as every protocol built on top of an open, distributed network, \ac{amm}-based \acp{dex} on Ethereum suffer from security vulnerabilities. Besides code auditing, a common practice that a protocol foundation adopts is to have the code vetted by a broader developer community and reward those who discover and/or fix bugs of the protocol with monetary prizes, commonly in fiat currencies, through a bounty program \cite{Breidenbach2018}.



\subsubsection{Explicit costs}

Interacting with \ac{amm} protocols incurs various costs, including charges for some form of \enquote{value} created or \enquote{service} performed and fees for interacting with the blockchain network. \ac{amm} participants need to anticipate three types of fees: liquidity withdrawal penalty, swap fee and gas fee.

\paragraph{Liquidity withdrawal penalty}

As introduced in \autoref{sec:genform} and further discussed in \autoref{sec:comp} of this paper, withdrawal of liquidity changes the shape of the conservation function and negatively affects the usability of the pool by elevating  slippage. Therefore, \acp{amm} such as DODO \cite{dodo2020whitepaper} levy a liquidity withdrawal penalty.

\paragraph{Swap fee}
\label{sec:swapfee}
Users interacting with the liquidity pool for token exchanges have to reimburse \acp{lp} for the supply of assets and for the divergence loss (see \autoref{sec:divergenceloss}). This compensation comes in the form of swap fees that are charged in every exchange trade and then distributed to liquidity pool shareholders. A small percentage of the swap fees may also go to the foundation of the \ac{amm} to further develop the protocol \cite{Xu2022a}.

\paragraph{Gas fee}
\label{sec:gasfee}
Every interaction with the protocol is executed in the form of an on-chain transaction, and is thus subject to a gas fee applicable to all transactions on the underlying blockchain.
In a decentralized network, validating nodes need to be compensated for their efforts, and transaction initiators must cover these operating costs. \revision{Interacting with more complex protocols will result in a higher gas fee due to the higher computational power needed for transaction verification.}

\subsubsection{Implicit costs}
\label{sec:implicitcosts}

Two essential implicit costs native to \ac{amm}-based \acp{dex} are slippage for exchange users and divergence loss for \acp{lp}.

\paragraph{Slippage}
\label{sec:slippage}

Slippage is defined as the difference between the spot price and the realized price of a trade. Instead of matching buy and sell orders, \acp{amm} determine exchange rates on a continuous curve, and {\em every} trade will encounter slippage conditioned upon the trade size relative to the pool size and the exact design of the conservation function. The spot price approaches the realized price for infinitesimally small trades, but they deviate more for bigger trade sizes.
This effect is amplified for smaller liquidity pools as every trade will significantly impact the relative quantities of assets in the pool, leading to higher slippage.
Due to continuous slippage, trades on \acp{amm} must be set with some slippage tolerance to be executed, a feature that can be exploited to perform e.g. sandwich attacks (see~\autoref{sec:appattaacks}).

\paragraph{Divergence loss}
\label{sec:divergenceloss}

For \acp{lp}, assets supplied to a protocol are still exposed to volatility risk, which comes into play in addition to the loss of time value of locked funds.
A swap alters the asset composition of a pool, which automatically updates the asset prices implied by the conservation function of the pool (\autoref{eq:trade}). This consequently changes the value of the entire pool. Compared to holding the assets outside of an \ac{amm} pool, contributing the same amount of assets to the pool in return for pool shares can result in less value with price movement, an effect termed \enquote{divergence loss} or \enquote{impermanent loss} (see \autoref{sec:comp}). This loss can be deemed \enquote{impermanent} because as asset price moves back and forth, the depreciation of the pool value continuously disappears and reappears and is only realized when assets are actually taken out of the pool. Well-devised \acp{amm} charge appropriate swap fees to ensure that \acp{lp} are sufficiently compensated for the divergence loss (see \autoref{sec:dynamicfee}).
Despite the fact that \enquote{impermanent loss} is a more widely used term on the Internet, we adhere to the more accurate term \enquote{divergence loss} in a scientific context. In fact, for the majority of \ac{amm} protocols, this \enquote{loss} only disappears when the current proportions of the pool assets equal exactly those at liquidity provision, which is rarely the case.

Since assets are bonded together in a pool, changes in prices of one asset affect all others in this pool. For an \ac{amm} protocol that supports single-asset supply, this forces \acp{lp} to be exposed to risk assets they have not been holding in the first place \revision{(see \autoref{sec:non-impact})}.

\input{exhibits/taxonomy}

\subsection{\ac{amm}-based \acp{dex} within \ac{defi}}

For brevity, we use \enquote{\ac{amm}} or \enquote{\ac{dex}} to refer to \ac{amm}-based \ac{dex} throughout the paper, unless indicated otherwise.
Nevertheless, it is to be noted that the term \enquote{\ac{amm}} emphasizes the algorithm of a protocol, whereas \enquote{\ac{dex}} emphasizes the use case, or application, of a protocol.
Within the context of blockchain-based \ac{defi}, there also exist orderbook-based \acp{dex} such as Gnosis and dYdX that do not rely on \ac{amm} algorithms.
\revision{
    Recently, \ac{dex} aggregators (\autoref{sec:dex-agg}) such as 1inch have emerged which incorporate both limited order books and \ac{amm} pools.
}
On the other hand, \ac{amm} algorithms are also not exclusively employed by \acp{dex}. \Ac{defi} applications such as lending platforms, \acp{nft}, stablecoins and derivatives all have protocols that make use of different \ac{amm} algorithms.

\Acp{amm} can also assume various forms (see \autoref{sec:ammwork}). Prediction markets for example commonly employs \ac{lmsr}, whereas \ac{cfmm} is the primary underpinning for \acp{dex}.
In particular, constant sum and constant product are the most representative forms of \ac{cfmm}, widely adopted by \ac{amm}-based \ac{dex} protocols.
\autoref{fig:taxonomy} illustrates \ac{amm}-based \ac{dex} within the broader taxonomy of \ac{defi} on blockchain.

The ensuing sections, \autoref{sec:formal} and \autoref{sec:comp}, focus on \ac{cfmm} mechanisms which have been adopted by major \acp{dex}, with their exact formulas derived in \autoref{appendix:formulas}. \autoref{subsec:implement} briefly presents other \ac{defi} applications with \ac{amm} implementations.

%% file: exhibits/taxonomy.tex
\begin{figure}[t]
    \centering
    \includegraphics[width=0.5\linewidth]{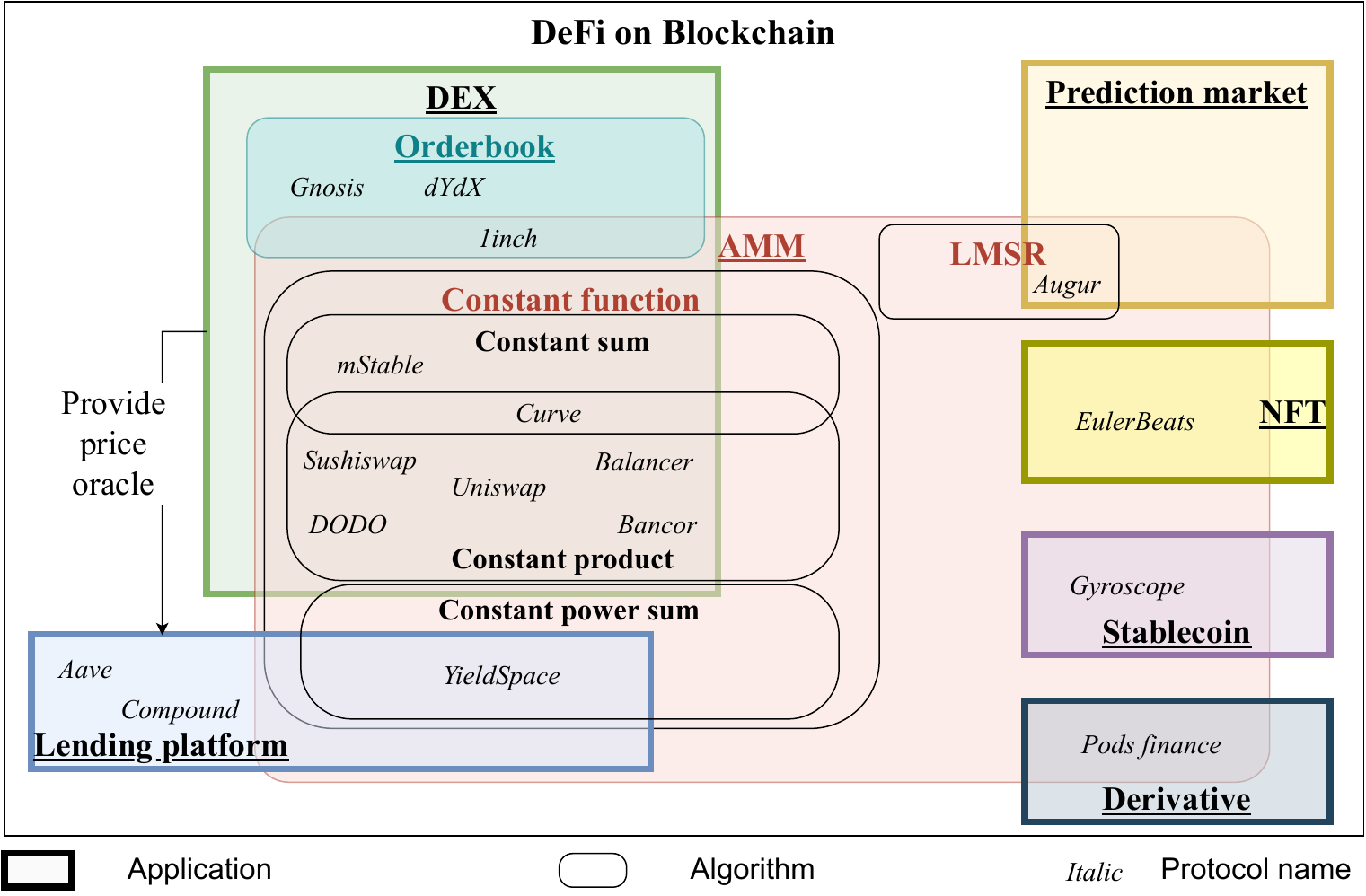}
    \caption{\ac{amm}-based \ac{dex} within the broader taxonomy of \ac{defi} on blockchain. \Ac{amm} as an algorithm and \ac{dex} as an application are not mutually inclusive.}
    \label{fig:taxonomy}
\end{figure}

%% file: sections/statespace.tex
\section{Formalization of Mechanisms}
\label{sec:formal}

Overall, the functionality of an \ac{amm} can be generalized formally by a set of few mechanisms. These mechanisms define how users can interact with the protocol and what the response of the protocol will be given particular user actions.

\subsection{State space representation}
\label{sec:statespace}
The functioning of any blockchain-based system can be modeled using state-space terminology. States and agents constitute the main system components; protocol activities are described as actions (\autoref{fig:mechanism}); the evolution of the system over time is modeled with state transition functions.
This can be generalized into a state transition function $f$ encoded in the protocol such that $\chi \xrightarrow[f]{a} \chi'$, where $a \in A$ represents an action imposed on the system while $\chi$ and $\chi'$ represent the current and future states of the system respectively.

The object of interest is the state $\chi$ of the liquidity pool which can be described with
\begin{align}
  \chi = (\{r_k\}_{k=1,...,n},\{p_k\}_{k=1,...,n},\invt, \Omega)
\end{align}
where $r_k$ denotes the quantity of token$_k$ in the pool, $p_k$ the current spot price of token$_k$, $\invt$ the conservation function invariant(s), and $\Omega$ the collection of protocol hyperparameters.
This formalization can encompass various \ac{amm} designs.

The most critical design component of an \ac{amm} is its conservation function which defines the relationship between different state variables and the invariant(s) $\invt$. The conservation function is protocol-specific as each protocol seeks to prioritize a distinct feature and target particular functionalities (see \autoref{sec:comp}).

The core of an \ac{amm} system state is the quantity of each asset held in a liquidity pool. Their sums or products are typical candidates for invariants.
Examples of a constant-sum market maker include mStable \cite{Andersson2020}. Uniswap \cite{Uniswap2020} represents constant-product market makers, while  Balancer \cite{martinelli2019whitepaper} generalizes this idea to a geometric mean. The Curve \cite{Egorov2019} conservation function is notably a combination of constant-sum and constant-product (see \autoref{sec:comp}).

\input{exhibits/mechanism}

\subsection{Generalized formulas}
\label{sec:genform}

In this section, we generalize \ac{amm} formulas necessary for demonstrating the interdependence between various \ac{amm} \revision{invariants} and state variables, as well as for computing slippage and divergence loss. Mathematical notations and their definitions can be found in \autoref{tab:token-exchange-notation}.

Hyperparameter set $\Omega$ is determined at pool creation and shall remain the same afterwards. While the value of hyperparameters might be changed through protocol governance activities, this does not and should not occur on a frequent basis.

Invariant $\invt$, despite its name, refers to the pool variable that stays constant only with swap actions \revision{(see \autoref{sec:path-binding})} but changes at liquidity provision and withdrawal. In contrast, trading moves the price of traded assets; specifically, it increases the price of the output asset relative to the input asset, reflecting a value appreciation of the output asset driven by demand \revision{(see \autoref{sec:demand-sensitivity})}. Liquidity provision and withdrawal, on the other hand, should not move the asset price \revision{(see \autoref{sec:non-impact})}. In particular, a {\em pure} liquidity provision and withdrawal activity requires a proportional change in reserves (\autoref{eq:liquidityChange}).

\input{exhibits/notations}

Formally, the state transition induced by  {\em pure}  liquidity change and asset swap can be expressed as follows.
\begin{align}
   & (\{r_k\}_{k=1,...,n},\{p_k\}_{k=1,...,n},\invt, \Omega)
  \xrightarrow[f] {\text{\tiny liquidity change}}
  (\{a \cdot r_k\}_{k=1,...,n},\{p_k\}_{k=1,...,n},\invt', \Omega) \text{, where } a > 0
  \label{eq:liquidityChange}
\end{align}

\begin{align}
  (\{r_k\}_{k=1,...,n},\{p_k\}_{k=1,...,n},\invt, \Omega)
  \xrightarrow[f]{\text{swap}}
   &
  (\{r'_k\}_{k=1,...,n},\{p'_k\}_{k=1,...,n},\invt, \Omega)
  \label{eq:trade}
\end{align}

\subsubsection{Conservation function}
\label{sec:form_cf}

An \ac{amm} conservation function, also termed ``bonding curve'', can be expressed explicitly as a relational function between \ac{amm} invariant and reserve quantities $\{r_k\}_{k=1,...,n}$:
\begin{align}
  \label{eq:conservationfunction}
  \invt = C(\{r_k\}
  )
\end{align}

A conservation function for each token pair, say $r_i$---$r_o$, must be concave, nonnegative and nondecreasing \cite{Angeris2021} (see also \autoref{fig:consfunc}).
For complex \acp{amm} such as Curve, it might be convenient to express the conservation function (\autoref{eq:conservationfunction}) implicitly in order to derive exchange rates between two assets in a pool:
\begin{align}
  \label{eq:implicitconservationfunction}
  Z(\{r_k\}; \invt
  ) = C(\{r_k\}
  ) - \invt = 0
\end{align}

Equation~\ref{eq:implicitconservationfunction} contains invariants $\invt$, whose value is determined by the initial liquidity provision (liquidity pool creation); afterwards, given the change in reserve quantity of one asset, the reserve quantity of the other asset can be solved.

\subsubsection{Spot exchange rate}
\label{sec:form_ser}

The spot exchange rate between token$_i$ and token$_o$ can be calculated as the slope of the $r_i$---$r_o$ curve (see examples in \autoref{fig:consfunc}) using partial derivatives of the conservation function $Z$.
\begin{align}
  _iE_o(\{r_k\}; \invt
  ) = \frac{
    \partial Z(\{r_k\}; \invt
    )/\partial r_o
  }{
    \partial Z(\{r_k\}; \invt
    )/\partial r_i
  }
\end{align}
Note that $_iE_o = 1 $ when $i=o$.

\subsubsection{Swap amount}
\label{sec:form_sa}

The amount of token$_o$ received $x_o$ (spent when $x_o<0$) given amount of token$_i$ spent $x_i$ (received when $x_i<0$) can be calculated following the steps below.

Note that $x_i > -r_i$ and $x_o > -r_o$. Their lower bound corresponds to the case when the received asset is depleted from the pool, i.e. its new reserve becomes 0 (see also \autoref{eq:standardinput} below). With common \ac{amm} protocols, $x_i, x_o$ theoretically often do not have a upper bound: if the reserve quantity is 1 unit, a trader can still sell 2 or more units into the pool, but mostly accompanied with a high slippage (see \autoref{fig:slippage} in the next section).

\paragraph{Update reserve quantities} Input quantity $x_i$ is simply added to the existing reserve of token$_i$; the reserve quantity of any token other than token$_i$ or token$_o$ stays the same:
\begin{align}
  \label{eq:standardinput}
  r_i' \coloneqq R_i(x_i; r_i)
       & = r_i + x_i                       \\
  r_j' & = r_j, \qquad \forall j \neq i, o
\end{align}

\paragraph{Compute new reserve quantity of token$_o$}

The new reserve quantity of all tokens except for token$_o$ is known from the previous step. One can thus solve $r_o'$, the unknown quantity of token$_o$, by plugging it in the conservation function:
\begin{align}
  Z(\{r'_k\}; \invt
  ) = 0
\end{align}

Apparently, $r_o'$ can be expressed as a function of the original reserve composition $\{r_k\}$, input quantity $x_i$, namely,
\begin{align}
  r_o' \coloneqq R_o(x_i, \{r_k\}; \invt
  )
\end{align}
\paragraph{Compute swapped quantity} The quantity of token$_o$ swapped is simply the difference between the old and new reserve quantities:
\begin{align}
  x_o \coloneqq X_o(x_i, \{r_k\}; \invt
  )
  = r_o - r_o'
\end{align}

\subsubsection{Slippage}
\label{sec:form_sl}

Slippage measures the deviation between effective exchange rate $\frac{x_i}{x_o}$ and the pre-swap spot exchange rate $_iE_o$, expressed as:
\begin{align}
  S(x_i, \{r_k\}; \invt
  ) = \frac{x_i/x_o}{_iE_o} -1
\end{align}

\subsubsection{Divergence loss}
\label{sec:form_dl}

Divergence loss describes the loss in value of all reserves in the pool compared to holding the reserves outside of the pool, after a price change of an asset (see \autoref{sec:divergenceloss}).
Based on the formulas for spot price and swap quantity established above, the divergence loss can generally be computed following the steps described below. In the valuation, we assign token$_i$ as the denominating currency for all valuations. While the method to be presented can be used for multiple token price changes through iterations, we only demonstrate the case where only the value of token$_o$ increases by $\rho$, while all other tokens' value stay the same. Token$_i$ is the numéraire. Designating one of the tokens in the pool as a numéraire can also be found in DeFi simulation papers such as \cite{Angeris2021}.


\paragraph{Calculate the original pool value} The value of the pool denominated in token$_i$ can be calculated as the sum of the value of all token reserves in the pool, each equal to the reserve quantity multiplied by the exchange rate with token$_i$:
\begin{align}
  V(\{r_k\}; \invt
  ) & = \sum_j {_iE_j}(\{r_k\}; \invt
  ) \cdot r_j
\end{align}

\paragraph{Calculate the reserve value if held outside of the pool} If all the asset reserves are held outside of the pool, then a change of $\rho$ in token$_o$'s value would result in a change of $\rho$ in token$_o$ reserve's value:
\begin{align}
  V_\text{held}(\rho; \{r_k\}, \invt) & = V(\{r_k\}; \invt) + [{_jE_o}(\{r_k\}; \invt) \cdot r_o] \cdot \rho
  \nonumber
\end{align}


\paragraph{Obtain re-balanced reserve quantities}
Exchange users and arbitrageurs constantly re-balance the pool through trading in relatively \enquote{cheap}, depreciating tokens for relatively \enquote{expensive}, appreciating ones.
As such, asset value movements are reflected in exchange rate changes implied by the dynamic pool composition.
Therefore,
the exchange rate between token$_o$ and each other token$_j$ ($j \neq o$) implied by new reserve quantities $\{r_k'\}$, compared to that by the original quantities $\{r_k\}$, can be expressed with Equation set \ref{eq:exchangerate}. At the same time, the equation for the conservation function must stand (\autoref{eq:newcf}).
\begin{align}
  \rho & = \frac{{_jE_o}(\{r_k'\}; \invt
    )
  }{{_jE_o}(\{r_k\}; \invt
    )
  } -1, \qquad \forall j \neq o
  \label{eq:exchangerate}
  \\
  0    & = Z(\{r_k'\}; \invt
  )
  \label{eq:newcf}
\end{align}
A total number of $n$-equations ($n-1$ with Equation set \ref{eq:exchangerate}, plus 1 with \autoref{eq:newcf}) would suffice to solve $n$ unknown variables $\{r_k'\}_{k= 1, ..., n}$, each of which can be expressed as a function of $\rho$ and $\{r_k\}$:
\begin{align}
  r'_k \coloneqq R_k (\rho, \{r_k\}; \invt)
\end{align}

\paragraph{Calculate the new pool value} The new value of the pool can be calculated by summing the products of the new reserve quantity multiplied by the new price (denominated by token$_i$) of each token in the pool:
\begin{align}
  V'(\rho, \{r_k\}; \invt) = \sum_j {_iE_j}(\{r_k'\}; \invt
  ) \cdot r_j'
\end{align}

\paragraph{Calculate the divergence loss} Divergence loss can be expressed as a function of $\rho$, the change in value of an asset in the pool:
\begin{align}
  L(\rho, \{r_k\}; \invt) & = \frac{
    V'(\rho, \{r_k\}; \invt)
  }{
    V_\text{held}(\rho; \{r_k\}, \invt)
  } -1
  \label{eq:curvediv}
\end{align}





%% file: exhibits/mechanism.tex
\begin{figure*}
    \begin{subfigure}{.48\textwidth}
        \includegraphics[width=\linewidth]{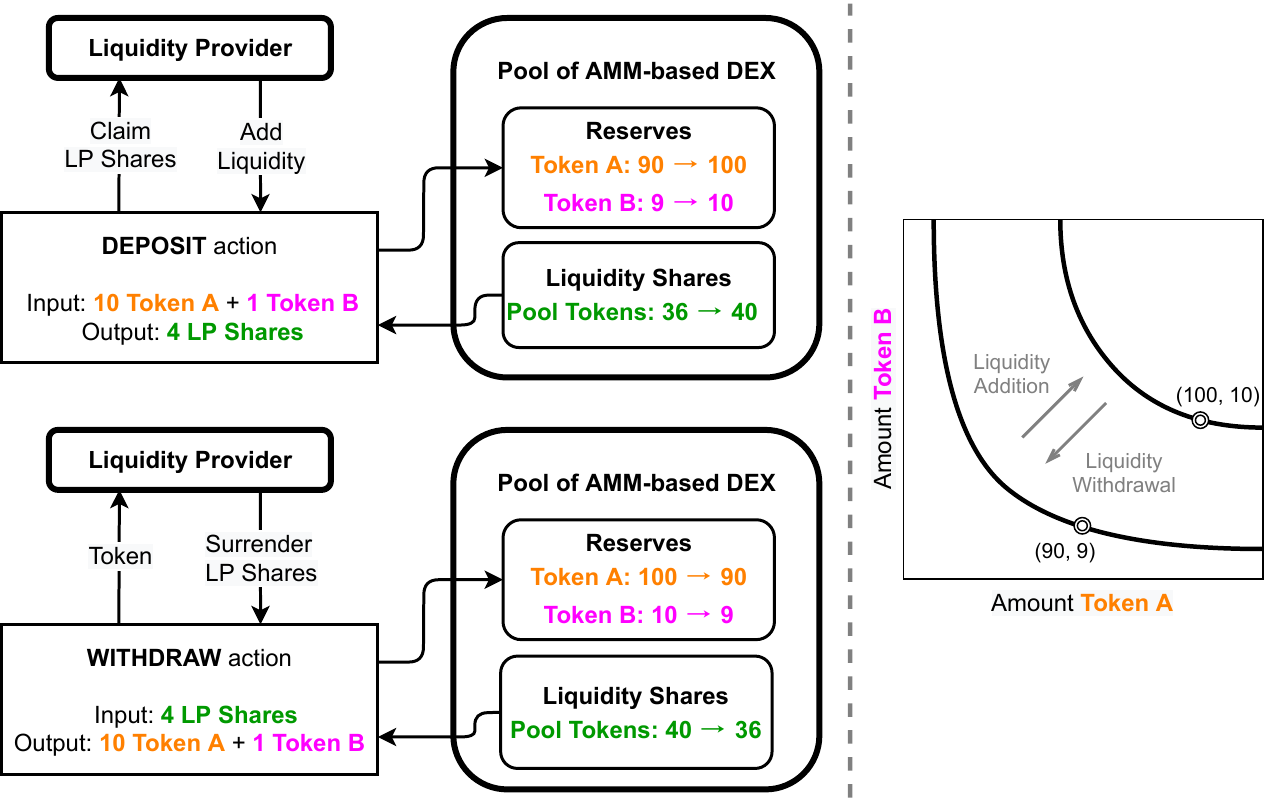}
        \caption{Liquidity provision and withdrawal.}
    \end{subfigure}
    \hspace{2mm}
    \begin{subfigure}{0.48\linewidth}
        \includegraphics[width=\linewidth]{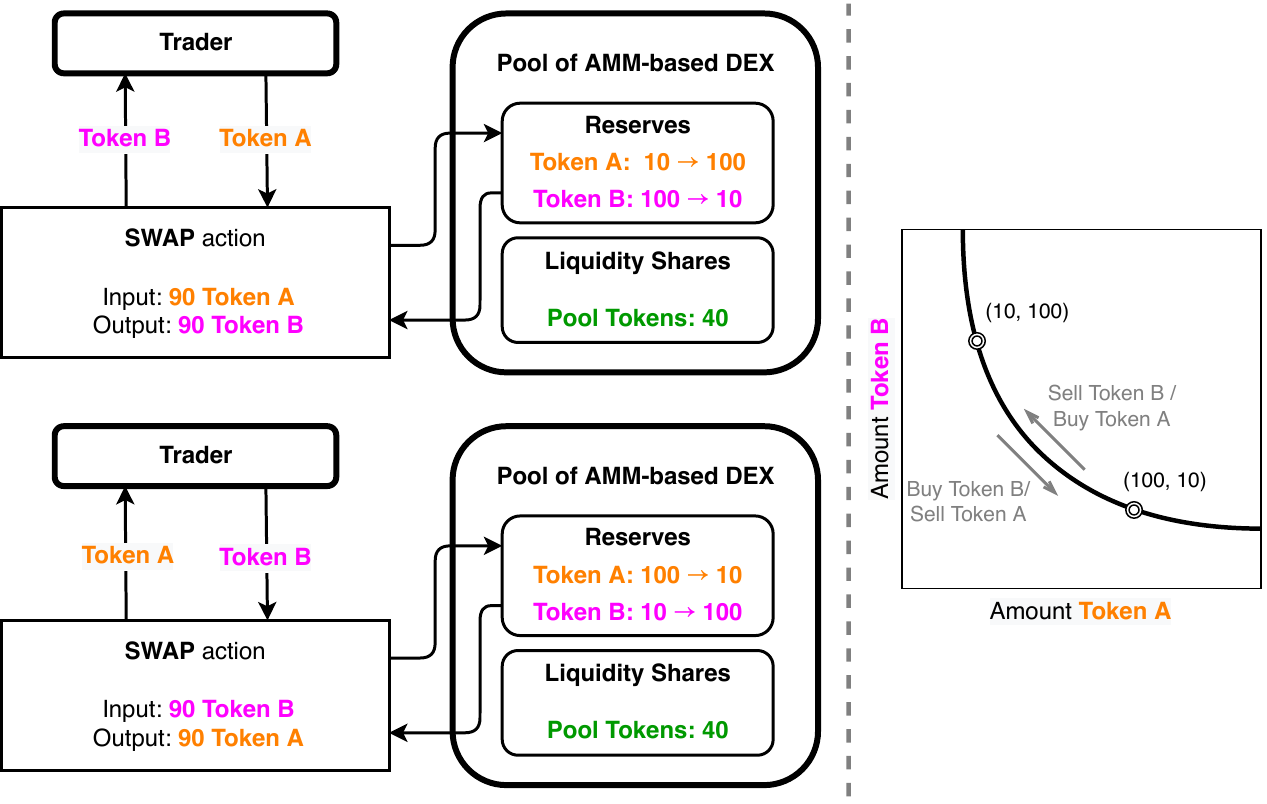}
        \caption{\revision{Swap}.}
        \label{fig:tokenexchange}
    \end{subfigure}
    \caption{Stylized \ac{amm} mechanisms for \revision{\acp{lp} (\autoref{sec:lp})} and traders (\autoref{sec:trader}).}
    \label{fig:mechanism}
\end{figure*}

%% file: exhibits/notations.tex
\begin{table}[tb]
  \caption{Mathematical notations for pool mechanisms}
  \scriptsize
  \begin{tabularx}{\linewidth}{Xlll}
    \toprule
    \textbf{Notation}          & \textbf{Definition}                                               & \textbf{Applicable protocols} \\
    \midrule
    \multicolumn{3}{l}{\textit{Preset hyperparameters, $\Omega$}}                                                                  \\
    $w_k$                      & Weight of asset reserve $r_k$                                     & Balancer                      \\
    $\mathcal{A}$              & Slippage controller                                               & Uniswap V3, Curve, DODO       \\
    \revision{$n$}             & \revision{Number of assets in a pool ($                                                           %
        n
        \begin{cases}
          =2 \text{ for asset-pair pools} \\
          >2 \text{ for multi-asset pools}
        \end{cases}
      $)
    }                          & Curve                                                                                             \\
    \midrule
    \multicolumn{3}{l}{\textit{\revision{Conservation function invariants,} $\invt$}}                                               \\
    $\const$                   & Conservation function constant                                    & Uniswap V2, Balancer, Curve   \\
    $\revision{\mathcal{R}_k}$ & Initial reserve of token$_k$                                      & Uniswap V3, DODO              \\
    \midrule
    \multicolumn{3}{l}{\textit{State variables}}                                                                                   \\
    $r_k$                      & Quantity of token$_k$ in the pool                                 & all                           \\
    $p_k$                      & Current spot price of token$_k$                                   & all                           \\
    \midrule
    \multicolumn{3}{l}{\textit{Process variables}}                                                                                 \\
    $x_i$                      & Input quantity added to token$_i$ reserve (removed if $x_i<0$)    & all                           \\
    $x_o$                      & Output quantity removed from token$_o$ reserve (added if $x_o<0$) & all                           \\
    $\rho$                     & Token value change                                                & all                           \\
    \midrule
    \multicolumn{3}{l}{\textit{Functions}}                                                                                         \\
    $C$                        & Conservation function                                             & all                           \\
    $Z$                        & Implied conservation function                                     & all                           \\
    $_iE_o$                    & token$_o$ price in terms of token$_i$                             & all                           \\
    $S$                        & Slippage                                                          & all                           \\
    $V$                        & Reserve value                                                     & all                           \\
    $L$                        & Divergence loss                                                   & Uniswap, Balancer, Curve      \\
    \bottomrule
  \end{tabularx}
  \label{tab:token-exchange-notation}
\end{table}

%% file: exhibits/comparison_table.tex
\begin{table}[tbhp]
    \centering
    \caption{Comparison table of discussed \ac{dex}: value locked, trade volume of the past 7 days,  the market share by the last 30 days volume, the governance token, the number of governance token holders and the fully diluted value, as on 21/09/2021. Data retrieved from \href{https://defipulse.com/}{DeFi Pulse} and \href{https://duneanalytics.com/hagaetc/dex-metrics}{Dune Analytics}.}
    \tiny
    \centering
    \begin{tabularx}{\linewidth}{@{}Xrrrlrr@{}}
        \toprule
        \textbf{Protocol} & \textbf{Value locked (\$bn)} & \textbf{Trade volume (\$bn)} & \textbf{Market (\%)} & \textbf{Governance token}                                        & \textbf{Governance token holders} & \textbf{Fully diluted value (\$bn)}
        \\
        \midrule
        Uniswap           & 6.15                         & 11.4                         & 66.7                 & \tokenaddress{UNI}{0x1f9840a85d5af5bf1d1762f925bdaddc4201f984}   & 269,923                           & 21.1                                \\
        Sushiswap         & 3.92                         & 2.9                          & 14.2                 & \tokenaddress{SUSHI}{0x6b3595068778dd592e39a122f4f5a5cf09c90fe2} & 71,007                            & 2.4                                 \\
        Curve             & 11.64                        & 1.8                          & 6.4                  & \tokenaddress{CRV}{0xD533a949740bb3306d119CC777fa900bA034cd52}   & 44,654                            & 4.0                                 \\
        Bancor            & 1.37                         & 0.4                          & 2.5                  & \tokenaddress{BNT}{0x1f573d6fb3f13d689ff844b4ce37794d79a7ff1c}   & 38,124                            & 0.8                                 \\
        Balancer          & 1.74                         & 0.5                          & 2.2                  & \tokenaddress{BAL}{0xba100000625a3754423978a60c9317c58a424e3d}   & 37,613                            & 1.0                                 \\
        DODO              & 0.07                         & 0.4                          & 2.1                  & \tokenaddress{DODO}{0x43Dfc4159D86F3A37A5A4B3D4580b888ad7d4DDd}  & 11,330                            & 1.2                                 \\
        \bottomrule
    \end{tabularx}
    \label{tab:comparison}
\end{table}

%% file: sections/amm_properties.tex
\subsection{Key common properties of \ac{amm}-based \acp{dex}}
\label{sec:ammprop}

In this section, we summarize key \emph{common} properties featured in \ac{amm}-based \acp{dex}. We also clarify that protocol-specific intricacies and real-life implementations may result in \enquote{violations} of certain properties listed below.

\subsubsection{Zero-impact liquidity change}
\label{sec:non-impact}

The price of assets in an \ac{amm} pool stays constant for {\em pure}, balanced liquidity provision and withdrawal activities. This feature describes when an \ac{lp} provides or withdraw liquidity, usually by linearly scaling up or down the existing reserves in the pool, no price impact shall occur (see \autoref{eq:liquidityChange}).

The asset spot price can remain the same only when assets are added to or removed from a pool proportionate to the current reserve ratio ($r_1 : r_2 : ... : r_n$). In any other case, a change of quantities in any pool would result in changes in relative prices of assets. To manage to uphold the invariances a disproportionate addition or removal can be treated as a combination of two actions: proportionate reserve change plus asset swap (see e.g. \autoref{sec:balancer}).

\subsubsection{Path-deterministic swap}
\label{sec:path-binding}

By its algorithm-based pricing nature, how a given swap  transitions an \ac{amm} pool's reserve balance can be deterministically computed.

In an idealized, frictionless market, an \ac{amm} pool's conservation function (see \autoref{sec:form_cf}) stipulates that the pool's invariant stays constant for {\em pure} swapping activities (see \autoref{eq:trade}). Figuratively speaking, absent additional liquidity provision/withdrawal, the coordinates of the reserve quantities in a liquidity pool would always slide up and down along the bonding curve (see \autoref{fig:tokenexchange}) through swaps.
In reality, swap fees (see \autoref{sec:swapfee}), when kept within a pool, cause invariant $\invt$ to become variant through trading. Also, as float numbers are not yet fully supported by Solidity \cite{Ethereum2020}---the language for Ethereum smart contracts, \ac{amm} protocols typically recalculate invariant $\invt$ after each trade to avoid the accumulation of rounding errors.

\subsubsection{Output-boundedness}
\label{sec:outputbound}

With an output-bounded \ac{amm}, there is always a sufficient quantity of output tokens for a swap, i.e. a user can never deplete one side of the pool reserve. An \ac{amm} with this feature usually constructs its bonding curve such that, when one reserve token is close to depletion (approaching 0), its price---denominated in the other reserve token of the pool---becomes astronomically high (approaching infinity) \cite{Bartoletti2021}.

Output-boundedness usually applies to continuous \acp{amm}. Hybrid \acp{amm} such as Uniswap V3 which incorporates bounded bonding curves, assimilating an order-book-like mechanism \cite{Chitra2021}, naturally do not carry this property.

\subsubsection{Liquidity sensitivity}
\label{sec:liquid-sensitivity}

An \ac{amm} is liquidity-sensitive when a fixed swap size (same input quantity $x_i$) makes a larger price impact, i.e. higher slippage (see \autoref{sec:form_sl}), in a deep liquidity pool than a thin liquidity pool \cite{Wang2020AutomatedDeFi,Othman2011}.

\subsubsection{Demand sensitivity}
\label{sec:demand-sensitivity}

An \ac{amm} is demand-sensitive when the average swap price (i.e. the effective exchange rate $\frac{x_i}{x_o}$) increases as the swap size (input quantity $x_i$) increases. Intuitively, this suggests that as with the increment of the demand in output token, its price denominated input token will be driven up.

A constant product \ac{amm} is both liquidity-sensitive and demand-sensitive, whereas, strickly speaking, a constant sum one is neither.

%% file: sections/comparison.tex
\section{Comparison of AMM protocols}
\label{sec:comp}

\ac{amm}-based \acp{dex} are home to billions of dollars' worth of on-chain liquidity. \autoref{tab:comparison} lists major \ac{amm} protocols, their respective value locked, as well as some other general metrics. Uniswap is undeniably the biggest \ac{amm} measured by trade volume and the number of governance token holders, although it is remarkable that Curve has more value locked within the protocol. The number of governance token holders of smaller protocols as Bancor and Balancer is relatively high compared to \coin{CRV} token holders, as they do approximately a third of the volume but have only slightly fewer governance token holders.

\subsection{Major \ac{amm} protocols}
\label{sec:major-amm}

This section focuses on the four most representative \acp{amm}: Uniswap (including V2 and V3), Balancer, Curve, and DODO. These protocols were selected based on their market share~\cite{obadiaa2019} on the Ethereum blockchain and the representativeness in their overall mechanism.

We describe the liquidity pool structures of those protocols in the main text. We also derive the conservation function, slippage, as well as divergence loss of those protocols. A summary of formulas can be found in \autoref{tab:functioncomparison}.  We refer our readers to Appendix \ref{appendix:formulas} for a detailed explanation and derivation of those formulas.
The protocols' conservation function, slippage, as well as divergence loss under different hyperparameter values are plotted in \autoref{fig:consfunc}, \ref{fig:slippage} and \ref{fig:divloss}, respectively.
We always use token$_1$ as price or value unit; namely, token$_1$ is the assumed numéraire.

\input{exhibits/consfunc}
\input{exhibits/slippage}
\input{exhibits/divloss}

\subsubsection{Uniswap V2}


The Uniswap protocol prescribes that a liquidity pool always consists of one pair of assets. Uniswap V2 implements a conservation function with a constant-product invariant \revision{(see \autoref{sec:univ2conservation})}, implying that the reserves of the two assets in the same pool always have equal value.

\revision{%
    Liquidity provision or withdrawal at Uniswap V2 must be balanced and makes
    (\autoref{sec:non-impact}).
    Swaps with Uniswap V2 are path-deterministic (\autoref{sec:path-binding}); however, due to the positive swap fee charged and then immediately deposited back into the pool \cite{Uniswap2022},  a trading action can be decomposed into asset swap and liquidity provision. This action is, therefore, no longer a {\it pure} asset swap and would thus move the value of $\const$ \cite{Senchenko2020}.
    Uniswap V2 carries the properties of
    (\autoref{sec:outputbound}),
    (\autoref{sec:liquid-sensitivity})
    and
    (\autoref{sec:demand-sensitivity}).
}

\subsubsection{Uniswap V3}
\label{sec:uniswap-v3}

Uniswap V3 enhances Uniswap V2 by allowing liquidity provision to be concentrated on a fraction of the bonding curve \cite{adams2021v3core} (see \autoref{sec:univ3conservation}), thus virtually amplifying the conservation function invariant and reducing the slippage.

\revision{The protocol}'s slippage controller $\mathcal{A}$ determines the degree of liquidity concentration.
Specifically,
$\mathcal{A}$ signifies how
concentrated the liquidity should be provided around the initial spot price: when $\mathcal{A} \rightarrow \infty$, the covered price range approaches $(0, \infty)$, and the LP's individual conservation function approximates a Uniswap V2 one (\revision{approximated with} $\mathcal{A} = 10000$ in \autoref{fig:unicons}); on the other extreme, when $\mathcal{A} \rightarrow 1$, the liquidity only supports swaps close to the initial exchange rate, and the conservation function approximates a constant-sum one (\autoref{fig:unicons}).

\revision{%
    Like V2, liquidity provision or withdrawal at Uniswap V3 must respect the existing ratio between two reserve assets, which results in
    (\autoref{sec:non-impact}).
    Different from V2 where fees are retained in the pool, Uniswap V3 deducts swap fees as a fraction of the input asset and credit that amount to \ac{lp}'s fee revenue balance; hence, swaps with V3 are path deterministic with no impact on the bonding curve invariants
    (\autoref{sec:path-binding}).
    By design, Uniswap V3 is not
    (\autoref{sec:outputbound})---a sufficiently large swap can deplete one reserve asset and leave the liquidity pool only with the other one in (see \autoref{fig:unicons}).
    Uniswap V3 also features {liquidity sensitivity} (\autoref{sec:liquid-sensitivity}).
    Thanks to its liquidity concentrating feature, Uniswap V3 is less
    (\autoref{sec:demand-sensitivity}) than V2---a swap with a fixed input quantity would experience a lower slippage at Uniswap V3 than at V2 with the same level of pre-swap reserves (see \autoref{fig:uniswapslippage}).%
}

\subsubsection{Balancer}
\label{sec:balancer}

The Balancer protocol allows each liquidity pool to have more than two assets~\cite{martinelli2019whitepaper}. Each asset reserve $r_k$ is assigned a weight $w_k$ at pool creation, where $\sum\limits_k w_k = 1$. Weights are pool hyperparameters and do not change with either liquidity provision/\revision{withdrawal} or asset swap. The weight of an asset reserve represents the value of the reserve as a fraction of the pool value.
Balancer can also be deemed a generalization of Uniswap; the latter is a special case of the former with $w_1 = w_2 = \frac{1}{2}$ \revision{for asset-pair pools} (\autoref{fig:balcons}).

\revision{%
    Balancer allows both balanced liquidity provision/withdrawal as well as single-asset liquidity change \cite{martinelli2019whitepaper}; the former does not cause price impact, whereas the latter does
    (\autoref{sec:non-impact}).
    When only one asset is provided instead of e.g. eight, the protocol would first execute seven trades to swap this one asset to arrive at a vector of quantities in current proportions and next add this vector to the liquidity pool.
    Consequently, this sequence of actions is no longer a {\it pure} liquidity provision/withdrawal and would thus move the asset spot price.
    As Uniswap V2, swap fees with Balancer are also retained in the pool \cite{Balancer2022}, leading to an update of the bonding curve after each swap
    (\autoref{sec:path-binding}).
    Balancer also features
    (\autoref{sec:outputbound}),
    liquidity sensitivity (\autoref{sec:liquid-sensitivity})
    and
    demand sensitivity
    (\autoref{sec:demand-sensitivity}).%
}

\subsubsection{Curve}

With the Curve protocol, formerly StableSwap \cite{Egorov2019}, a liquidity pool \revision{typically} consists of two or more assets with the same peg, for example, \coin{USDC} and \coin{DAI}, or \coin{wBTC} and \coin{renBTC}.
Curve approximates Uniswap V2 when its constant-sum component (\autoref{sec:curve-cons-func}) has a near-0 weight, i.e. $\mathcal{A} \rightarrow 0$ (\autoref{fig:curvcons}).

\revision{%
    Like Balancer, Curve allows both proportionate and disproportionate liquidity change to the pool, depending on which the \ac{lp}'s action can induce either zero or some price impact
    (\autoref{sec:non-impact}).
    As Uniswap V2 and Balancer, Curve updates its invariant after each trade to account for swap fees retained in the pool
    (\autoref{sec:path-binding}).
    Curve is also
    (\autoref{sec:outputbound}),
    liquidity sensitive (\autoref{sec:liquid-sensitivity})
    and
    demand sensitive
    (\autoref{sec:demand-sensitivity}).
}

\subsubsection{DODO}
\label{sec:dodointro}

DODO supports customized pools \cite{DODO2021a}
where a pool creator provides reserves on both sides of the trading pair with arbitrary quantities, which determines the pool's initial {\em equilibrium state}.
Unlike conventional \acp{amm} such as Uniswap, Balancer and Curve where the exchange rate between two assets in a pool is derived purely from the conservation function, DODO does it the other way around. Resorting to external market data as a major determinant of the exchange rate, DODO has its conservation function (see \autoref{sec:dodoconservation}) derived from its exchange rate formula (see \autoref{sec:dodoexchange}).

Specifically, the pool exhibits an arbitrage opportunity---namely a gap between the price offered by the pool and that from the external market---as soon as the reserve ratio between the two assets in the pool deviates from its {\em equilibrium state}. Price alignment by arbitrageurs always pulls the reserve ratio back to its equilibrium state set by the \ac{lp}, thus eliminating any divergence loss.
Due to this feature, DODO differentiates itself from other \acp{amm} and terms their pricing algorithm as \enquote{\acl{pmm}}, or \acs{pmm}.

In DODO, a higher slippage controller $\mathcal{A} \in (0,1)$ results in a greater slippage around the market price---i.e. the equilibrium price.
Specifically,
when $\mathcal{A} \rightarrow 1$, the DODO bonding curve resembles Uniswap V2 (\revision{approximated with} $\mathcal{A} = 0.99$ in \autoref{fig:dodocons}); and with a high slippage around the market price (\autoref{fig:dodolippage}), the pool exhibits a strong tendency to fall back to the equilibrium state.
When $\mathcal{A} \rightarrow 0$, the DODO bonding curve resembles a constant sum one (\revision{approximated with} $\mathcal{A} = 0.01$ in \autoref{fig:dodocons}); and with a near-flat slippage (\autoref{fig:dodolippage}), the algorithm's force to pull the reserve ratio back to equilibrium is at its weakest due to little arbitrage profitability exhibited.

\revision{%
    As with Balancer and Curve, \acp{lp} can provide/withdraw both balanced and unbalanced reserves to a DODO pool
    (\autoref{sec:non-impact}).
    Similar to Uniswap V3, swap fees with DODO are recorded outside of the liquidity pool. DODO's bonding curve is thus redrawn not after each swap, but after each price change reported by the oracle
    (\autoref{sec:path-binding}).
    DODO is also
    (\autoref{sec:outputbound}).
    Although relying on external price feeds for the construction of its conservation function, DODO is still
    liquidity sensitive (\autoref{sec:liquid-sensitivity})
    and
    demand sensitivite
    (\autoref{sec:demand-sensitivity})
    since the size of a trade relative to the pool depth determines the magnitude of slippage and the price movement local to the pool.
}

%% file: exhibits/consfunc.tex
\begin{figure}[tbhp]
  \centering
  \begin{subfigure}[b]{0.248\linewidth}
    \centering
    \includegraphics[height=0.113\paperheight, trim={20 20 20 20}]{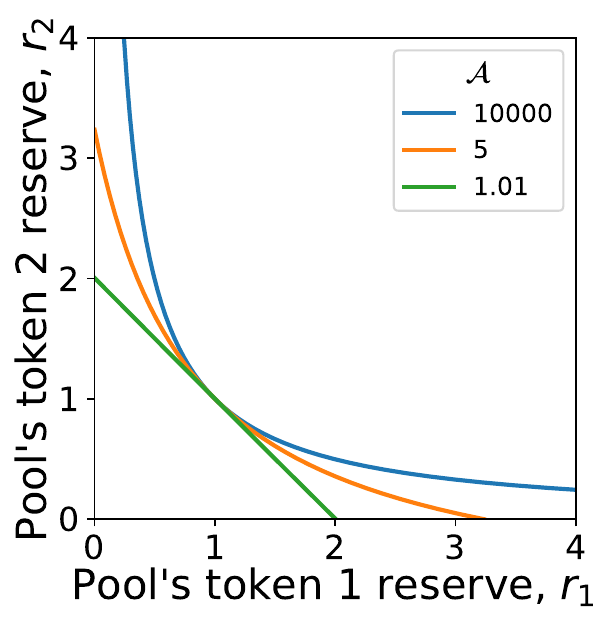}
    \caption{\scriptsize Uniswap, \autoref{eq:unicons}}
    \label{fig:unicons}
  \end{subfigure}%
  \begin{subfigure}[b]{0.248\linewidth}
    \centering
    \includegraphics[height=0.113\paperheight, trim={20 20 20 20}]{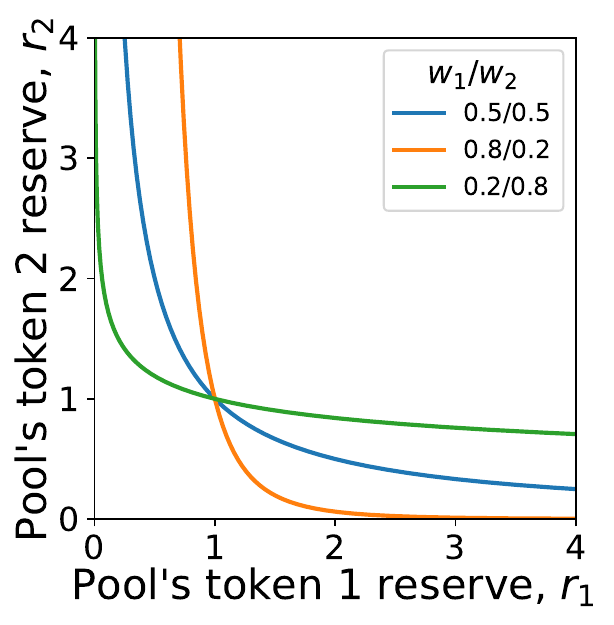}
    \caption{\scriptsize Balancer, \autoref{eq:balcons}}
    \label{fig:balcons}
  \end{subfigure}%
  \begin{subfigure}[b]{0.248\linewidth}
    \centering
    \includegraphics[height=0.113\paperheight, trim={20 20 20 20}]{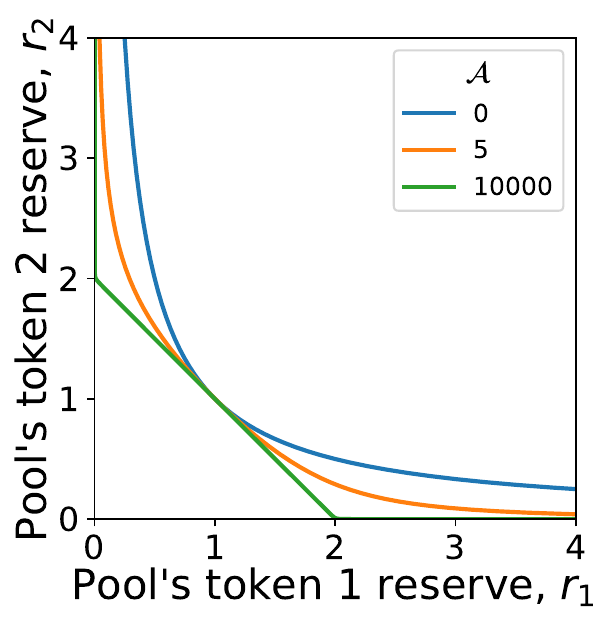}
    \caption{\scriptsize Curve, \autoref{eq:curvcons}}
    \label{fig:curvcons}
  \end{subfigure}%
  \begin{subfigure}[b]{0.248\linewidth}
    \centering
    \includegraphics[height=0.113\paperheight, trim={20 20 20 20}]{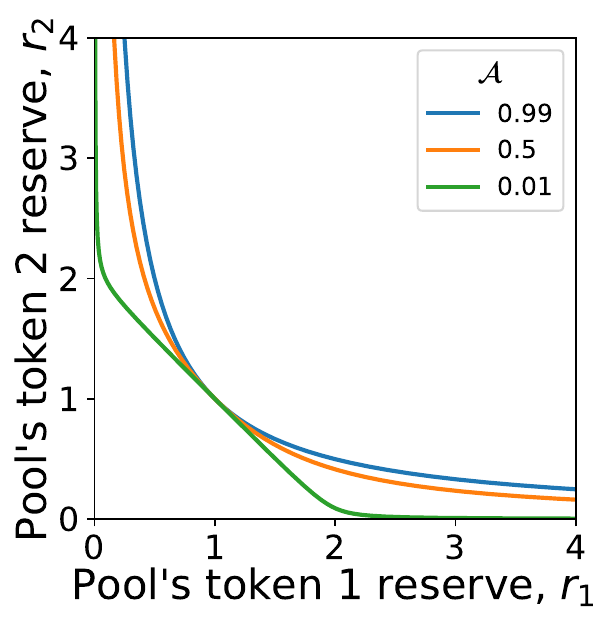}
    \caption{\scriptsize DODO, \autoref{eq:dodocons}}
    \label{fig:dodocons}
  \end{subfigure}
  \caption{Conservation function (see \autoref{sec:form_cf}) of \acp{amm} with initial reserves of token$_1$ and token$_2$ both equal to 1, namely, $\const = \revision{\mathcal{R}_1} = \revision{\mathcal{R}_2} = 1$.}
  \label{fig:consfunc}
\end{figure}

%% file: exhibits/slippage.tex
\begin{figure}[tbhp]
  \centering
  \begin{subfigure}[b]{0.248\linewidth}
    \centering
    \includegraphics[height=0.113\paperheight, trim={20 20 20 20}]{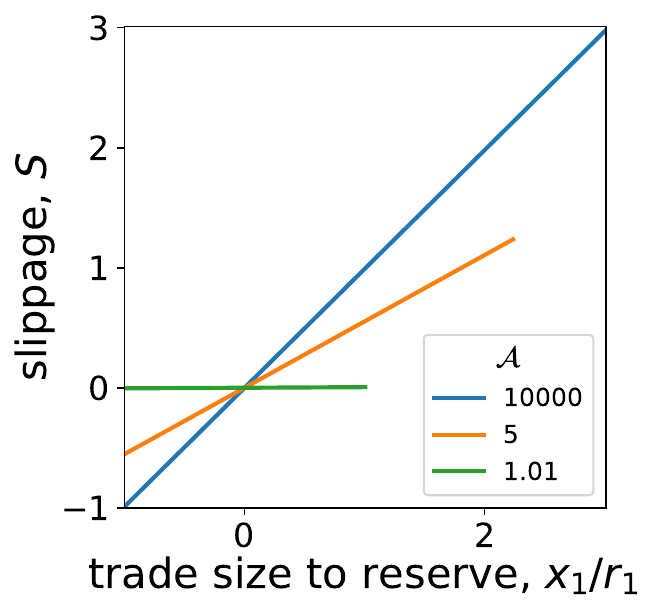}
    \caption{\scriptsize Uniswap, \autoref{eq:uniswapv3slippage}}
    \label{fig:uniswapslippage}
  \end{subfigure}%
  \begin{subfigure}[b]{0.248\linewidth}
    \centering
    \includegraphics[height=0.113\paperheight, trim={20 20 20 20}]{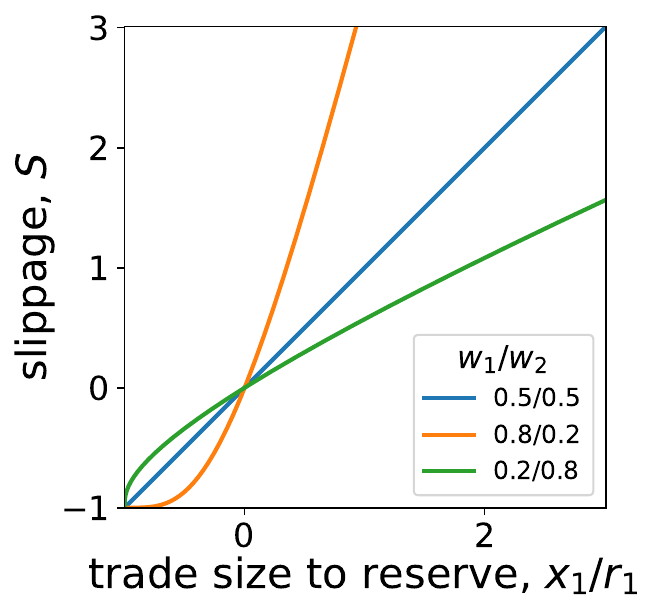}
    \caption{\scriptsize Balancer, \autoref{eq:balancerslippage}}
    \label{fig:balancerslippage}
  \end{subfigure}%
  \begin{subfigure}[b]{0.248\linewidth}
    \centering
    \includegraphics[height=0.113\paperheight, trim={20 20 20 20}]{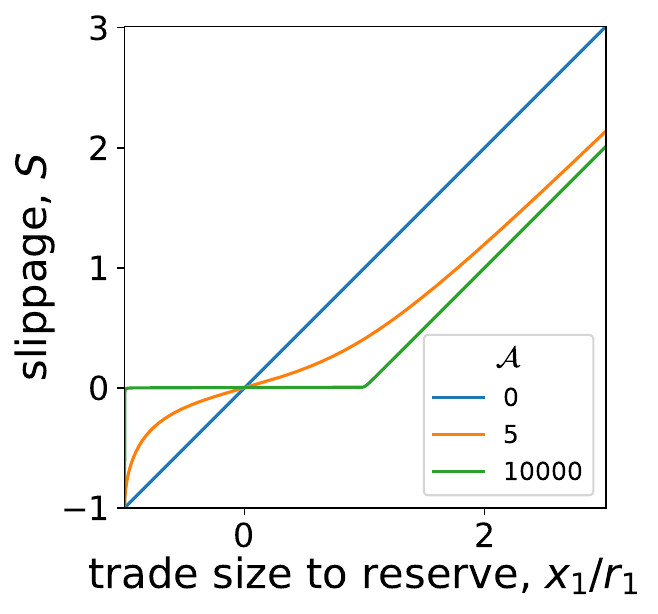}
    \caption{\scriptsize Curve, \autoref{eq:curveslippage}}
    \label{fig:curveslippage}
  \end{subfigure}%
  \begin{subfigure}[b]{0.248\linewidth}
    \centering
    \includegraphics[height=0.113\paperheight, trim={20 20 20 20}]{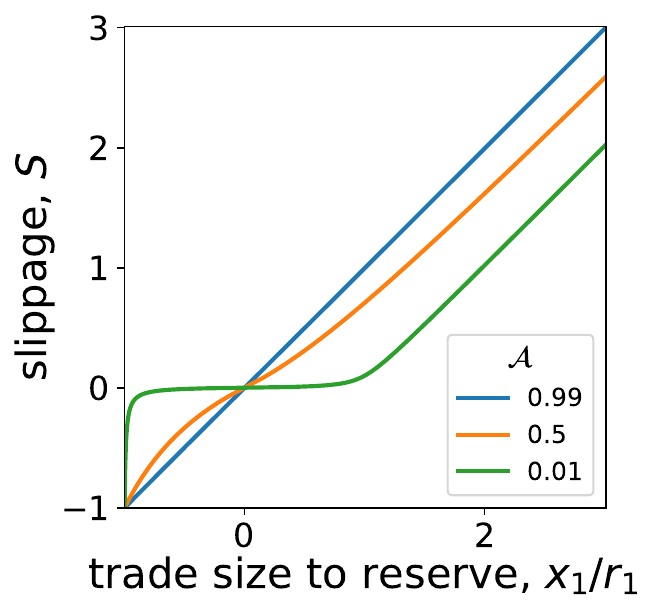}
    \caption{\scriptsize DODO, \autoref{eq:dodoslippage}}
    \label{fig:dodolippage}
  \end{subfigure}
  \caption{Slippage (see \autoref{sec:form_sl}) of AMMs, depicted with $\frac{x_1}{r_1} \in [-1,3]$, corresponding to the after-trade token$_1$ reserve quantity $r_1 \in [0, 4]$, which is the x-axis of \autoref{fig:consfunc}.}
  \label{fig:slippage}
\end{figure}

%% file: exhibits/divloss.tex
\begin{figure}[tbhp]
  \centering
  \begin{subfigure}[b]{0.248\linewidth}
    \centering
    \includegraphics[height=0.113\paperheight, trim={20 20 20 20}]{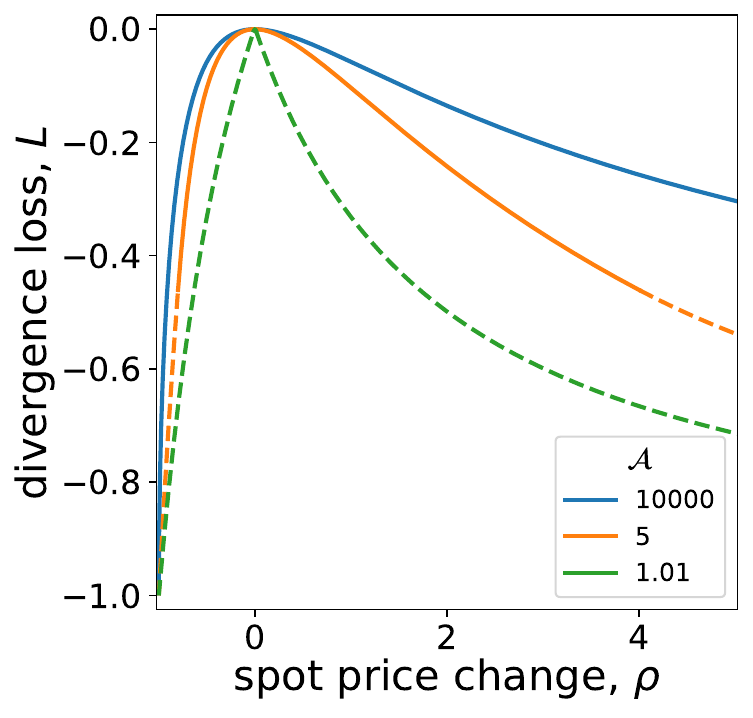}
    \caption{\scriptsize Uniswap, \autoref{eq:uniswapv3div}}
    \label{fig:uniswapdiv}
  \end{subfigure}%
  \begin{subfigure}[b]{0.248\linewidth}
    \centering
    \includegraphics[height=0.113\paperheight, trim={20 20 20 20}]{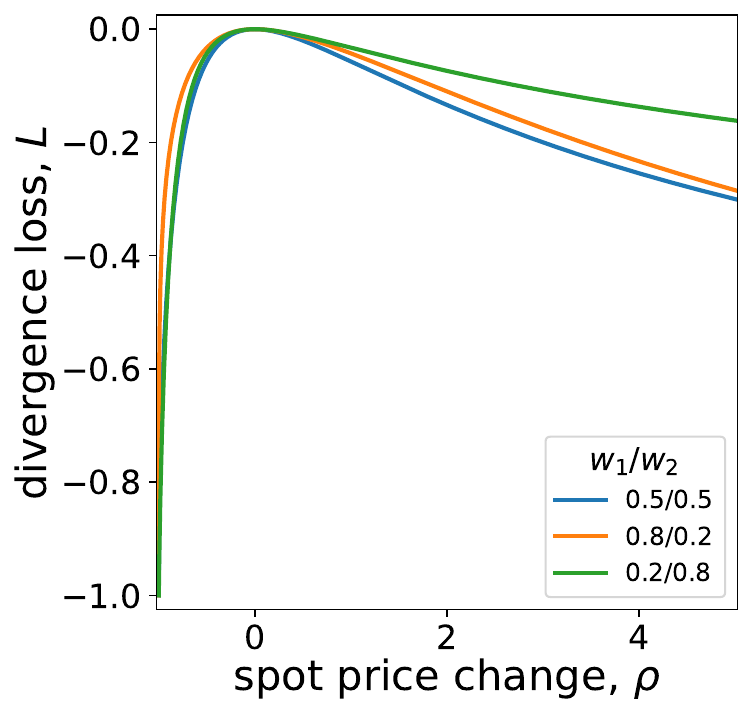}
    \caption{\scriptsize Balancer, \autoref{eq:balancerdiv}}
    \label{fig:balancerdiv}
  \end{subfigure}%
  \begin{subfigure}[b]{0.248\linewidth}
    \centering
    \includegraphics[height=0.113\paperheight, trim={20 20 20 20}]{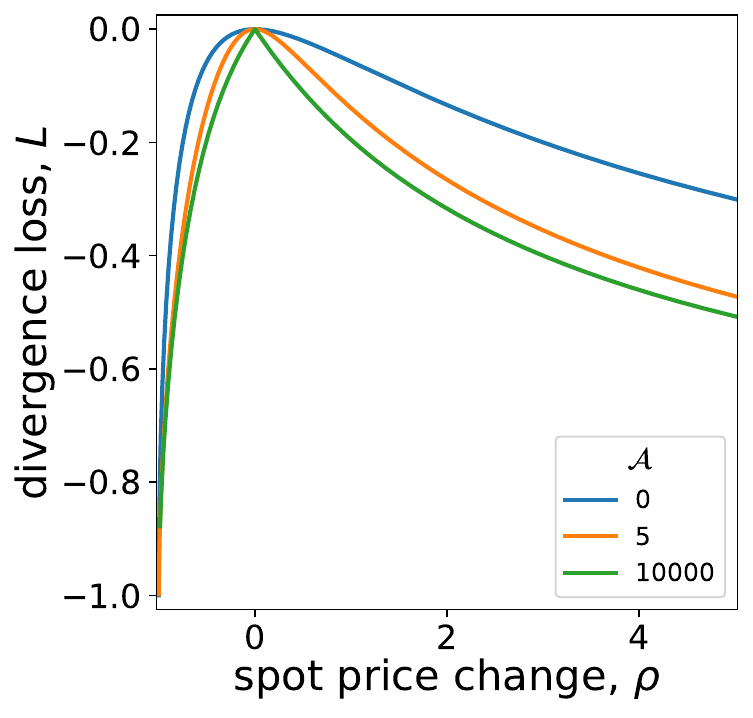}
    \caption{\scriptsize Curve, \autoref{eq:curvediv}}
    \label{fig:curvediv}
  \end{subfigure}%
  \begin{subfigure}[b]{0.248\linewidth}
    \centering
    \includegraphics[height=0.113\paperheight, trim={20 20 20 20}]{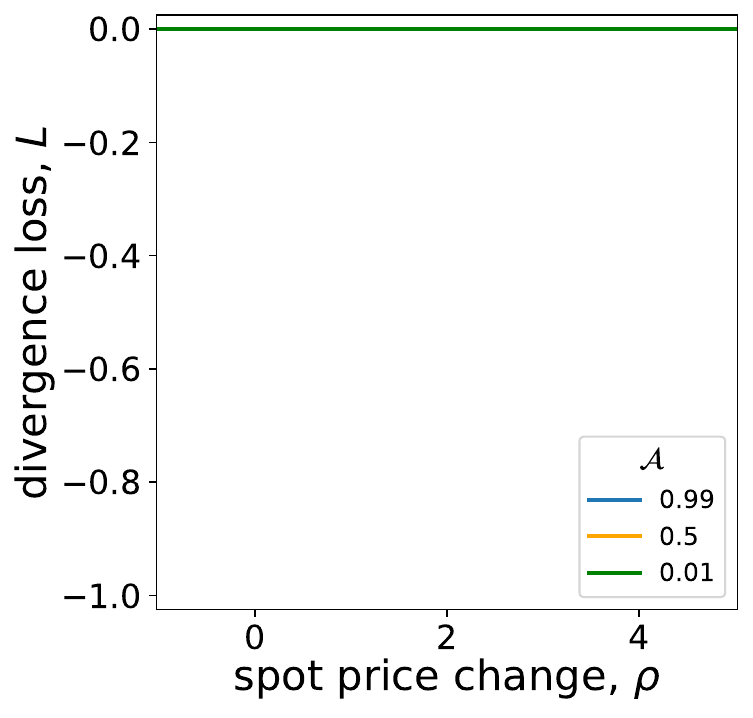}
    \caption{\scriptsize DODO, $L \equiv 0$ at equilibrium}
    \label{fig:dododiv}
  \end{subfigure}
  \caption{Divergence loss (see \autoref{sec:form_dl}) of AMMs}
  \label{fig:divloss}
\end{figure}

%% file: sections/otherAMMprotocols.tex
\subsubsection{Other \ac{amm}-based \ac{dex}s}
\label{subsec:other}

\paragraph{Sushiswap}
\label{subsubsec:sushi}

Sushiswap is a fork of Uniswap V2 (see \autoref{sec:vampireattack}). Though the two mainly differ in governance token structure and user experience, Sushiswap share the same conservation function, slippage and divergence loss functions as Uniswap.

\paragraph{Kyber Network}
Currently in its 3.0 version, the \ac{dex} uses a Dynamic Market Maker (DMM) mechanism, which allows for dynamic conservation functions based on amplified balances, called \enquote{virtual balances} \cite{nguyen2021damm}. This is supposed to result in higher capital efficiency for \acp{lp} and better slippage for traders. Also, the trading fees are adjusted automatically to market conditions. A volatile market causes increased fees, to offset impermanent loss for \acp{lp}.



\paragraph{Bancor}

While Bancor's white paper \cite{Hertzog2018} gives the impression that a different conservation function is applied, a closer inspection of their transaction history and smart contract leads to the conclusion that Bancor is using the same formula as Balancer (confirmed by a developer in the Bancor Discord community). As the majority of Bancor pools consist of two assets, one of which is usually \coin{BNT}, with the reserve weights of 50\%--50\%, Bancor's swap mechanism is equivalent to Uniswap. Bancor V2.1 now allows single-sided asset exposure, and provides divergence loss insurance \cite{bancor2020v2} (see \autoref{sec:dli}).

%% file: sections/discussion.tex


Each \ac{amm} has its quirks. Uniswap V2 implements an rudimentary bonding curve that achieves a low gas fee; Uniswap V3 allows for concentrated liquidity provision which improves capital efficiency; Balancer supports more than 2 assets in a pool; Curve is suitable for swapping assets with the same peg; DODO proactively reduces divergence loss by leveraging external price feeds.
\revision{
    As discussed in \autoref{sec:major-amm}, common \acp{amm} can be predominantly seen as a generalization, or an extension, of the most fundamental constant-product protocol that is applied by Uniswap V2.
}

\revision{
    When hyperparameters such as reserve weights $w_k$ and slippage controller $\mathcal{A}$ are assigned with certain values, various \acp{amm} can be reduced to the basic form equivalent to Uniswap V2 (illustrated with {\color{pythonblue} blue curves} in \autoref{fig:consfunc}, \ref{fig:slippage} and \ref{fig:divloss}).
    In fact, the majority of top \acp{amm}---including Sushiswap, PancakeSwap, VVS Finance, Quickswap and BiSwap---are a simple clone of the Uniswap protocol \cite{DefiLlama} with some adjustment in the fee and reward structure.}

\revision{When deciding on a new conservation function, \ac{amm} developers and designers must consider the trade-off between different features and properties (\autoref{sec:ammprop}).
    For example,
    seeking liquidity insensitivity (\autoref{sec:liquid-sensitivity}) and demand insensitivity (\autoref{sec:liquid-sensitivity}) for low slippage leads to higher divergence loss (see \autoref{sec:implicitcosts}): given a range of price movement, traders would be able to swap out an asset with a larger quantity, whereas \acp{lp} would suffer a bigger divergence loss.
    In the extreme case like Uniswap V3, the trader-favoring feature sacrifices the output-bounded property (\autoref{sec:outputbound}), which is to the detriment of \acp{lp}, leaving them completely \enquote{rekt}\footnote{\Ac{defi} jargon for \enquote{wrecked}, in this context, meaning exposed to a single, undiversified asset has has depreciated in value.} in the case of significant price swings.
    Seemingly capable of achieving both low slippage and zero divergence loss at equilibrium by setting its $\mathcal{A}$ low, DODO appears to be an exception. Nevertheless, it is to be noted that with a small $\mathcal{A}$, DODO's \ac{pmm} algorithm is less effective in restoring the pool to its equilibrium state (see \autoref{sec:dodointro}), thus still exposing \acp{lp} to divergence loss risks in non-equilibrated states.
}

In a similar vein, users interacting with an \ac{amm}-based \ac{dex}, including both traders and \acp{lp}, form a zero-sum game. They should understand the protocol design, and beware of embedded hidden costs such as slippage and divergence loss, which impose economic risks on their funds.

%% file: sections/additionalAMMfeatures.tex
\subsection{Additional features of \ac{amm}-based \acp{dex}}

\subsubsection{Time component}

A time component refers to the ability to change traditionally fixed hyperparameters over time. Balancer V1 and V2 implement this (\autoref{tab:overview}), by allowing liquidity pool creators to set a scheme that changes the weights of two pool assets over time. This implementation is called a Liquidity Bootstrapping Pool (see \autoref{sec:balancerlbp}).

\subsubsection{Dynamic swap fee}
\label{sec:dynamicfee}

Dynamic fees are introduced by Kyber 3.0 to reduce the impact of divergence loss for \acp{lp}. The idea is to increase swap fees in high-volume markets and reduce them in low-volume markets. This should result in more protection against divergence loss, as during periods of sharp token price movements during a high-volume market, \acp{lp} absorb more fees. In low-volume and -volatility markets, trading is encouraged by lowering the fees.

\subsubsection{Divergence loss insurance}
\label{sec:dli}

Popularized by Bancor V2.1, \acp{lp} are insured against divergence loss after 100 days in the pool, with a 30-day cliff at the beginning. Bancor achieves this by using an elastic \coin{BNT} supply that allows the protocol to co-invest in pools and pay for the cost of impermanent loss with swap fees from its co-investments \cite{bancor2021ilprotection}. This insurance policy is earned over time, 1\% each day that liquidity is staked in the pool.

\input{exhibits/descriptive_table}

%% file: exhibits/descriptive_table.tex
\begin{table*}[tbhp]

   \setlength{\tabcolsep}{3.7pt}
   \centering
   \caption{Overview of major existing \ac{amm}-based \ac{dex} on Ethereum, Solana, Polkadot, Tezos, EOS, Polygon and \revision{BNB Chain}. CP: Constant product, CS: Constant sum, OP: Oracle price component, CC: Capital concentration, T: Time component.%
   }
   \tiny
   \begin{tabularx}{\linewidth}{@{}X@{\hspace{1pt}}llccccclllll@{}}
      \toprule
                            &                                 &             & \multicolumn{3}{c}{\ac{amm}} & \multicolumn{3}{c}{\ac{amm} add-ons} &             &            &
      \\
      \cmidrule(lr){4-6} \cmidrule(lr){7-9}

      \textbf{\ac{dex}}     &
                            & \textbf{Pool structure}         & \textbf{CP} & \textbf{CS}                  & \textbf{OP}                          & \textbf{CC} & \textbf{T} & \textbf{Divergence loss compensation} & \textbf{Chain}            & \textbf{Mainnet launch}                  \\
      \midrule
      Uniswap V1            & \cite{Adams2018}                & asset-pair  & \CIRCLE                      & \Circle                              & \Circle     & \Circle    & \Circle                               & ---                       & Ethereum                       & 11/2018 \\
      Uniswap V2            & \cite{Core2020}                 & asset-pair  & \CIRCLE                      & \Circle                              & \Circle     & \Circle    & \Circle                               & ---                       & Ethereum                       & 05/2020 \\
      Uniswap V3            & \cite{adams2021v3core}          & asset-pair  & \CIRCLE                      & \Circle                              & \Circle     & \CIRCLE    & \Circle                               & ---                       & Ethereum                       & 05/2021 \\
      Balancer V1           & \cite{martinelli2019whitepaper} & multi-asset & \CIRCLE                      & \Circle                              & \Circle     & \Circle    & \CIRCLE                               & ---                       & Ethereum                       & 03/2020 \\
      Balancer V2           & \cite{martinelli2021balancerv2} & multi-asset & \CIRCLE                      & \Circle                              & \Circle     & \Circle    & \CIRCLE                               & ---                       & Ethereum                       & ---     \\
      Curve                 & \cite{Egorov2019}               & multi-asset & \CIRCLE                      & \CIRCLE                              & \Circle     & \Circle    & \Circle                               & ---                       & Ethereum                       & 01/2020 \\
      DODO                  & \cite{dodo2020whitepaper}       & various     & \CIRCLE                      & \Circle                              & \CIRCLE     & \Circle    & \Circle                               & ---                       & Ethereum, \revision{BNB Chain} & 09/2020 \\
      Bancor V1             & \cite{Hertzog2018}              & asset-pair  & \CIRCLE                      & \Circle                              & \Circle     & \Circle    & \Circle                               & ---                       & Ethereum, EOS                  & 06/2017 \\
      Bancor V2             & \cite{bancor2020v2}             & asset-pair  & \CIRCLE                      & \Circle                              & \CIRCLE     & \Circle    & \Circle                               & ---                       & Ethereum, EOS                  & 04/2020 \\
      Bancor V2.1           & \cite{Bancor2020}               & asset-pair  & \CIRCLE                      & \Circle                              & \Circle     & \Circle    & \Circle                               & Divergence loss insurance & Ethereum, EOS                  & 10/2020 \\
      SushiSwap             & \cite{sushi2020}                & asset-pair  & \CIRCLE                      & \Circle                              & \Circle     & \Circle    & \Circle                               & ---                       & Ethereum                       & 08/2020 \\
      Mooniswap             & \cite{mooniswap2020whitepaper}  & asset-pair  & \CIRCLE                      & \Circle                              & \Circle     & \Circle    & \CIRCLE                               & ---                       & Ethereum                       & 08/2020 \\
      mStable               & \cite{Andersson2020}            & asset-pair  & \Circle                      & \CIRCLE                              & \Circle     & \Circle    & \Circle                               & ---                       & Ethereum                       & 07/2020 \\
      Kyber 3.0             & \cite{Kyber2021whitepaper}      & multi-asset & \CIRCLE                      & \Circle                              & \Circle     & \CIRCLE    & \Circle                               & Dynamic swap fee          & Ethereum, Tezos                & 03/2021 \\
      Saber                 & \cite{saber2021}                & multi-asset & \CIRCLE                      & \CIRCLE                              & \Circle     & \Circle    & \Circle                               & ---                       & Solana                         & 06/2021 \\
      HydraDX               & \cite{hydradx2021}              & multi-asset & \CIRCLE                      & \Circle                              & \CIRCLE     & \CIRCLE    & \Circle                               & ---                       & Polkadot                       & ---     \\
      Uranium Finance       & \cite{Uranium.finance}          & asset-pair  & \CIRCLE                      & \Circle                              & \Circle     & \Circle    & \Circle                               & ---                       & \revision{BNB Chain}           & 05/2021 \\
      QuickSwap             & \cite{QuickSwapOfficial2020}    & asset-pair  & \CIRCLE                      & \Circle                              & \Circle     & \Circle    & \Circle                               & ---                       & Polygon                        & 10/2020 \\
      \revision{Burgerswap} & \cite{burgerswap2022}           & asset-pair  & \CIRCLE                      & \Circle                              & \Circle     & \Circle    & \Circle                               & ---                       & BNB Chain                      & 10/2020 \\

      \bottomrule
   \end{tabularx}
   \label{tab:overview}
\end{table*}

%% file: sections/implementations.tex
\subsection{Other \ac{defi} protocols with \ac{amm} implementations}
\label{subsec:implement}

\acp{amm} form the basis of other \ac{defi} applications (see \autoref{fig:taxonomy}) that implement existing or invent newly designed bonding curves, facilitating the functionalities of these implementing protocols. In this section, we present a few examples of projects that use \ac{amm} designs under the hood.

\subsubsection{Gyroscope}

Gyroscope \cite{gyroscope2021} is a stablecoin backed by a reserve portfolio that tries to diversify \ac{defi} tail risks. Gyro Dollars can be minted for a price near \$1 and can be redeemed for around \$1 in reserve assets, as determined through a new \ac{amm} design that balances risk in the system.
Gyroscope includes a Primary-market \ac{amm} (P-\ac{amm}), through which Gyro Dollars are minted and redeemed, and a Secondary-market \ac{amm} (S-\ac{amm}) for Gyro Dollar trading.
Similar to Uniswap V3 where a price range constraint is imposed, the P-\ac{amm} yields a mint quote and a redeem quote that serves as a price range constraint for the S-\ac{amm} to decide upon concentrated liquidity ranges~\cite{gyroscope2021amm}.

\subsubsection{EulerBeats}

EulerBeats \cite{eulerbeats2021} is a protocol that issues limited edition sets of algorithmically generated art and music, based on the Euler number and Euler totient function. The project uses self-designed bonding curves to calculate burn prices of music/art prints, depending on the existing supply. The project thus implements a form of \ac{amm} to mint and burn NFTs price-efficiently.

\subsubsection{Pods Finance}

Pods \cite{pods2021} is a decentralized non-custodial options protocol that allows users to create calls and or puts and trade them in the Options \ac{amm}. Users can participate as sellers and buy puts and calls in a liquidity pool or act as \acp{lp} in such a pool. The specific \ac{amm} is one-sided and built to facilitate an initially illiquid options market and price option algorithmically using the Black-Scholes pricing model. Users can effectively earn fees by providing liquidity, even if the options are out-of-the-money, reducing the cost of hedging with options.

\subsubsection{Balancer \acf{lbp}}
\label{sec:balancerlbp}

\ac{lbp} are pools where controllers can change the parameters of the pool in controlled ways, unlike immutable pools described in \autoref{sec:comp}. The idea of an \ac{lbp} is to launch a token fairly, by setting up a two-token pool with a project token and a collateral token. The weights are initially set heavily in favor of the project token, then gradually \enquote{flip} to favor the collateral coin by the end of the sale. The sale can be calibrated to keep the price more or less steady (maximizing revenue) or declining to the desired minimum (e.g., the initial offering price) \cite{balancer2021lbp}.

\subsubsection{YieldSpace}

The YieldSpace paper \cite{Niemerg2020} introduces an automated liquidity provision for fixed yield tokens. A formula called the \enquote{constant power sum invariant} incorporates time to maturity as input and ensures that the liquidity provision offers a constant interest rate---rather than price---for a given ratio of its reserves. \coin{fyTokens} are synthetic tokens that are redeemable for a target asset after a fixed maturity date \cite{robinson2020yield}. The price of a \coin{fyToken} floats freely before maturity, and that price implies a particular interest rate for borrowing or lending that asset until the \coin{fyToken}'s maturity. Standard \ac{amm} protocols as discussed in \autoref{sec:comp} are capital-inefficient. By introducing the concept of a constant power sum formula, the writers want to build a liquidity provision formula that works in \enquote{yield space} instead of \enquote{price space}.

\subsubsection{Notional Finance}

Notional Finance \cite{notionalfinance2021} is a protocol that facilitates fixed-rate, fixed-term crypto-asset lending and borrowing. Fixed interest rates provide certainty and minimize risk for market participants, making this an attractive protocol among volatile asset prices and yields in DeFi. Each liquidity pool in Notional refers to a maturity, holding \coin{fCash} tokens attached to that date. For example, \coin{fDai} tokens represent a fixed amount of \coin{DAI} at a specific future date. The shape of the Notional \ac{amm} follows a logit curve, to prevent high slippage in normal trading conditions. Three variables parameterize the AMM: the scalar, the anchor, and the liquidity fee \cite{notional2020amm}. The first and second mentioned allowing for variation in the steepness of the curve and its position in a xy-plane, respectively. By converting the scalar and liquidity fee to a function of time to maturity, fees are not increasingly punitive when approaching maturity.

\subsubsection{Gnosis \ac{cmm}}

The Gnosis \ac{cmm} \cite{gnosiscmm} allows users to set multiple limit orders at custom price brackets and passively provide liquidity on the Gnosis Protocol. The mechanism used is similar to the Uniswap V3 structure, although it allows for even more possibilities to market makers by allowing them to choose price upper and lower limits and a number of brackets within that price range. Uniswap V3 allows \acp{lp} to solely choose the upper and lower limits. Because users deposit funds to the assets at different price levels specifically, the protocol behaves more like a central limit order book than an \ac{amm} pool.

\subsubsection{\ac{dex} aggregators}
\label{sec:dex-agg}
\revision{\ac{dex} aggregators are a type of emerging \ac{defi} protocols that connect to various other \acp{dex} and can also have their own liquidity pools \cite{Ushida2021}. They offer traders superior swap rates through routing across liquidity pools from different \acp{dex} with one single user interface \cite{Raikwar2021}. 1inch and Paraswap are two major \acp{dex} aggregators for \ac{evm}-compatible chains that incorporate \acp{amm} such as Uniswap and Curve. Rango \cite{Rango2022} is an example of cross-chain \acp{dex} aggregators incorporate \acp{amm}, \ac{dex} aggregators and bridges to facilitate token swaps across both \ac{evm} and non-\ac{evm} blockchains.
}

%% file: sections/layer2.tex
\subsection{\acp{amm} on Layer 2 solutions}
\label{sec:layer2}

The growth and success of \ac{defi} on Ethereum have put a strain on the Ethereum network's ability to process transactions, leading to increasing gas \cite{ethereumgasprice2021}. As the Ethereum Network becomes busier, user experience decreases because of increasing gas prices and decreasing transaction speed. Users aim to outbid each other by increasing the gas prices. Also, transaction speed decreases, which results in poor user experience for certain types of \acp{dapp}. And as the network gets busier, gas prices increase as transaction senders aim to outbid each other. In an attempt to prevent these consequences, \enquote{layer 2 solutions} (L2) are being developed. These solutions handle transactions outside of the Ethereum network, but still rely on the decentralized security model of the mainnet \cite{ethereum2021layertwo}. Examples of layer 2 technologies include Plasma, Sidechains, Optimistic Rollups and ZK-Rollups. For a more comprehensive reading on this topic, we direct the reader to \cite{ethereum2021scaling}. Examples of Ethereum layer 2 solutions are Polygon \cite{polygon2021}, Arbitrum \cite{arbitrum2021}, Optimism \cite{optimism2021} and Starknet \cite{starknet2021}.

Characteristics of layer 2 solutions, such scaling and security, have been well-documented across different sources \cite{gudgeon2020layer2} \cite{hafid2020scaling} \cite{jourenko2019layer2} and are out of scope for this \ac{sok}. The advantages and disadvantages of transacting on layer 2 solutions are not specifically related to \acp{amm}, but to all protocols on these technologies. Therefore, we focus on the security issues and user experience of interacting with \acp{amm} on layer 2.

Daian et al. \cite{daian2020flash} note that abstraction achieved by layer 2 exchange systems is not sufficient to prevent sandwich attacks and a report of Delphi Digital \cite{delphi2020rollups} concludes that there is still front-running risk when a protocol wants to aggregate liquidity across layer 1 and layer 2 pools. The front-running and sandwich attacks does not seem to be solved by layer 2 solutions. \cite{konstantopoulos2021rollups} proposes a simple solution for front-running attacks on Optimistic Rollups technology.

One of the most important advantages of deploying a protocol on layer 2 is the reduction in gas fees. This dramatically enhances the user experience, and opens up ways to introduce new forms of decentralized exchanges. One example is ZKSwap \cite{zkswap2021}, allowing users to trade with zero gas fees. Sushiswap and Curve have deployed their contracts on Polygon and QuickSwap is a fork of Uniswap on that same layer 2 solution \cite{nasdaq2021layer2}. As a result, users are now able to use the same products on layer 2 solutions, with drastically reduces costs and allows faster transactions. In 2021, dYdX launched its order book-based \acp{dex} on StarkEx \cite{dydx2021layer2}.

In sum, \acp{amm} on layer 2 solutions result in faster transactions and a reduction of costs due to zero or decreased gas costs, ultimately enhancing the user experience.

%% file: sections/securityprivacy.tex
\section{Security and privacy concerns}
\label{sec:security}

The previous sections focus on implicit economic costs---including slippage and divergence loss---imposed on the funds of users interacting with \ac{amm}-based \acp{dex}. Besides those risks, security and privacy matters are also to be taken into account when using \ac{amm}-based \acp{dex}.

In particular, as a complex, distributed system with a variety of software and hardware components interacting with each other, \ac{amm}-based \acp{dex} are prone to exhibit attack interfaces \cite{Li2020ASystems,Lin2017AChallenges.,Zhang2019SecurityBlockchain,Massacci2021}.
With conventional exchanges, the success of market manipulation is uncertain as each trade must be agreed upon between the sell and buy sides. In contrast, \ac{amm}-based \acp{dex} are subject to
atomic, risk-free exploits on the protocol's technical structure such as its algorithmic pricing scheme~\cite{werner2020sokDeFi}.
Built on top of public blockchain infrastructures featuring transparency and traceability, \ac{amm}-based \acp{dex} also expose their users to privacy risks.

In this section, we define a taxonomy (illustrated in \autoref{fig:security_taxonomy}) to enumerate potential security and privacy concerns of \ac{amm}-based \acp{dex},
expounding their root causes and possible mitigation solutions.

\input{exhibits/security_taxonomy}

\subsection{Associated attacks}
\label{sec:associated_attacks}
We identify three classes of attacks according to the architectural layer on which they occur: infrastructure-layer attacks, middleware-layer attacks, and application-layer attacks. Sometimes, a certain attack (e.g. frontrunning) can target multiple layers simultaneously.
We present known historical attacks affecting \acp{amm} in \autoref{tab:attack}.

\subsubsection{Infrastructure-layer attacks}
\label{sec:infraattack}

\revision{The proper operation of \acp{dex} are based upon healthy and stable blockchain infrastructures} (\ie validators, network, full nodes, etc.).
However, since the birth of the blockchain systems, various attacks have threatened their normal operations,
potentially affecting the robustness and user experience of \acp{dex}.

\paragraph{Block timestamp manipulation}
\label{sec:blockseq}
A timestamp field is set by miners during the validation process.
However, malicious miners can manipulate the block timestamps within constraints to win rewards from certain smart contracts~\cite{CryptoMarketPool2020BlockAttack}, or to tamper with the execution order of \ac{dex} transactions packed in different blocks \cite{Huang2019}.

To mitigate the negative impact of such manipulation, \ac{dex} contracts should be timestamp independent~\cite{Antonopoulos2018MasteringDapps}. For example, smart contract engineers should avoid using block timestamps as program inputs or make sure a contract function can tolerate variations by a certain time period (e.g. 15 seconds~\cite{Narayanan2016BitcoinIntroduction}) and still maintain integrity~\cite{Mense2018}. Besides, \acp{dex} should choose to be built on a blockchain that applies rigorous constraints to the timestamps of committed blocks or uses external timestamp authorities to assert a block creation time~\cite{Szalachowski2018}.

\paragraph{Transaction sequence manipulation}
\label{sec:txseq}

While transactions within a block share the same timestamp, miners can order transactions, and choose to include or exclude certain transactions at their discretion. Malicious miners can abuse their \enquote{power} to prioritize transactions in their favor, profiting from the \ac{mev}, \revision{which is the value that is extractable by miners directly from smart
    contracts during the validation or mining process~\cite{zhou2021just,Qin2021a}} This can be further facilitated by open-source software such as \revision{Flashbots~\cite{daian2020flash}}.

To prevent transaction sequence manipulation, \acp{dex} should first be built upon reputable, frequently-used blockchain systems, as they feature high miner/validator participation, making transaction sequence manipulation difficult.
Besides, this attack can be mitigated through an enforced transaction sequencing rule
that relies on a trusted third party to assign sequential numbers to transactions~\cite{Eskandari2020}. We also discuss how \acp{dex} and their transactions can practice transaction sequencing from application-layer in~\autoref{sec:frontrunning}, and how privacy-preserving blockchain and \acp{dex} are resistant to this attack in~\autoref{sec:privacy}.

\paragraph{Other infrastructure-layer attacks}
Aiming to perturb operations of blockchain systems~\cite{Saad2019}, many other attacks do not target \ac{amm}-based \ac{dex} specifically, but can indirectly affect the service of \ac{dex}.
For example, attackers can launch spam or \ac{ddos} attacks towards the blockchain system~\cite{Greene2018,Perez2020d},
thereby increasing the latency or even hindering the accessibility of \ac{dex} services;
\ac{bdos} attacks exploit the reward mechanism to discourage miner participation, thereby causing a blockchain to a halt with significantly fewer resources~\cite{Mirkin2020};
the 51\% attack~\cite{Saad2019}, the most classic blockchain attack, is able to tamper with the blockchain in any way by controlling more than 50\% of the network's mining hash rate;
network attacks can destroy the network connections between the users and the blockchain system through \ac{dns} hijacking~\cite{Ramdas2019} or \ac{bgp} hijacking~\cite{Apostolaki2017}.

In short, \ac{amm}-based \acp{dex} should be built upon distributed ledgers with active service, community maintenance, and upgrades,
as well as modular security designs. Only by ensuring each module of the blockchain system and the interactions between them are secure, can the relative security of the entire blockchain system be ensured.

\subsubsection{Middleware-layer attacks}

An \ac{amm}-based \ac{dex} usually consists of various smart contracts, in which each serves as a middleware that bridges some application-layer functions with blockchain infrastructures, and collectively
support operations of \acp{dex}.
However, the smart contracts and complex collaborations between them can also lead to potential system vulnerabilities~\cite{tsankov2018securify}.
Attackers can exploit such attack interfaces to steal tokens from a \ac{dex} or even paralyze it.

\paragraph{Reentrancy attack} Reentrancy attack can happen when two or more entities (e.g. smart contract, side-chain) call or execute certain functions in specific sequences or frequencies.
Ever since July 2016, reentrancy attacks have captured the attention of the crypto industry, when the \ac{dao}, an Ethereum smart contract, was executed maliciously with such an attack, causing a \$50 million economic loss in tokens~\cite{daoattack2016}.
Afterwards, despite the emergence of various proposals addressing the reentrancy problems~\cite{luu2016making,tsankov2018securify,albert2020taming}, reentrancy attacks persist, and became particularly threatening to \ac{amm}-based \acp{dex}. In January 2019, an audit identified a reentrancy vulnerability in Uniswap~\cite{uniswap-audit}, which was then exploited by hackers to steal \$25 million worth of tokens in April 2020~\cite{uniswap2020april}
Hackers
performed this attack
by leveraging
a subtle interaction between two contracts that were secure in isolation, and a third malicious contract~\cite{cecchetti2021compositional}.
In March 2021, \$3.8 million worth of tokens were stolen from DODO through a series of attacks, which began with reentrancies via the \texttt{init()} function in a liquidity pool smart contract, followed by frontrunning and honeypot attacks~\cite{DODO2021}.


The security community has proposed a variety of approaches to tackle reentrancy attacks~\cite{huang2019smart}.
For example, Rodler \etal~\cite{Rodler2019a} protect existing smart contracts on Ethereum in a backwards compatible way based on run-time monitoring and validation;
Das~\etal~\cite{das2021resource} propose Nomos, a reentrancy-aware language that enforces security using resource-aware session types;
Cecchetti~\etal~\cite{cecchetti2021compositional} formalize a general definition of reentrancy and leverage information flow control to solve this problem in general.
However, with the increasing complexity of \ac{amm}, it can become more difficult for developers to reason about the reentrant interface,
thus making reentrancy attacks a more intractable problem for \ac{amm}-based \acp{dex}.

\paragraph{Other middleware-layer attacks}
On the middleware layer, there are many other attacks and threats that can affect normal operations of smart contracts~\cite{sayeed2020smart}, such as replay attack~\cite{ramanan2021blockchain}, exception mishandling~\cite{Praitheeshan2019}, integer underflow/overflow attacks~\cite{sun2021mutation}, etc.
These threats are not specifically targeted at \acp{dex}, but can potentially be harmful to \ac{dex} operation.
The security community has proposed a variety of approaches to secure smart contract from these threats, such as Smartshield~\cite{zhang2020smartshield}, Zether~\cite{bunz2020zether}, and NeuCheck~\cite{lu2019neucheck}. Users may also purchase insurance cover to hedge smart contract risks \cite{Cousaert2022}.
To fundamentally counter those attacks, smart contract coders must strictly abide by software development specifications and conduct thorough security tests.

\subsubsection{Application-layer attacks}
\label{sec:appattaacks}

\paragraph{Oracle attack}

A flash loan is a feature provided by lending platforms where an uncollateralized borrow position can be created as long as the borrowed funds can be repaid within one transaction \cite{Xu2022d}. Flash loans can be used to repay at discount debts that are liquidable without having to acquire borrowed assets in the first place.
In this kind of attack, adversaries manipulate lending platforms that use a \ac{dex} as their sole price oracle (see \autoref{fig:taxonomy}).

Following \autoref{flashloanattack}, an attacker profits with $\Delta_3$ token$_A$ less any transaction fees incurred. Utilizing continuous slippage native to an \ac{amm}-based \ac{dex} (see \autoref{sec:form_sl}), the attack temporarily distorts the price of token$_A$ relative to token$_B$. After the prices are arbitraged back, the attack would leave the loan taken from step~\ref{step:borrow} undercollateralized, jeopardizing the safety of lenders' funds on the lending platform.
Examples of such attacks are exploits on Harvest finance \cite{harvest2020attack}, Value DeFi \cite{peckshield2020valuedefi} and Cheese bank \cite{pirus2020cheesbank}.

This broken design can generally be fixed by either providing time-weighted price feeds, or using external decentralized oracles. The first solution ensures that a price feed cannot be manipulated within the same block, while the second solution aggregates price data from multiple independent data providers that add a layer of security behind the aggregation algorithm, making sure that prices are not easily manipulated \cite{smartcontent2021TWAP}.

\paragraph{Rug pull}

A rug pull involves the abandonment of a project by the project foundation after collecting investor's funds~\cite{Xia2021a}. \revision{One way of doing this, is to lure people into buying the coin with no value through a \ac{dex}, subsequently swapping this coin for \coin{ETH} or another cryptocurrency with value, as shown in \autoref{rugpull}.} \acp{dex} allow users to deploy markets without audit and for free (barring the gas costs), which makes them an excellent target to scam investors. One method is to create a coin with the same name as an existing one. This attracts a lot of attention since everyone wants to pick up the coin at the lowest price possible. The coin is being bought up, and the original \ac{lp} swaps his fake coin for \coin{ETH}.
In other cases, the creators of the scam token reach out to several prominent people, creating false hype. Once potential buyers see that major players have purchased the token, they start buying themselves, before realizing that the token cannot be swapped back for \coin{ETH}. Sometimes, the attackers let people trade the coin back for \coin{ETH}, but only for a short period since they are running the risk of losing money. \cite{Xia2021a} research data on scam tokens on Uniswap and confirm that rug pulls commonly find their victims through \acp{dex}.

In August 2020, an rug puller extracted 3 \coin{ETH} by imitating the well-known \coin{AMPL} token~\cite{ampleforth2021home} with a scam token \coin{TMPL}. The token's transaction history shows the provision of 150 \coin{ETH} and \coin{TMPL} tokens to a Uniswap V2 pool by the attacker, who removed 153.81 \coin{ETH} only 35 minutes later~\cite{tmpl2021etherscan}.

To protect themselves from being rugged,
investors should exercise caution and always confirm a project's credibility before investing in its \ac{ido}~\cite{BybitLearn,rug-pull-businessreview}.
Usually, reputable \acp{ido} feature high liquidity and a pool lock~\cite{rug-pull-coinmarketcap} that disables withdrawal for a fixed period, so that \acp{lp} are unable to quickly empty the pool once it has absorbed a sufficient amount of valuable assets from investors~\cite{MudraManager}.

\paragraph{Frontrunning}
\label{sec:frontrunning}

Frontrunning is often enabled through access to privileged market information about upcoming transactions and trades~\cite{Eskandari2020}.
Since all transactions are visible for a very short period of time before being committed to a block,
it is possible for a user to observe and react to a transaction while it is still in the mempool.
Those who place their trade immediately before someone else's are called frontrunners~\cite{daian2020flash,Eskandari2020}.
Frontrunners attempt to get the best price of a new coin before selling them onto the market.
They can buy up a great portion of the supply of a new token to create exorbitant prices.
Due to the hype, this does not stop retail traders from further buying.
The frontrunner, who is the seller with the most significant supply, can swap the purchased token for popular coins (e.g. \coin{ETH}) paid by retail traders.
For example, a considerable amount of \acp{ido} on Polkastarter are frontran on Uniswap~\cite{frontrunningnews}.

Frontrunning can also be achieved through transaction sequence manipulation (see \autoref{sec:txseq}) and by exploiting the general mining mechanism. Most mining software, including the vanilla Go-Ethereum (geth), the most popular command-line interface for running Ethereum node, sorts transactions based on their gas price and nonce \cite{Zhou2021High-Frequency,Eskandari2020,daian2020flash}. This feature can be exploited by malicious users of \ac{dex} who broadcast a transaction with a higher gas price than the target one to distort transaction ordering and thus achieve frontrunning.

\revision{Frontrunning can be avoided with various approaches~\cite{Baum2021a}.} Normal exchange users can set a low slippage tolerance to avoid suffering from a price elevated by front-runners. However, an overly low slippage tolerance may lead a transaction to fail, especially when the trade size is large, resulting in a waste of gas fee \cite{sandwhichattack2021}.
\acp{dex} can enforce transaction sequencing to fundamentally solve frontrunning. Some exchanges, such as EtherDelta~\cite{Narayanan2016BitcoinIntroduction} and 0xProject~\cite{0xProject}, utilize centralized time-sensitive functionalities in off-chain order books~\cite{Eskandari2020}. In addition, transactions can specify the sequence by including the current state of the contract as the only state to execute on~\cite{Eskandari2020}, thereby preventing some types of frontrunning attacks. Frontrunning can further be tackled by addressing privacy issues (see \autoref{sec:privacy}) and transaction sequence manipulation on the infrastructure layer (see \autoref{sec:txseq}).


\paragraph{Backrunning}
\label{sec:backrunning}

Backrunners place their trade immediately after someone else's trade. The attacker needs to fill up the block with a large number of cheap gas transactions to definitively follow the target's transaction. Compared to frontrunning which only requires a single high valued transaction and is detrimental to the user being frontrun, backrunning is disastrous to the whole network by hindering the throughput with useless transactions \cite{Livnev2020}.


Backrunning attacks can be mitigated through anti-\ac{bdos} solutions such as A$^2$MM \cite{Zhou2021A2MM}.
Theoretically, defense solutions of \ac{l7ddos} attacks~\cite{feng2020application,xie2008monitoring,wang2017skyshield} can also be adopted to tackle backrunning problems, as they all aim to secure usability of web services to legitimate users.
In addition, backrunning can be addressed by confidentiality-enhancing solutions that hide the content of a transaction before it is committed to a block (see \autoref{sec:privacy}).

\paragraph{Sandwich attacks}
\label{sec:sandwich}

Combining front- and back-running, an adversary of a sandwich attack places his orders immediately before and after the victim's trade transaction. The attacker uses front-running to cause victim losses, and then uses back-running to pocket benefits. While there are endless examples of sandwich attacks, Zhou \etal~\cite{Zhou2021High-Frequency} detail two types that can occur on an \ac{amm}: an \ac{lp} attacking an exchange user (see \autoref{sandwichlp}) by exploiting \ac{amm}'s liquidity sensitive property (\autoref{sec:liquid-sensitivity}), and one exchange user attacking another \revision{(see \autoref{sandwichprice})} by taking advantage of \ac{amm}'s demand sensitive property (\autoref{sec:demand-sensitivity}). The latter is particularly common. \autoref{fig:sandwich_example} shows two examples of such attacks on Uniswap V2 within a time frame of 3 minutes.

\input{exhibits/sandwich_example}

Considering swap fee (see \autoref{sec:swapfee}), gas fee (see \autoref{sec:gasfee}) and slippage (see \autoref{sec:slippage}), sandwich attacks are only profitable if the size of the target trade exceeds a certain threshold, a value that depends on the \ac{dex}'s design and the pool size. \acp{dex} can thus prevent sandwich attacks by disallowing transactions above the threshold \cite{Zhou2021A2MM}. Naturally, sandwich attacks can also be curbed by deterring either frontrunning (see \autoref{sec:frontrunning}) or backrunning (see \autoref{sec:backrunning}).

\paragraph{Vampire attack}
\label{sec:vampireattack}

A vampire attack targets an \ac{amm} by creating a more attractive incentive scheme for \acp{lp}, thereby siphoning out liquidity from the target \ac{amm} \cite{jakub2020vampire} to the detriment of the protocol foundation (see \autoref{sec:protocolfoundation}).
In September 2020, Sushiswap gained \$830 million of liquidity through a vampire attack \cite{dale2020sushiswapvampire}, where Sushiswap users were incentivized to provide Uniswap \ac{lp} tokens into the Sushiswap protocol for rewards in \coin{SUSHI} tokens \cite{sushi2020}.
A migration of liquidity from Uniswap to protocol Sushiswap was executed by a smart contract that took the Uniswap \ac{lp} tokens deposited in Sushiswap, redeeming them for liquidity on Uniswap which was then transferred to Sushiswap and converted to Sushiswap \ac{lp} tokens.

Legal approaches such as applying a restrictive license to the protocol code base---as done later by Uniswap with its V3 new release~\cite{Foxley2021}---can be employed to hinder vampire attacks.

\subsection{Privacy concerns}
\label{sec:privacy}

\revision{Most blockchain systems are open, traceable, and transparent, which can raise severe privacy concerns to \acp{dex} that built upon them. Besides, \ac{amm} protocols will reveal real-time \acp{dex} information to the public, bringing additional privacy concerns to \ac{amm}-based \acp{dex}.} In this section, we introduce the privacy issues that users may face in using \ac{amm}-based \ac{dex} and discuss their possible solutions.

\subsubsection{Transaction inspection}
\label{sec:transinspect}

The transparency and openness of public blockchains, where most \ac{amm}-based \acp{dex} are built, allow transactions to be observable to everyone.
However, this characteristic enables malicious parties to inspect transactions, thereby seeking profits or even disrupting the market~\cite{Eskandari2020}.
The inspection activities can occur before or at the moment that the transaction is committed to the blockchain \revision{by miners, validators, or even any third parties.}
For example, the aforementioned frontrunning, backrunning, sandwich attacks, transaction sequence manipulation, and block timestamp manipulation in Section~\ref{sec:associated_attacks} are all based on transaction inspections~\cite{Goldreich1994DefinitionsSystems}.
In fact, major \ac{amm}-based \acp{dex} are fraught with bots, constantly monitoring transactions for possible profit opportunities~\cite{Narayanan2016BitcoinIntroduction}.

\subsubsection{Identity tracing}

Although most blockchains and \ac{amm}-based \revision{\acp{dex}} usually feature a certain degree of anonymity, the linkabilities of transactions and accounts over time still enable attackers to dig identities information of users~\cite{Zhang2019SecurityBlockchain}.
\revision{Actually, with some off-chain information (e.g. social network posts, public speak, location~\cite{DuPont2015TowardLocation}), eavesdroppers can launch de-anonymization inference attacks to bridge virtual accounts with real-world individuals or uncover the true identities of traders by linking the transactions of an account together and matching relevant information~\cite{Narayanan2016BitcoinIntroduction}.}

\subsubsection{Behavioral model inference}

\revision{By collecting data from corresponding blocks and analyzing historical transactions of an account, any third parties infer an account's behavioral model, understanding its active phase, trading frequency, or even preferences.} Such activity is called behavioral model inference, which not only compromises user privacy, \revision{but can also make preparations for launching honeypot attacks~\cite{Qin2021a} and phishing scams~\cite{Xia2021a,Chen2020PhishingEcosystem.,Phillips2020TracingWebsites,Wu2020WhoEmbedding}.}

\subsubsection{\revision{\ac{amm}-specific privacy concerns}}

\revision{While it is possible to anable privacy-preservance with non-\ac{amm}-based \acp{dex} \cite{govindarajan2022privacy,Baum2021} by hiding all the information such as the transaction, order book, and trading volume, it is challenging to make \ac{amm}-based \acp{dex} fully privacy-preserving.
    Due to their path-deterministic property (see \autoref{sec:path-binding}), information on swaps with an \ac{amm} pool is often reverse-engineerable \cite{Angeris2021ReplicatingMakers}.
    Some researchers even argue that complete privacy is impossible once \acp{dex} apply ordinary implementations of \acp{cfmm} (e.g., Uniswap, Balancer, Curve, etc.) under reasonable adversarial models~\cite{Angeris2021AMakers}. The transparency and openness allow any third party to capture rich data about the \ac{amm}-based \acp{dex}, such as the overall trading situation per block, asset pool changes over time, or popular tokens on the platform. Under this circumstance, even though the \acp{dex} can hide the detailed information about a transaction, we can still estimate the transaction amount and transaction currencies according to the \ac{amm} protocols.}

\subsubsection*{\bf Solution}
%

The confidentiality of \ac{amm}-based \ac{dex}'s pipeline can be enhanced from various angles to limit public access to certain information.
Such information includes but is not limited to the transaction amount, asset type, user/pool balance, user identification, order of transactions, \revision{\acp{mev},} or protocol-related information.
Hiding all the information from both the public and computation parties can fundamentally resolve the privacy concerns, thereby eliminating all the associated attacks and privacy disclosure.
However, it may break the market visibility to traders, cause difficulties to governance and regulation, and make the whole system inefficient to operate.
On the other hand, achieving partial confidentiality may be sufficient to prevent many attacks and provide enough privacy protections to traders.
Thus, most existing solutions focus on enhancing the user privacy of certain components in \ac{amm}-based \acp{dex}.

On the infrastructure layer, many privacy-preserving blockchain solutions have been proposed to increase confidentiality~\cite{Bunz2018Bulletproofs:More,Bernabe2019Privacy-preservingChallenges} and anonymity~\cite{Miers2013Zerocoin:Bitcoin,Noether2015RingMonero.,Sasson2014Zerocash:Bitcoin} for transactions. These solutions can be based on \ac{zkp}~\cite{Goldreich1994DefinitionsSystems}, homomorphic encryption~\cite{Gentry2009FullyLattices}, or identity obfuscation~\cite{Goldwasser2007OnObfuscation}, aiming to break the linkability of transactions, encrypt transaction content, or anonymize users accounts. \ac{amm}-based \ac{dex} built upon these blockchains not only can protect user privacy, but also can defend against attacks discussed in
\autoref{sec:infraattack} and
\autoref{sec:appattaacks}.
\revision{Besides, developers can choose to leverage \ac{pbs}~\cite{pbs} to hide transaction details from miners or validators. Although \ac{pbs} can only achieve miner/validator-specific privacy, it can increase the censorship resistance of transactions against miners and validators, thereby defending against frontrunning attacks effectively and efficiently. Ferveo~\cite{bebel2022ferveo} is a similar approach, which is a fast protocol for Mempool Privacy to avoid transaction censorship.}

Even though the blockchain infrastructures are transparent, confidentiality can be achieved through privacy designs on upper layers.
On the middleware layer, developers can leverage techniques such as Hawk~\cite{Kosba2016Hawk:Contracts}, Ekiden~\cite{Cheng2019Ekiden:Contracts}, and Submarine Commitments~\cite{Breidenbach2018} to develop privacy-enhanced smart contracts for \ac{amm}-based \ac{dex}.
On the application layer, privacy-enhanced \acp{dex} are proposed to resolve the privacy concerns. For example, P2DEX~\cite{Baum2021} harnesses \ac{mpc}~\cite{Cramer2015SecureComputation} to realize efficient privacy-preserving \ac{dex}; ZKSwap~\cite{zkswap2021}, an \ac{amm}-based \ac{dex} utilizing ZK-Rollup~\cite{gluchowski2019zk} technology, not only provides users with extra privacy protections when withdrawal, deposit, and transfer of tokens by leveraging \ac{zkp}, but also significantly reduce gas fees for transactions;
ZEXE~\cite{Bowe2020}, a ledger-based system that enables users to realize privacy-preserving
analogues of popular applications, such as \ac{dex}. ZEXE can make it easy to satisfy two main privacy properties. First,
transactions hide all information about the offline computations.
Second, transactions can be validated in constant time by anyone, regardless of the offline computation;
\revision{Zswap~\cite{Engelmann2022}, a zk-SNARKs-based transaction scheme that enables multiple assets and atomic exchanges with sparse homomorphic commitments and Zcash friendly simulation-extractable \ac{nizk} proofs.}

Furthermore, on-chain privacy-preserving services and products are on the rise.
For example, Blank~\cite{Blank2021}, a non-custodial Ethereum browser extension wallet, offers
transaction obfuscation; Enigma~\cite{Kisagun2019} builds a network of \enquote{secret nodes} that can perform computations on encrypted data without the necessity to expose original raw data.


\revision{However, as mentioned above, even details of an individual transaction will not be disclosed, it is still possible to infer rough information of transactions through the asset pool state changes and \ac{amm} protocols~\cite{Angeris2021AMakers}.}
As the \ac{amm} protocols and asset pool state should be accessible to the public, any third party can keep tracking the asset pool changes and then deduce how many assets and which assets were traded during each block.
\revision{From the perspective of computation parties (\ie miner), they can fetch even more granular pool state changes to infer detailed transaction information in the mempool.
    Hence, most existing privacy-preserved \ac{dex} designs are hardly compatible with \ac{amm} protocols.
    To overcome this problem, developers can use non-constant function market makers or apply fuzzification to conservation functions. However, this will also bring some side effects to traders, as it can incur larger slippages to the final trading prices. In addition, privacy-enhanced \ac{amm}-based \acp{dex} can increase the difficulty of market supervision,} creating obstacles for governance, regulation, and financial control. Therefore, how to reach a balance between privacy protection and policy compliance is a question worth exploring for the entire crypto industry.

%% file: exhibits/security_taxonomy.tex
\begin{figure}[tbhp]
    \centering
    \includegraphics[width=0.7\linewidth]{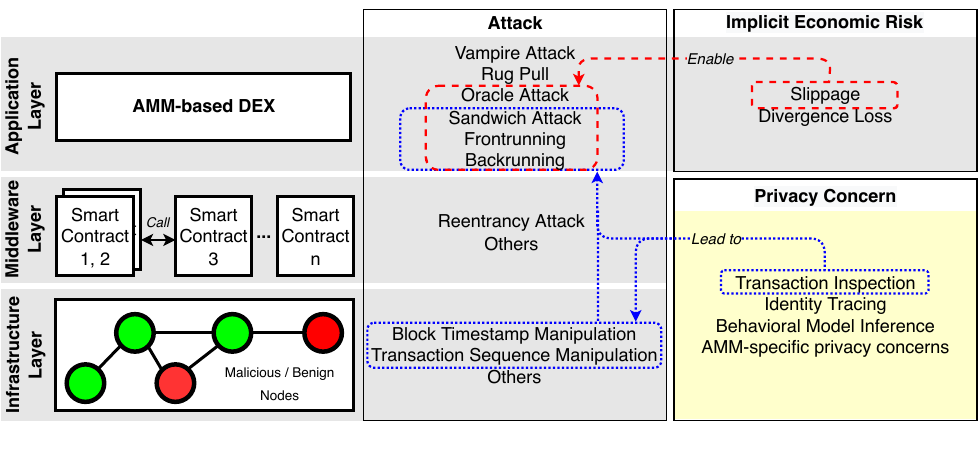}

\caption{\revision{Architectural layers of an \ac{amm}-based \ac{dex} with its implicit economic risks, attacks, privacy concerns, and their relationships.}}
    \label{fig:security_taxonomy}
\end{figure}

%% file: exhibits/sandwich_example.tex
\begin{figure}[tbhp]
    \centering
    \frame{\includegraphics[width=0.8\linewidth]{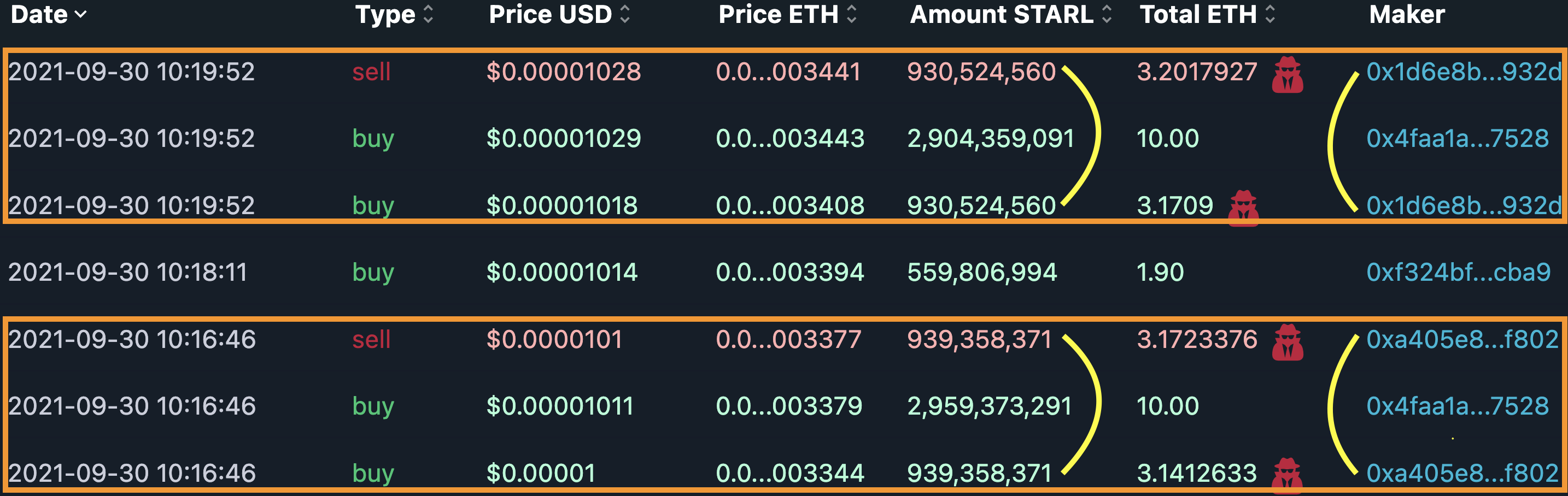}}
    \caption{Two sandwich price attacks (see Attack Algorithm~\ref{sandwichprice}), marked with {\color{orange}orange \Squarepipe} on 30/09/2021, with the \href{https://www.dextools.io/app/ether/pair-explorer/0xa5e9c917b4b821e4e0a5bbefce078ab6540d6b5e}{STARL/ETH} pair on Uniswap V2. The attack conducted at 10:16:46
        by \href{https://etherscan.io/address/0xa405e822d1c3a8568c6b82eb6e570fca0136f802}{0xa405e8...f802} resulted in a profit of 0.0310743 \coin{ETH};
        the attack conducted at 10:19:52 by \href{https://etherscan.io/address/0x1d6e8bac6ea3730825bde4b005ed7b2b39a2932d}{0x1d6e8b...932d} resulted in a profit of 0.0308927 \coin{ETH}.}
    \label{fig:sandwich_example}
\end{figure}

%% file: sections/future.tex
\section{Avenues of future research}
\label{sec:future}

While \acp{dex} with \ac{amm} protocols are maturing, there still exists room for further development, expansion, and exploration. In this section, according to existing literature, our observations about recent trends, and major problems to be solved in this field, \revision{we discuss avenues for future research from the perspectives of security, privacy, system design, protocol design, and governance. \autoref{fig:future_research} shows a graphical illustration of future research avenues.}

\input{exhibits/future_research}

\subsection{Finance and system security risk}

Compared with order-book-based \ac{dex}, an \ac{amm}-based \ac{dex} leverages bonding curves to determine asset prices without reaching any agreements between buyers and sellers. However, this feature can also bring severe security risks and leave many unsolved problems. For example, once attackers successfully manipulate the asset prices in \ac{amm}-based \ac{dex}, they can theoretically drain all the associated assets from the asset pool, causing severe damage to the trading market. This matter is worthy of deep investigation in the future.

\revision{In addition, as a nascent type of trading platforms built on complicated distributed systems (i.e., blockchain),} \ac{amm}-based \acp{dex} operate upon various components run by many different parties. Therefore, any flaws in this system can pose threats to the trading market, potentially causing both economic losses and system dysfunction. As discussed in \autoref{sec:security}, in recent years, \ac{amm}-based \acp{dex} suffered from a variety of attacks for different reasons, such as oracle attacks caused by flaws of oracle components in blockchains, vampire attacks caused by the lack of market regulation, flash loan attacks caused by unsound trading rules, and reentrancy attacks caused by inappropriate function calls of smart contracts.
Patching those vulnerabilities without bringing significant changes to existing systems requires joint efforts from both computer science, economics, and finance researchers.

\subsection{Finance and system privacy concern}

Privacy is another major concern in \ac{amm}-based \acp{dex} (\autoref{sec:privacy}). At present, most \ac{dex} systems are built on top of public blockchains such as Ethereum to record transactions in plain text, which enables everyone to observe detailed transactional information. An attacker could inspect an \ac{amm} protocol, access the associated transactions from the blockchain, and launch attacks such as frontrunning, backrunning, and sandwich attacks. \revision{Even most blockchain systems provide a certain degree of anonymity (i.e. pseudonymous), recent studies have shown that deanonymization attacks can link transactions to users' accounts and reveal users' real identities. Besides, the AMM algorithm itself can leak rich information about the \ac{dex}, enabling any third parties to estimate the detailed transaction information.}

As discussed in \autoref{sec:privacy}, some general solutions such as Zerocash~\cite{Sasson2014Zerocash:Bitcoin}, Hawk~\cite{Kosba2016Hawk:Contracts}, and ZEXE~\cite{Bowe2020} have been proposed to achieve ledger privacy in \ac{dex}. However, they would substantially decrease the real-time efficiency in \ac{amm}-based \ac{dex}, thereby very expensive to deploy in \ac{amm}-based \ac{dex}. Researchers can try to optimize the system design and the cryptographic algorithm efficiency of those approaches, thus preserving the swap velocity while enhancing user privacy. In addition, existing privacy-preserving blockchains and \acp{dex} are not compatible with \ac{amm} protocols. Even transaction information is protected on blockchains, people can still deduce the transaction's asset type and exchange amount from asset pool transformations. Therefore, a new \ac{amm}-oriented \ac{dex} that can preserve user privacy at certain degrees is worthy of study. \revision{Possible research directions include adding stochasticity to the \ac{amm} protocol, developing non-constant function market makers, or applying fuzzification to current \acp{amm}.} Moreover, \revision{existing privacy solutions} could also result in governance issues; since these solutions provide full privacy for transactions, it is almost impossible for law enforcement to investigate cryptocurrency-related crimes in \ac{dex} such as theft, money laundering, and illegal transactions in dark markets. Finally, introducing privacy in \ac{amm}-based \ac{dex} may affect asset prices, asset liquidity, and market predictability across different markets, thus requiring new economic models and protocols to analyze the reward and cost for \ac{amm} economics.

\revision{%
    \subsection{Speed and scalability issues}
    As \ac{amm}-based \acp{dex} operate on top of blockchain systems, their transaction speed and scalability are limited by the growth speed and throughput of the blockchain networks.}

\revision{%
    On the one hand, each \ac{dex} transaction takes time to be validated on the blockchain network before it takes effect. The validation speed depends on the miners or validators, rather than the \ac{dex}. Compared with centralized exchanges that will immediately finishing processing user transactions, \acp{dex} will incur a processing delay ranging from a couple of seconds to several hours. Although some blockchain networks are specifically designed for velocity requirements (e.g., Tezos, XRPL, EOSIO, etc.), their delays are still measured in seconds~\cite{Perez2020d}.
}%

\revision{
    On the other hand, as the information of \ac{dex} transactions must be recorded in blocks, the total number of transactions that a \ac{dex} can accept per batch is limited. Although many multi-layer blockchains (e.g., layer 2 blockchains) were proposed to resolve the throughput issue~\cite{sguanci2021layer}, their throughput is still an order of magnitude behind centralized exchanges.
}

\revision{
    Therefore, a vital future research direction is to keep improving the validation speed and throughput of blockchain systems, thereby increasing the speed and scalability of the \ac{amm}-based \acp{dex} built upon them. Resolving this issue not only can scale \ac{amm}-based \acp{dex} to a large group of users, but also can ensure that each transaction can be processed timely. New types of data structure, synchronization mechanism, or validation approaches should be proposed to conquer this problem.
}

\subsection{\revision{Limited trading functionality}}
\revision{Compared to centralized exchanges nowadays, the trading functionalities of \ac{amm}-based \acp{dex} are limited to buying and selling. Due to the nature of \ac{amm}-based \acp{dex}, the lacks of stop-loss ability, margin trading, and put and call options greatly restrict users' financial operations on the platforms.
    Besides, it is difficult to implement additional functionalities to existing \ac{amm} protocols.
}

\revision{To enhance the trading functionality of \ac{amm}-based \acp{dex}, researchers need to bring some structural changes to currently \ac{amm}-based \acp{dex}. For example, researchers can try to bridge \ac{amm} and order book to provide put and call options; leveraging smart-contract-based lending can also help \ac{amm}-based \acp{dex} to provide trading margins to users.}


\subsection{\revision{New design for \ac{amm}}}

We find that the majority of \ac{amm}-based \ac{dex} use a \ac{cfmm} algorithm that can be seen as a variation of Uniswap's constant product protocol.
\revision{It would be worth exploring novel bonding curves or new \ac{amm} designs that can, e.g., balance slippage and impermanent loss in different ways.}

Current \ac{amm} implementation can be mainly seen with \ac{dex} for spot markets. While still at its infancy, \ac{dex} for financial derivative markets have also been witnessing an increase in the adoption of \ac{amm} algorithms, such as
Siren \cite{Siren2021} and Hegic \cite{Wintermute2020} for options,
the Perpetual Protocol for perpetual contracts \cite{Protocol2021}, and Tracer \cite{Garner2021} for swaps. As those protocols are still at their early development phase, they have not been thoroughly tested by the market or scrutinized by academia. As the derivative \ac{dex} using \ac{amm} become more mature, it would be of scholars' and practitioners' interest to see a systematization of these protocols and an investigation on how they advance from the basic \acp{amm} for spot markets.

\subsection{Governance}

Due to its decentralized and censorship-resistant nature, \ac{dex} participants have the liberty to do whatever is permissible by the smart contract code, sowing the seeds for malicious and fraudulent behavior. In such context, governance schemes are essential to ensure the proper operation of \ac{amm}-based \acp{dex}.
However, centralization of voting power is frequently observable with \ac{amm}-based \ac{dex} due to the often concentrated nature of protocol token distribution.
\revision{
    Furthermore, regulatory concerns often go hand in hand with governance issues. For example, the Uniswap community has been debating whether or not to turn the fee switch on for the \coin{UNI} buyback program \cite{UniswapGovernance}. In fear of legal consequences in the case of \coin{UNI} being categorized as a security token, the protocol has yet to enable the function due to its similarity to a regular stock buyback exercise, leaving the \coin{UNI} token value unbacked by any monetary flow.
}

Whether and how the governance mechanism can be improved for more sustainable development of a protocol \revision{in a legally compliant way} thus merit further research.



%


%% file: exhibits/future_research.tex
\begin{figure}[tbhp]
    \centering
    \includegraphics[width=0.8\linewidth]{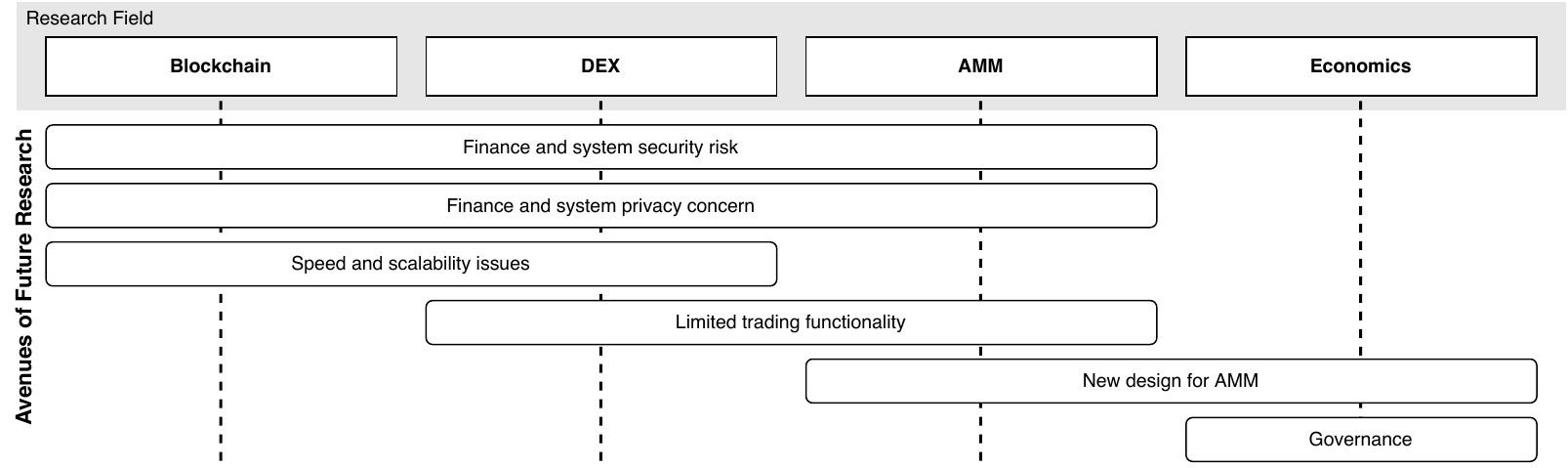}
    \caption{A graphical illustration of future research avenues.}
    \label{fig:future_research}
\end{figure}

%% file: sections/relatedwork.tex
\section{Related work}
\label{sec:related}

In \autoref{tab:litreview}, we summarize most \acsp{sok}, surveys and tutorials that investigate the \ac{dapp} ecosystem on blockchain. Our work differs from those existing works in the following aspects:

\begin{enumerate}
  \item we have a clear, focused study subject: \ac{dex}-based \ac{amm}, while the majority of related survey studies either examines other \ac{defi} applications (e.g. lending protocols \cite{Bartoletti2020sokLendingPools}, yield aggregators \cite{Cousaert2021,Xu2022g}) or have a broader coverage;
  \item we use a comprehensive set of methods to generalize and systematize \ac{dex}-based \ac{amm} protocols, including taxonomization, state space modeling, numerical simulation and empirical investigation, while most existing related survey studies use only a subset of the aforementioned methods;
  \item we examine an array of aspects of \ac{dex}-based \ac{amm}, including their architectural design and internal mechanism, financial economics as well as as associated security and privacy concerns, while most of the related survey papers only cover some of these aspects.
\end{enumerate}

In the following, we discuss more literature related to our study in various ways.

\subsection{AMM-based DEX on blockchain}

Our work is first and foremost related to the literature body covering \ac{amm}-based \ac{dex} on blockchain.

\subsubsection{Protocol mechanism}

Angeris \etal~\cite{Angeris2021} discuss arbitrage behavior and price stability in constant product and constant mean markets.
Lo \etal~\cite{Lo2020UniswapExchange} empirically evidence that the simplicity of Uniswap ensures the ratio of reserves to match the trading pair price.
Despite historical oracle attacks associated with \acp{amm} (see \autoref{sec:security}), Angeris \etal~\cite{Angeris2021,Angeris2020} show that \ac{cfmm} users are incentivized to correctly report the price of an asset, suggesting the suitability for those \acp{amm} to
act as a decentralized price oracle for other \ac{defi} protocols.
Angeris \etal~\cite{Angeris2021ReplicatingMakers} present a method for constructing \ac{cfmm} whose portfolio value functions match an arbitrary payoff.
Richardson \etal~\cite{Richardson2020c} detail the mechanism of the Bancor \ac{amm} and its potential implementation in carbon trading markets.

\subsubsection{Security}

Qin et al.~\cite{Qin2021} conduct empirical analyses on various \ac{amm} attacks, including transaction (re)ordering and front-running, and demonstrate the profitability in performing transaction replay through a simple trading bot.
Mitigating solutions for front-running attacks in \ac{defi} are surveyed in Baum et al.~\cite{Baum2021a}.
Security risk in terms of attack vectors in high-frequency trading on \acp{dex} are discussed in Zhou et al.~\cite{Zhou2021High-Frequency}, and Qin et al.~\cite{Qin2020a}.
Flash loan attacks with the aid of \acp{amm} on Ethereum are described in Cao et al.~\cite{cao2021flashot}, Perez et al.~\cite{Perez} and Wang et al.~\cite{Wang2020FlashLoan}.
Victor et al.~\cite{Victor2021DetectingExchanges} detect self-trading and wash trading activities on order-book based DEXs.
Gudgeon et al.~\cite{gudgeon2020} explore design weaknesses and volatility risks in \ac{amm} \acp{dex}.

\subsubsection{Privacy}

Angeris \etal~\cite{Angeris2021AMakers}
argue that privacy is impossible with typical \acp{cfmm} and propose several mitigating strategies.
Baum et al.~\cite{Baum2021a} examine input privacy in the context of \ac{amm}.
Stone \etal~\cite{Stone2021TrustlessBridges} describe a protocol that allows trustless, privacy-preserving cross-chain cryptocurrency transfers but is yet susceptible to vampire attacks.

\subsection{\ac{dex} and \ac{amm} in the context of market microstructure}

As two core topics of market microstructure \cite{Garman1976MarketMicrostructure}, decentralized exchange and market-making have been intensively covered in financial economics long before the emergence of blockchain.

\subsubsection{DEX}
\label{sec:dexwork}

Existing literature primarily suggests the higher efficiency of \ac{dex} markets over centralized ones.
Perraudin et al.~\cite{Perraudin1996} investigate decentralized forex markets and conclude that \acp{dex} are efficient when different market makers can transact with each other and that decentralized markets are more immune to crashes than centralized ones.
Nava~\cite{Nava2015EfficiencyMarkets} analyzes quantity competition in the decentralized oligopolistic market and suggest perfect competition can be approximated in large rather than small \ac{dex} markets.
Malamud et al.~\cite{Malamud2017} develop an equilibrium model of general \ac{dex} and prove that decentralized markets can more efficiently allocate risks to traders with heterogeneous risk appetites than centralized ones.

\subsubsection{AMM}
\label{sec:ammwork}

The concept of automated market making can be traced back to Hanson's \acf{lmsr} \cite{Hanson2003CombinatorialDesign, Hanson2012LogarithmicAggregation}.
\ac{lmsr} has since been refined and compared to alternative market-making strategies.

Othman \etal~\cite{Othman2013} address non-sensibility to liquidity and non-profitability of \ac{lmsr} market making. They propose a bounded, liquidity-sensitive \ac{amm} that runs with a profit by levying transaction cost to subsidize liquidity, a strategy later widely implemented by blockchain-based \acp{dex} with \ac{amm} protocols to compensate for divergence loss (see \autoref{sec:divergenceloss}) experienced by \acp{lp}.
Brahma \etal~\cite{Brahma2012} propose a Bayesian market maker for binary markets which exhibit better convergent behavior at equilibrium than \ac{lmsr}.

Jumadinova \etal~\cite{Jumadinova2012} compare \ac{lmsr} with different \ac{amm} strategies, including myopically optimizing market-maker, reinforcement learning market maker and utility-maximizing market maker. Simulating empirical market data, they find that reinforcement-learning-based \ac{amm} outperforms other strategies in terms of maintaining low spread while simultaneously obtaining high utilities.
Slamka~\cite{Slamka2013} compare \ac{lmsr} with dynamic parimutuel market (DPM), dynamic price adjustments (DPA) and an \ac{amm} by the Hollywood stock exchange (HSX) in the context of prediction markets. They show that \ac{lmsr} and DPA generate the highest forecast accuracy and lowest losses for market operators. Today, \ac{lmsr} has become the de facto \ac{amm} for prediction markets~\cite{Wang2020AutomatedDeFi} and was adopted by the Ethereum-based betting platform Augur~\cite{augur}.

Wang~\cite{Wang2020AutomatedDeFi} compares mathematical models for \acp{amm}, including \ac{lmsr}, liquidity sensitive \ac{lmsr} (LS-LMSR) and common \acp{cfmm}, and proposes constant circle/ellipse-based cost functions for superior computational efficiency.
Capponi \etal~\cite{Capponi2021TheMaker} analyze the market microstructure of constant-product \acp{amm}, and predict that \acp{amm} will be used more for low-volatility tokens.

\subsection{State space modeling framework}

Foundational concepts of the design approach used in tokenized economic systems which \acp{amm} are an example of, are presented in the following stream of work: The conceptual engineering framework for modeling, analysis and design of blockchain based infrastructure is introduced in Zargham \etal~\cite{Zargham2018}. A formalization of the blockchain as a state machine is presented in Shorish~\cite{Shorish2018}. The extension of this framework to dynamical stochastic games is presented in Zhang~\cite{Zhang2020}. A formal discussion on how the evolution of a dynamical system can be constrained to uphold desired system properties is conducted in Zargham \etal~\cite{Zargham2020a} using bonding curves as an example. A theoretical framework on estimation properties of aggregated agent signals into systemic statistics in dynamic economic games is provided in Zargham \etal~\cite{Zargham2020b}.

%% file: sections/conclusion.tex
\section{Conclusion}
\label{sec:conc}


\revision{The \ac{defi} ecosystem is a relatively new concept, and innovations within the space are being developed at an incredible speed.
    As an integral part of that ecosystem, \ac{amm}-based \ac{dex} are an incredible innovation sprung up by the trustless, verifiable and censorship-resistant \acl{dlt}.}


In this paper, we systematize the knowledge around \ac{amm}-based \ac{dex} and use state-space representation to formalize and generalize the \ac{amm} algorithms.
We apply our protocol design framework to major exchanges---Uniswap, Balancer, Curve and DODO---and comment on various other exchanges such as Sushiswap, Kyber Network and Bancor.
We examine the implied economic risks in \acp{amm} including slippage and divergence loss, and establish a taxonomy covering security and privacy issues associated with \acp{amm}. In particular, \ac{amm}-based \ac{dex} can be the target of a plethora of infrastructure-, middleware- and application-layer attacks.
Future research into \ac{amm} mechanisms can build upon this systematization of knowledge, establish unique ways for differentiating \ac{amm} innovations, and expand on our security taxonomy that can help the development of more robust \ac{amm}-based \acp{dex}.

%% file: sections/comparison_app.tex
\section{Formulas of major AMM-based DEX}
\label{appendix:formulas}

All formulas provided here adhere to the state space representation introduced in \autoref{sec:statespace} proving that this generalized framework can be used to present and discuss various \ac{amm} protocols. All important properties such as the \textit{conservation function}, \textit{state updates} and \textit{metrics} can be framed within the defined taxonomy and allow an unified analysis of various specifications. The survey of knowledge presented in this paper has been couched in the context introduced in \autoref{sec:formal} to facilitate a qualitative comparison among different protocols without presuming to provide a methodology sufficiently abstract for engineering rigor. This task is on the agenda of researchers in the interdisciplinary space of \textit{Token Engineering} where methodologies are being developed which allow to perceive such interactions on top of blockchain protocols as \textit{Economic Games} \cite{Zargham2020b} in the context of \textit{Generalized Dynamical Systems}. Those methodologies currently under construction will in the future allow quantitative comparisons within unified frameworks where numerical experiments and system identification techniques support system designers in the construction, design, analysis and maintenance of \textit{classes of protocols} (such as e.g. the \ac{amm}-protocol-class as one representative from the \ac{defi} space) and where the framework both allows to derive system properties from a given representation or alternatively find a system representation conditioned on desired system requirements.

\subsection{Uniswap V2}
\label{app:uniswapv2}

\subsubsection{Conservation function}
\label{sec:univ2conservation}
The product of reserve quantity of token$_1$, $r_1$, and reserve quantity of token$_2$, $r_2$, stays constant with swapping:
\begin{align}
  \label{eq:uniconsv2}
  \const = r_1 \cdot r_2
\end{align}

\subsubsection{Spot exchange rate}

Given the {\em equal value assumption} encoded in the pool smart contract, the implied spot price of assets in a liquidity pool can be derived based on the ratio between their reserve quantities. Specifically, denominated in token~1, the price of token$_2$ can be expressed as:
\begin{align}
  \label{eq:uniswapprice}
  {_1E_2} = \frac{r_1}{r_2}
\end{align}

\subsubsection{Swap amount}
Based on the Uniswap conservation function (\autoref{eq:unicons}), the amount of token$_2$ received $x_2$ (spent when $x_2<0$) given amount of token$_1$ spent $x_1$ (received when $x_1<0$) can be calculated following the steps described in \autoref{sec:form_sa}:
\begin{align}
  \label{eq:uniswapinput}
  r'_1 &
  = r_1 + x_1
  \nonumber \\
  r'_2 &
  = \frac{\const}{r'_1}
  \nonumber \\
  x_2  &
  = r_2 - r'_2
\end{align}



\subsubsection{Slippage}
The slippage that a Uniswap user experiences when swapping $x_1$ token$_1$ with $x_2$ token$_2$ can be expressed as:
\begin{align}
  S(x_1) = \frac{x_1/x_2}{{_1E_2}} -1 = \frac{x_1}{r_1}
  \label{eq:uniswapslippage}
\end{align}

\autoref{fig:uniswapslippage} illustrates the relationship between Uniswap slippage and normalized token$_1$ reserve change $\frac{x_1}{r_1}$.

\subsubsection{Divergence loss}
\label{sec:univ2div}
Given the equal value assumption with Uniswap, the reserve value of token~1, $V_1$, equals exactly half of original value of the entire pool $V$ (token$_1$ being numéraire):
\begin{align}
  \frac{V}{2} & = V_1 = V_2 = r_1
\end{align}

Should a \ac{lp} have held $r_1$ token$_1$ and $r_2$ token$_2$, then when token$_2$ appreciates by $\rho$ (depreciates when $\rho<0$), the total value of the original reserve composition $V_\text{held}$ becomes:
\begin{align}
  V_\text{held} & = V + V_2 \cdot \rho = r_1 \cdot (2+\rho)
\end{align}

With $r_1$ token$_1$ and $r_2$ token$_2$ locked in a liquidity pool from the beginning, their quantity ratio would have been updated through users' swapping to result in token$_2$'s price change of $\rho$. The equal value assumption still holds, and the updated pool value $V'$ becomes:
\begin{align}
  \frac{V'}{2} & = V'_1  = V'_2 = r'_1 = r_1 \cdot \sqrt{1+ \rho}
\end{align}

Note that $r'_2  = \frac{r_2}{\sqrt{1+ \rho}}$ and $p' = \frac{(1+\rho) r_1}{r_2}$, which preserves the invariance of $\const$, and reflects the change in token$_2$'s spot exchange rate against token$_1$.

As illustrated in \autoref{fig:unicons}, the divergence loss due to liquidity provision as opposed to holding can thus be expressed as a function of price change:
\begin{align}
  L(\rho) & = \frac{V'}{V_\text{held}} -1 = \frac{\sqrt{1+\rho}}{1+\frac{\rho}{2}} -1
  \label{eq:unidiv}
\end{align}

\subsection{Uniswap V3}

\subsubsection{Conservation function}
\label{sec:univ3conservation}
The conservation function of a Uniswap V3 pool is an aggregate of all the individual LP's conservation functions, each dependent on the exchange rate range that the LP wants to provide his liquidity for.

Suppose an \ac{lp} supplies $\revision{\mathcal{R}_1}$ token$_1$ and $\revision{\mathcal{R}_2}$ token$_2$, with the restriction that his liquidity is only provided for users swapping within a specific range of exchange rates: $[\frac{\revision{\mathcal{R}_1}}{\revision{\mathcal{R}_2} \cdot \mathcal{A}}, \frac{\revision{\mathcal{R}_1} \cdot \mathcal{A}}{\revision{\mathcal{R}_2}}]$ where $\mathcal{A} > 1$ and the initial exchange rate equals $\frac{\revision{\mathcal{R}_1}}{\revision{\mathcal{R}_2}}$.

The shape of the conservation function is then {\em identical} to liquidity provision of the following amounts under Uniswap V2:
\begin{align}
  r_1^\text{equiv} = \frac{\revision{\mathcal{R}_1}}{1-\frac{1}{\sqrt{\mathcal{A}}}}
  \quad
  \text{and}
  \quad
  r_2^\text{equiv} = \frac{\revision{\mathcal{R}_2}}{1-\frac{1}{\sqrt{\mathcal{A}}}}
  \nonumber
\end{align}

The bonding curve of a Uniswap V3 pool is equivalent to that of a Uniswap V2 one moving left along the x-axis by $(r_1^\text{equiv} - \revision{\mathcal{R}_1})$ and down along the y-axis by $(r_2^\text{equiv} - \revision{\mathcal{R}_2})$. Thus, Uniswap V3 conservation function can be expressed as:
\begin{align}
  [r_1 + (r_1^\text{equiv} - \revision{\mathcal{R}_1})] \cdot
  [r_2 + (r_2^\text{equiv} - \revision{\mathcal{R}_2})] = r_1^\text{equiv} \cdot r_2^\text{equiv}
  \nonumber \\
  \left(
  r_1 + \frac{\revision{\mathcal{R}_1}}{\sqrt{\mathcal{A}}-1}
  \right) \cdot
  \left(
  r_2 + \frac{\revision{\mathcal{R}_2}}{\sqrt{\mathcal{A}}-1}
  \right) = \frac{
    \mathcal{A} \cdot \revision{\mathcal{R}_1} \cdot \revision{\mathcal{R}_2}}{(\sqrt{\mathcal{A}} -1)^2}
  \label{eq:unicons}
\end{align}
where $0 \leq r_1 \leq \revision{\mathcal{R}_1} \cdot (\sqrt{A} + 1)$ and $0 \leq r_2 \leq \revision{\mathcal{R}_2} \cdot (\sqrt{A} + 1)$.\footnote{\autoref{eq:unicons} is equivalent to $(x+\frac{L}{\sqrt{pb}})(y+L\sqrt{pa}) = L^2$, equation (2.2) from page 2 of the Uniswap V3 whitepaper \cite{adams2021v3core}. This can be seen by equating their notation with ours as follows:
  $L \coloneqq \frac{\sqrt{\mathcal{A} \cdot C_1 \cdot C_2}}{\sqrt{\mathcal{A}} -1}$,
  $x \coloneqq r_1$,
  $y \coloneqq r_2$,
  $pa \coloneqq \frac{C_1 \cdot \mathcal{A}}{C_2}$,
  $pb \coloneqq \frac{C_1}{C_2 \cdot \mathcal{A}}$.}


\subsubsection{Exchange rate}

\begin{align}
  \label{eq:uniswapv3ex}
  {}_1E_2 = \frac{
    r_1 + \frac{\revision{\mathcal{R}_1}}{\sqrt{\mathcal{A}}-1}
  }{
    r_2 + \frac{\revision{\mathcal{R}_2}}{\sqrt{\mathcal{A}}-1}
  }
\end{align}

Note that when token$_1$ is depleted, i.e. $r_1 = 0$, then
\begin{align}
  r_2 + \tfrac{\revision{\mathcal{R}_2}}{\sqrt{\mathcal{A}}-1} = \tfrac{\mathcal{A} \cdot \revision{\mathcal{R}_2}}{
    \sqrt{\mathcal{A}} -1
  }
  \nonumber \\
  {}_1E_2 = \tfrac{\revision{\mathcal{R}_1}}{\revision{\mathcal{R}_2} \cdot \mathcal{A}
  }
  \nonumber
\end{align}
Similarly, when token$_2$ is depleted, i.e. $r_2 = 0$, then ${}_1E_2 =  \frac{\revision{\mathcal{R}_1} \cdot \mathcal{A}}{\revision{\mathcal{R}_2}}$. Be reminded that
$[\frac{\revision{\mathcal{R}_1}}{\revision{\mathcal{R}_2} \cdot \mathcal{A}}, \frac{\revision{\mathcal{R}_1} \cdot \mathcal{A}}{\revision{\mathcal{R}_2}}]$ is exactly the pre-specified exchange rate range that the liquidity supports.

\subsubsection{Swap amount} The swap amount can be derived from the conservation function \autoref{eq:unicons}:
\begin{align}
  \label{eq:uniswapv3input}
  r'_1 &
  = r_1 + x_1
  \nonumber \\
  r'_2 &
  =
  \tfrac{\revision{\mathcal{R}_1} \revision{\mathcal{R}_2}}{(1-\frac{1}{\sqrt{\mathcal{A}}})^2} \Big/
  \left(
  r'_1 + \tfrac{\revision{\mathcal{R}_1}}{\sqrt{\mathcal{A}}-1}
  \right) - \tfrac{\revision{\mathcal{R}_2}}{\sqrt{\mathcal{A}}-1}
  \nonumber \\
  x_2  &
  = r_2 - r'_2
\end{align}


\subsubsection{Slippage} The slippage should have the same magnitude as in Uniswap V2, but with $r_1$ amplified by an increase of $\frac{\revision{\mathcal{R}_1}}{\sqrt{\mathcal{A}}-1}$:
\begin{align}
  S(x_1) = \frac{x_1/x_2}{{_1E_2}} -1 = \frac{x_1}{r_1 +  \frac{\revision{\mathcal{R}_1}}{\sqrt{\mathcal{A}}-1}
  }
  \label{eq:uniswapv3slippage}
\end{align}

Again, when $\mathcal{A} \rightarrow \infty$, the slippage function approximates a Uniswap V2 one; when $\mathcal{A} \rightarrow 1$, slippage is restrained as long as there exists liquidity for both assets (\autoref{fig:uniswapslippage}).

\subsubsection{Divergence loss}
\label{sec:unidivergenceloss}

Using the intermediary results from \autoref{sec:univ2div}, we can easily derive $V_\text{held}$, $r_1'$ and $r_2'$, and subsequently $V'$:
\begin{align}
  V_\text{held} & = C_1 \cdot (2+\rho)
  \\
  r'_1          & = r_1^\text{equiv}
  \sqrt{1+\rho}
  - \tfrac{\revision{\mathcal{R}_1}}{\sqrt{\mathcal{A}}-1}
  = \tfrac{\revision{\mathcal{R}_1} \cdot (\sqrt{1+\rho} -
    \frac{1}{\sqrt{\mathcal{A}}})
  }{1-\frac{1}{\sqrt{\mathcal{A}}}}
  \nonumber                                                                                                                                            \\
  r'_2          & = \tfrac{r_2^\text{equiv}}{\sqrt{1+\rho}}
  - \tfrac{\revision{\mathcal{R}_2}}{\sqrt{\mathcal{A}}-1}
  =
  \tfrac{ \revision{\mathcal{R}_2}
    (
    \frac{1}{\sqrt{1+\rho}} -\frac{1}{\sqrt{\mathcal{A}}}
    )
  }{
    1-\frac{1}{\sqrt{\mathcal{A}}}
  }
  \nonumber                                                                                                                                            \\
  V'            & = V_1' + V_2' = r_1' + \tfrac{\revision{\mathcal{R}_1} (1+\rho) r'_2}{\revision{\mathcal{R}_2}} = \tfrac{ \revision{\mathcal{R}_1} (
    2 \sqrt{1+\rho} - \frac{2+\rho}{\sqrt{\mathcal{A}}}
    )
  }{1-\frac{1}{\sqrt{\mathcal{A}}}}
\end{align}

When $-1 \leq \rho \leq \frac{1}{\mathcal{A}} -1$, then token$_1$ becomes depleted, and the \ac{lp} is left with token$_2$:
\begin{align}
  V' & = \frac{\revision{\mathcal{R}_1} (1+\rho)}{\revision{\mathcal{R}_2}} \cdot r'_2   =
  \revision{\mathcal{R}_1} \cdot (1+\rho) \cdot \left(\sqrt{\mathcal{A}} + 1 \right)
\end{align}

When $\rho \geq \mathcal{A} -1$, then token$_2$ becomes depleted, and the \ac{lp} is left with token$_1$ only:
\begin{align}
  V' & = r'_1   =
  \revision{\mathcal{R}_1}  \cdot \left(\sqrt{\mathcal{A}} + 1 \right)
\end{align}

The divergence loss can thus be calculated as:
\begin{align}
  L(\rho)  = \frac{V'}{V_\text{held}} -1
  = \begin{cases}
      \frac{(\rho + 1) \cdot \sqrt{\mathcal{A}}  - 1}{2+\rho},
       & -1 \leq \rho \leq \frac{1}{\mathcal{A}} -1
      \\
      \frac{\frac{\sqrt{1+\rho}
        }{
          1+\frac{\rho}{2}
        } - 1
      }
      {1- \frac{1}{\sqrt{\mathcal{A}}}},
       & \frac{1}{\mathcal{A}} -1 \leq \rho \leq \mathcal{A} -1
      \\
      \frac{\sqrt{\mathcal{A}}-1-\rho}{2+\rho},
       & \rho \geq \mathcal{A} -1
    \end{cases}
  \label{eq:uniswapv3div}
\end{align}


\subsection{Balancer}

\subsubsection{Conservation function}

Balancer implements a conservation function with a weighted-product invariant (\autoref{fig:balcons}). Specifically, the product of reserve quantities each raised to the power of its weight stays constant with swapping:
\begin{align}
  \label{eq:balcons}
  \const = \prod_{k}r_k^{w_k}
\end{align}

\subsubsection{Spot exchange rate} Given the quantity ratio $r_1 : r_2$ between token$_1$ and 2 and the implicit assumption on their value ratio $w_1 : w_2$, the price of token$_2$ denominated by token$_1$ can be expressed as:

\begin{equation}
  \label{eq:balancer}
  {_1E_2}  = \frac{r_1 \cdot w_2}{r_2 \cdot w_1}
\end{equation}

\subsubsection{Swap amount}
We investigate the case when a user swaps token$_1$ for token$_2$, while the reserves of all other assets remain untouched in the pool.
Based on the Balancer conservation function (\autoref{eq:balcons}), the amount of token$_2$ received $x_2$ (spent when $x_2<0$) given amount of token$_1$ spent $x_1$ (received when $x_1<0$) can be calculated following the steps described in \autoref{sec:form_sa}:
\begin{align}
  \label{eq:balancerinput}
  r'_1 & = r_1 + x_1
  \nonumber                                                 \\
  r'_2 & = r_2\left(\frac{r_1}{r'_1}\right)^\frac{w_1}{w_2}
  \nonumber                                                 \\
  x_2  & = r_2 - r'_2
\end{align}



\subsubsection{Slippage}

The slippage that a Balancer user experiences when swapping $x_1$ token$_1$ with $x_2$ token$_2$ can be expressed as:
\begin{align}
  S(x_1) = \frac{x_1/x_2}{{_1E_2}} -1 = \frac{
  \frac{x_1}{r_1} \cdot \frac{w_1}{w_2}
  }{
  1- \left(\frac{r_1}{r'_1}\right)^{\frac{w_1}{w_2}}
  }-1
  \label{eq:balancerslippage}
\end{align}

\autoref{fig:balancerslippage} illustrates the relationship between Uniswap slippage and normalized token$_1$ reserve change $\frac{x_1}{r_1}$.

\subsubsection{Divergence loss}

Given the constant value ratio assumption with Balancer, the value of the entire pool $V$ can be expressed by the reserve quantity of token$_1$, $r_1$ divided by its weight $w_1$ (token$_1$ being numéraire):
\begin{align}
  V & =
  \frac{V_1}{w_1} =
  \frac{V_2}{w_2}
  = \frac{V_k}{w_k} = \frac{r_1}{w_1}
\end{align}

If token$_2$ appreciates by $\rho$ (depreciates when $\rho<0$) while all other tokens' prices remain unchanged, the total value of the original reserve composition, when held outside of the pool, $V_\text{held}$ becomes:
\begin{align}
  V_\text{held} & = V + V_2 \cdot \rho
  = V \cdot (1+ w_2 \cdot \rho)
\end{align}

With $r_1$ token$_1$ and $r_2$ token$_2$ locked in a liquidity pool from the beginning, their quantity ratio would have been updated through users' swapping to result in token$_2$'s price change of $\rho$. The value ratio between the pool, token$_1$ and token$_2$, remains $1:w_1:w_2$, and the updated pool value $V'$ becomes:
\begin{align}
  V' & = \frac{V'_1}{w_1} =
  \frac{r_1'}{w_1} =
  \frac{r_1 \cdot (1+ \rho)^{w_2}}{w_1}
  = V \cdot (1+ \rho)^{w_2}
\end{align}

The exchange rate range corresponds the \ac{lp}'s range requirement. Specifically, when $r'_2  = \frac{r_2}{(1+ \rho)^{1- w_2}}$
and
$r'_k = r_k \cdot (1+\rho)^{w_2}$ for $k\neq 2$, reflecting the assumed scenario that only the value of token$_2$ appreciates by $\rho$, while the value of all other tokens against token$_1$ remains unchanged.

As illustrated in \autoref{fig:balancerdiv}, the divergence loss due to liquidity provision as opposed to holding can thus be expressed as a function of price change:
\begin{align}
  L(\rho) & =
  \frac{V'}{V_\text{held}} -1 = \frac{(1+ \rho)^{w_2}}{1+w_2 \cdot \rho} -1
  \label{eq:balancerdiv}
\end{align}

\subsection{Curve}

\subsubsection{Conservation function}
\label{sec:curve-cons-func}

As assets from the same pool are connected to the same peg by design, the ideal exchange rate between them should always equal 1. Theoretically, this could be achieved by a constant-sum invariant. Nevertheless, Curve seeks to allow an exchange rate to deviate from 1, in order to reflect the supply-demand dynamic, while simultaneously keeping the slippage low.

Curve achieves this by interpolating between two invariants, constant sum and constant product \cite{Egorov2019}, with hyperparameter $\mathcal{A}$ as the interpolating factor (\autoref{eq:curvcons}).\footnote{Note that $\mathcal{A}$ here is equivalent to $A \cdot n^n$ in Curve's white paper \cite{Egorov2019}.} When $\mathcal{A} \rightarrow 0$, the conservation function boils down to a constant-product one, as with Uniswap; when $\mathcal{A} \rightarrow +\infty$, the conservation function is essentially a constant-sum one with constant exchange rate equal to 1 (\autoref{fig:curvcons}).
\begin{align}
  \label{eq:curvcons}
  \mathcal{A} \left(\tfrac{\sum\limits_k r_k}{\const}-1\right) & = \tfrac{
    \left(\frac{\const}{n}\right)^n
  }{\prod\limits_k r_k} -1
\end{align}

\subsubsection{Spot exchange rate}

Rearrange \autoref{eq:curvcons} and let
\begin{align}
  Z(r_1,r_2)
  =
  \tfrac{
    \left(\frac{\const}{n}\right)^n
  }{r_1 r_2 \prod\limits_{k \neq 1,2} r_k} -1 - \mathcal{A} \left(\tfrac{r_1 + r_2 + \sum\limits_{k \neq 1,2} r_k}{\const}-1\right)
  \nonumber
\end{align}

Following \autoref{sec:genform}, the spot exchange rate can be calculated as:
\begin{align}
  _1E_2 = \tfrac{
    \partial Z(r_1,r_2)/\partial r_2
  }{
    \partial Z(r_1,r_2)/\partial r_1
  }
  = \tfrac{
    r_1 \cdot \left[\mathcal{A} \cdot r_2 \cdot \prod\limits_k r_k + \const \cdot \left(\frac{\const}{n}\right)^n\right]
  }{
    r_2 \cdot \left[\mathcal{A}  \cdot r_1 \cdot \prod\limits_k r_k + \const \cdot \left(\frac{\const}{n}\right)^n\right]
  }
  \label{eq:curvexchange}
\end{align}


\subsubsection{Swap amount}

We investigate the case when a user swaps token$_1$ for token$_2$, while the reserves of all other assets remain untouched in the pool.
Based on the Curve conservation function (\autoref{eq:curvcons}), the amount of token$_2$ received $x_2$ (spent when $x_2<0$) given amount of token$_1$ spent $x_1$ (received when $x_1<0$) can be calculated following the steps below:
\begin{align}
  \label{eq:curveinput}
  r'_1 & = r_1 + x_1
  \nonumber                                                                                                       \\
  r'_2 & = \tfrac{\sqrt{\frac{4 \const \left(\frac{\const}{n}\right)^n}{\mathcal{A} \cdot \prod\limits_{k\neq 2}'
      }+\left[\left(1-\frac{1}{\mathcal{A}}\right) \const -\sum\limits_{k\neq 2}'\right]^2}+\left(1-\frac{1}{\mathcal{A}}\right) \const - \sum\limits_{k\neq 2}'
  }{2}
  \nonumber                                                                                                       \\
  x_2  & = r_2 - r'_2
\end{align}
where
$
  \prod\limits_{k\neq 2}' = r'_1 \cdot \prod\limits_{k \neq 1,2} r_k
  \quad \text{and} \quad
  \sum\limits_{k\neq 2}' = r'_1 + \sum\limits_{k \neq 1,2} r_k
$.

\subsubsection{Slippage}

As illustrated in \autoref{fig:curveslippage}, the slippage that a Curve user experiences when swapping $x_1$ token$_1$ with $x_2$ token$_2$ can be expressed as:
\begin{align}
  S(x_1) & = \tfrac{x_1/x_2}{_1E_2} -1
  \label{eq:curveslippage}
  = \tfrac{
    \frac{
      x_1 \cdot \left[\mathcal{A}  \cdot r_1 \cdot \prod\limits_k r_k + \const \cdot \left(\frac{\const}{n}\right)^n\right]
    }{
      r_1 \cdot \left[\mathcal{A} \cdot r_2 \cdot \prod\limits_k r_k + \const \cdot \left(\frac{\const}{n}\right)^n\right]
    }
  }{1 -
    \frac{\sqrt{\frac{4 \const \left(\frac{\const}{n}\right)^n}{a \cdot \prod\limits_{k\neq 2}'
        }+\left[\left(1-\frac{1}{a}\right) \const -\sum\limits_{k\neq 2}'\right]^2}+\left(1-\frac{1}{a}\right) \const - \sum\limits_{k\neq 2}'
    }{2 r_2}
  }-1
\end{align}

\subsubsection{Divergence loss}

Curve's divergence loss in full form cannot be easily presented in a concise and comprehensible fashion.
Therefore, for Curve, we use the generalized method to calculate its divergence loss as described in \autoref{sec:genform}. The divergence loss in the case of a 2-asset pool is presented in \autoref{fig:curvediv}.

\subsection{DODO}
\label{app:dodo}

\subsubsection{Spot exchange rate}
\label{sec:dodoexchange}


The exchange rate between the two assets in a DODO pool is set by the market rate with an adjustment based on the pool composition.
We denote the market exchange rate as $P$, namely $1 \text{ token}_2 = P \text{ token}_1$, and the initial reserve for token$_1$ and token$_2$ as $\revision{\mathcal{R}_1}$ and $\revision{\mathcal{R}_2}$ respectively. The formula \autoref{eq:dodoexchange} sets the exchange rate $_1E_2$ higher than the market rate $P$---i.e. token$_2$ exhibits higher price in the pool than in the market, when the reserve of token$_1$ $r_1$ exceeds its initial state $\revision{\mathcal{R}_1}$, and sets $_1E_2$ lower than $P$---i.e. token$_1$ more expensive than its market value, when $r_1$ falls short of $\revision{\mathcal{R}_1}$. Formally,
\begin{align}
  \label{eq:dodoexchange}
  {_1E_2} & =
  \begin{cases}
    {P} \left[ 1 + \mathcal{A}\left( \left(\frac{\revision{\mathcal{R}_2}}{r_2}\right)^2-1\right) \right],         & r_1 \geq \revision{\mathcal{R}_1}
    \\
    {P} \Bigg/ \left[ 1 + \mathcal{A}\left( \left(\frac{\revision{\mathcal{R}_1}}{r_1}\right)^2-1\right) \right] , & r_1 \leq \revision{\mathcal{R}_1}
  \end{cases}
\end{align}


\subsubsection{Conservation function}
\label{sec:dodoconservation}

DODO's conservation function can be derived from its exchange formula \autoref{eq:dodoexchange}. In particular, the initial state of token$_1$ and token$_2$ reserves, $\revision{\mathcal{R}_1}$ and $\revision{\mathcal{R}_2}$ can be regarded as the two invariants of the conservation function.
This aligns with the definition according to our framework (\autoref{sec:formal}), as $\revision{\mathcal{R}_1}$ and $\revision{\mathcal{R}_2}$ remain constant with swapping activities, but get updated with liquidity provision or withdrawal.
\begin{align}
  \label{eq:dodocons}
        & r_1-\revision{\mathcal{R}_1}  = \int_{r_2}^{\revision{\mathcal{R}_2}}
  {P} \left[ 1 + \mathcal{A}\left( \left(\tfrac{\revision{\mathcal{R}_2}}{\delta}\right)^2-1\right) \right]
  \, d\delta
  = \,  & P \cdot
  (\revision{\mathcal{R}_2} - r_2)
  \cdot \left[1 + \mathcal{A} \cdot \left(\tfrac{\revision{\mathcal{R}_2}}{r_2} -1 \right) \right],
        & \quad r_1 \geq \revision{\mathcal{R}_1}                                                                                                \\
        & r_2-\revision{\mathcal{R}_2}  = \int_{r_1}^{\revision{\mathcal{R}_1}}
  \frac{ 1 + \mathcal{A}\left( \left(\frac{\revision{\mathcal{R}_1}}{\delta}\right)^2-1\right) }
  {P}
  \, d\delta
  = \,  & \frac{(\revision{\mathcal{R}_1} - r_1) \cdot \left[1 + \mathcal{A} \cdot \left(\frac{\revision{\mathcal{R}_1}}{r_1} -1 \right) \right]
  }{P}, & \quad  r_1 \leq \revision{\mathcal{R}_1}
\end{align}

In the special case of $\mathcal{A} = 1$, when $C_1 = P \cdot C_2$, i.e. liquidity provided on both assets are of equal value, then DODO's conservation function is equivalent to Uniswap, with $r_1 \cdot r_2 = C_1 \cdot P \cdot C_2$. This can be observed from \autoref{fig:consfunc}, where the DODO's conservation function curve with $\mathcal{A} \rightarrow 1$ appears identical to that of Uniswap.

\subsubsection{Swap amount}

The swap amount can be derived directly from the DODO conservation function (\autoref{eq:dodocons}):
\begin{align}
   & r_1' = r_1 + x_1
  \nonumber
  \\
   & r_2' =
  \begin{cases}
    \tfrac{
      {
          \revision{\mathcal{R}_1}- r_1'+ P \cdot \revision{\mathcal{R}_2} \cdot (1-2\mathcal{A}) +
          \atop
          \sqrt{
            \left[\revision{\mathcal{R}_1}-r_1' + P \cdot \revision{\mathcal{R}_2} \cdot (1-2\mathcal{A})\right]^2 + 4 \mathcal{A} \cdot (1-\mathcal{A}) \cdot (P \cdot \revision{\mathcal{R}_2})^2
          }
        }
    }{2 P \cdot (1- \mathcal{A})},
          & r_1' \geq \revision{\mathcal{R}_1} \\
    \revision{\mathcal{R}_2} + \frac{(\revision{\mathcal{R}_1} - r_1') \cdot \left[1 + \mathcal{A} \cdot \left(\frac{\revision{\mathcal{R}_1}}{r_1'} -1 \right) \right]
    }{P}, & r_1' \leq \revision{\mathcal{R}_1}
  \end{cases}
  \nonumber           \\
   & x_2 = r_2 - r_2'
  \label{eq:dodoswap}
\end{align}



\subsubsection{Slippage}

As illustrated in \autoref{fig:dodolippage}, the slippage that a DODO user experiences when swapping $x_1$ token$_1$ with $x_2$ token$_2$ can be expressed as:
\begin{align}
  \label{eq:dodoslippage}
   & S(x_1) =
  \begin{cases}
    \frac{2 \cdot (1-\mathcal{A}) \cdot x_1
    }{
      {r_1'-\revision{\mathcal{R}_1}+\revision{\mathcal{R}_2} \cdot P -
          \atop
          \sqrt{
            \left[\revision{\mathcal{R}_1}-r_1' + P \cdot \revision{\mathcal{R}_2} \cdot (1-2\mathcal{A})\right]^2 + 4 \mathcal{A} \cdot (1-\mathcal{A}) \cdot (P \cdot \revision{\mathcal{R}_2})^2
          }
        }
    } -1,
                                                                                                                                & r_1' \geq \revision{\mathcal{R}_1} \\
    \frac{x_1
    }{(r_1'-\revision{\mathcal{R}_1}) \cdot \left[1+A \cdot \left( \frac{\revision{\mathcal{R}_1}}{r_1'}-1 \right) \right]} -1, & r_1' \leq \revision{\mathcal{R}_1}
  \end{cases}
\end{align}

\subsubsection{Divergence loss}

DODO eliminates the kind of divergence loss seen in previously discussed protocols by setting the ratio between the reserve assets supplied by the \ac{lp} as the pool's equilibrium state (see \autoref{sec:dodointro}).


%% file: exhibits/landscape_table.tex
\begin{landscape}
    \input{exhibits/formula_comparison}

\end{landscape}

%% file: exhibits/formula_comparison.tex
\begin{table}[H]
  \centering
  \caption{Function comparison table of Uniswap, Balancer, Curve and DODO. Formulas are derived in~\autoref{app:uniswapv2}--\autoref{app:dodo}. Conservation functions are visualized in \autoref{fig:consfunc}, slippage functions in \autoref{fig:slippage} and divergence loss functions in \autoref{fig:divloss}.
    Python implementation is available on \href{https://github.com/xujiahuayz/bondingcurves}{GitHub}.}
  \label{tab:functioncomparison}
  \tiny
  \setlength{\tabcolsep}{1pt} 
  \renewcommand{\arraystretch}{1.5} 
  \begin{tabular}{l|l|l|l|l|l}
    \toprule
                                                                                                    & \textbf{Uniswap V2}                                                                          & \textbf{Uniswap V3}                                                                                        & \textbf{Balancer}                                                                                            & \textbf{Curve}                   & \textbf{DODO}                 \\
    \midrule
    \textbf{\begin{tabular}[l]{l}Conservation function
                \\ \\ $\begin{aligned}
            Z(\{r_k\}; \invt) = 0
          \end{aligned}$\end{tabular}}                                              & $\begin{aligned}
                                                                                                                \const = r_1 \cdot r_2
                                                                                                              \end{aligned}$                                                                       &
    $\begin{aligned} \left(
         r_1 + \frac{\revision{\mathcal{R}_1}}{\sqrt{\mathcal{A}}-1}
         \right) \cdot
         \left(
         r_2 + \frac{\revision{\mathcal{R}_2}}{\sqrt{\mathcal{A}}-1}
         \right) \\ = \frac{
           \mathcal{A} \cdot \revision{\mathcal{R}_1} \cdot \revision{\mathcal{R}_2}}{(\sqrt{\mathcal{A}} -1)^2}\end{aligned}
    $
                                                                                                    &
    $\begin{aligned} \const = \prod_{k}r_k^{w_k} \end{aligned}$                                     & $\begin{aligned} \mathcal{A} \left(\tfrac{\sum\limits_k r_k}{\const}-1\right) & = \tfrac{
                  \left(\frac{\const}{n}\right)^n
                }{\prod\limits_k r_k} -1\end{aligned}$   &
    $\begin{aligned}
         \begin{cases}
          r_1-\revision{\mathcal{R}_1} = P \cdot
          (\revision{\mathcal{R}_2} - r_2)
          \cdot \left[1 + \mathcal{A} \cdot \left(\tfrac{\revision{\mathcal{R}_2}}{r_2} -1 \right) \right], & r_1 \geq \revision{\mathcal{R}_1}
          \\
          r_2-\revision{\mathcal{R}_2}  = \frac{(\revision{\mathcal{R}_1} - r_1) \cdot \left[1 + \mathcal{A} \cdot \left(\frac{\revision{\mathcal{R}_1}}{r_1} -1 \right) \right]
          }{P},                                                                                             & r_1 \leq \revision{\mathcal{R}_1}
        \end{cases}  \end{aligned}$                                                                                                                                                                                                                                                                                                                                                                                                                                                    \\

    \hline
    \textbf{\begin{tabular}[l]{l}Spot exchange rate\\ \\ $\begin{aligned} _iE_o(\{r_k\}; \invt
            ) \\
            = \frac{
              \partial Z(\{r_k\}; \invt
              )/\partial r_o
            }{
              \partial Z(\{r_k\}; \invt
              )/\partial r_i
            }\end{aligned}$ \end{tabular}} & $\begin{aligned}
                                                                                                           \frac{r_1}{r_2}
                                                                                                         \end{aligned}$                                                                             & $\begin{aligned} \frac{
                                                                                                                                                                                                           r_1 + \frac{\revision{\mathcal{R}_1}}{\sqrt{\mathcal{A}}-1}
                                                                                                                                                                                                         }{
                                                                                                                                                                                                           r_2 + \frac{\revision{\mathcal{R}_2}}{\sqrt{\mathcal{A}}-1}
                                                                                                                                                                                                         }\end{aligned}$                                                                               & $\begin{aligned} \frac{r_1 \cdot w_2}{r_2 \cdot w_1} \end{aligned}$                                          & $\begin{aligned} \tfrac{
                                                                                                                                                                                                                                                                                                                                                                                                                           r_1 \cdot \left[\mathcal{A} \cdot r_2 \cdot \prod\limits_k r_k + \const \cdot \left(\frac{\const}{n}\right)^n\right]
                                                                                                                                                                                                                                                                                                                                                                                                                         }{
                                                                                                                                                                                                                                                                                                                                                                                                                           r_2 \cdot \left[\mathcal{A}  \cdot r_1 \cdot \prod\limits_k r_k + \const \cdot \left(\frac{\const}{n}\right)^n\right]
                                                                                                                                                                                                                                                                                                                                                                                                                         }\end{aligned}$ & $\begin{aligned} \begin{cases}
          {P} \left[ 1 + \mathcal{A}\left( \left(\frac{\revision{\mathcal{R}_2}}{r_2}\right)^2-1\right) \right],         & r_1 \geq \revision{\mathcal{R}_1}
          \\
          {P} \Bigg/ \left[ 1 + \mathcal{A}\left( \left(\frac{\revision{\mathcal{R}_1}}{r_1}\right)^2-1\right) \right] , & r_1 \leq \revision{\mathcal{R}_1}
        \end{cases} \end{aligned}$ \\
    \hline
    \multirow{3}{*}{\textbf{\begin{tabular}[l]{l}
                                  Post-swap token$_1$ reserve $r'_1$
                                  \\ \\ \\
                                  Post-swap token$_2$ reserve $r'_2$
                                  \\ \\  \\  Swap amount $x_2$
                                \end{tabular}}
    }                                                                                               & \multicolumn{5}{c}{\(r_1 + x_1\)}                                                                                                                                                                                                                                                                                                                                                           \\ \cdashline{1-6}
                                                                                                    & $\begin{aligned}\frac{\const}{r'_1} \end{aligned}$                                           &

    $\begin{aligned} \frac{\tfrac{\revision{\mathcal{R}_1} \revision{\mathcal{R}_2}}{(1-\frac{1}{\sqrt{\mathcal{A}}})^2}}{ \left(
           r'_1 + \tfrac{\revision{\mathcal{R}_1}}{\sqrt{\mathcal{A}}-1}
           \right)} - \tfrac{\revision{\mathcal{R}_2}}{\sqrt{\mathcal{A}}-1}\end{aligned}$

                                                                                                    & $\begin{aligned} r_2\left(\frac{r_1}{r'_1}\right)^\frac{w_1}{w_2} \end{aligned}$             & $\begin{aligned} \tfrac{\sqrt{\frac{4 \const \left(\frac{\const}{n}\right)^n}{\mathcal{A} \cdot \prod\limits_{k\neq 2}'
                                                                                                                                                                                                              }+\left[\left(1-\frac{1}{\mathcal{A}}\right) \const -\sum\limits_{k\neq 2}'\right]^2}+\left(1-\frac{1}{\mathcal{A}}\right) \const - \sum\limits_{k\neq 2}'
                                                                                                                                                                                                          }{2}\end{aligned}$ & $\begin{aligned} \begin{cases}
          \tfrac{
            {
                \revision{\mathcal{R}_1}- r_1'+ P \cdot \revision{\mathcal{R}_2} \cdot (1-2\mathcal{A}) +
                \atop
                \sqrt{
                  \left[\revision{\mathcal{R}_1}-r_1' + P \cdot \revision{\mathcal{R}_2} \cdot (1-2\mathcal{A})\right]^2 + 4 \mathcal{A} \cdot (1-\mathcal{A}) \cdot (P \cdot \revision{\mathcal{R}_2})^2
                }
              }
          }{2 P \cdot (1- \mathcal{A})},
                & r_1' \geq \revision{\mathcal{R}_1} \\
          \revision{\mathcal{R}_2} + \frac{(\revision{\mathcal{R}_1} - r_1') \cdot \left[1 + \mathcal{A} \cdot \left(\frac{\revision{\mathcal{R}_1}}{r_1'} -1 \right) \right]
          }{P}, & r_1' \leq \revision{\mathcal{R}_1}
        \end{cases} \end{aligned}$                                                                                                                                                                                    \\
    \cdashline{1-6}                                                                                 & \multicolumn{5}{c}{\(r'_2 - r_2\)}                                                                                                                                                                                                                                                                                                                                                          \\
    \hline
    \textbf{\begin{tabular}[l]{l}
                Slippage \\ \\
                $
                  \begin{aligned}
            S(x_i, \{r_k\}; \invt) \\
            = \frac{x_i/x_o}{_iE_o} -1
          \end{aligned}
                $
              \end{tabular}}
                                                                                                    & $\begin{aligned} \frac{x_1}{r_1} \end{aligned}$                                              &
    $\begin{aligned} \frac{x_1}{r_1 +  \frac{\revision{\mathcal{R}_1}}{\sqrt{\mathcal{A}}-1}
         }\end{aligned}$                                      & $\begin{aligned} \frac{
                                                                 \frac{x_1}{r_1} \cdot \frac{w_1}{w_2}
                                                                 }{
                                                                 1- \left(\frac{r_1}{r'_1}\right)^{\frac{w_1}{w_2}}
                                                                 }-1\end{aligned}$
                                                                                                    & $\begin{aligned} \tfrac{
                                                                                                             \frac{
                                                                                                               x_1 \cdot \left[\mathcal{A}  \cdot r_1 \cdot \prod\limits_k r_k + \const \cdot \left(\frac{\const}{n}\right)^n\right]
                                                                                                             }{
                                                                                                               r_1 \cdot \left[\mathcal{A} \cdot r_2 \cdot \prod\limits_k r_k + \const \cdot \left(\frac{\const}{n}\right)^n\right]
                                                                                                             }
                                                                                                           }{1 -
                                                                                                             \frac{\sqrt{\frac{4 \const \left(\frac{\const}{n}\right)^n}{a \cdot \prod\limits_{k\neq 2}'
                                                                                                                 }+\left[\left(1-\frac{1}{a}\right) \const -\sum\limits_{k\neq 2}'\right]^2}+\left(1-\frac{1}{a}\right) \const - \sum\limits_{k\neq 2}'
                                                                                                             }{2 r_2}
                                                                                                           }-1\end{aligned}$                                                                     & $\begin{aligned} \begin{cases}
          \frac{2 \cdot (1-\mathcal{A}) \cdot x_1
          }{
            {r_1'-\revision{\mathcal{R}_1}+\revision{\mathcal{R}_2} \cdot P -
                \atop
                \sqrt{
                  \left[\revision{\mathcal{R}_1}-r_1' + P \cdot \revision{\mathcal{R}_2} \cdot (1-2\mathcal{A})\right]^2 + 4 \mathcal{A} \cdot (1-\mathcal{A}) \cdot (P \cdot \revision{\mathcal{R}_2})^2
                }
              }
          } -1,
                                                                                                                                      & r_1' \geq \revision{\mathcal{R}_1} \\
          \frac{x_1
          }{(r_1'-\revision{\mathcal{R}_1}) \cdot \left[1+\mathcal{A} \cdot \left( \frac{\revision{\mathcal{R}_1}}{r_1'}-1 \right) \right]} -1, & r_1' \leq \revision{\mathcal{R}_1}
        \end{cases} \end{aligned}$                                                                                                                                                                                                                                         \\
    \hline
    \begin{tabular}[l]{l}
      {\bf Divergence loss}
      \\ \\
      $\begin{aligned}
           L(\rho, \{r_k\}; \invt) \\
           = \frac{
             V'(\rho, \{r_k\}; \invt)
           }{
             V_\text{held}(\rho; \{r_k\}, \invt)
           } -1
         \end{aligned}$
    \end{tabular}
                                                                                                    & $\begin{aligned} \frac{\sqrt{1+\rho}}{1+\frac{\rho}{2}} -1 \end{aligned}$ & $\begin{aligned}  \begin{cases}
          \frac{(\rho + 1) \cdot \sqrt{\mathcal{A}}  - 1}{2+\rho},
           & -1 \leq \rho \leq \frac{1}{\mathcal{A}} -1
          \\
          \frac{\frac{\sqrt{1+\rho}
            }{
              1+\frac{\rho}{2}
            } - 1
          }
          {1- \frac{1}{\sqrt{\mathcal{A}}}},
           & \frac{1}{\mathcal{A}} -1 \leq \rho \leq \mathcal{A} -1
          \\
          \frac{\sqrt{\mathcal{A}}-1-\rho}{2+\rho},
           & \rho \geq \mathcal{A} -1
        \end{cases} \end{aligned}$                                                & $\begin{aligned} \frac{(1+ \rho)^{w_2}}{1+w_2 \cdot \rho} -1 \end{aligned}$ & Complex                          & 0 at equilibrium              \\
    \bottomrule
  \end{tabular}
\end{table}

%% file: sections/attack_app.tex
\section{Selected Attack Algorithms}
\label{app:attack_alg}

In this appendix, we show four algorithms of selected attacks.

\input{exhibits/attack_oracle}
\input{exhibits/attack_rugpull}
\input{exhibits/attack_lp}
\input{exhibits/attack_frontback}

%% file: exhibits/attack_oracle.tex
\begin{algorithm}[tbh]
  \small
  \floatname{algorithm}{Attack Algorithm}
  \caption{\revision{Flash-loan-funded price oracle attack}}
  \label{flashloanattack}
  \begin{algorithmic}[1]
    \STATE {\bf Take a flash loan} to borrow $x_A$ token$_A$ from a lending platform, whose value is equivalent to $x_B$ token$_B$ at market price.
    \STATE {\bf Swap} $x_A$ token$_A$ for $x_B - \Delta_1$ token$_B$ on an \ac{amm}, pushing
    the new price of token$_A$ in terms of token$_B$ down to $\frac{x_B - \Delta_2}{x_A}$,
    where $\Delta_2 > \Delta_1 > 0$ due to slippage.
    \STATE {\bf Borrow} $x_A+ \Delta_3$ token$_A$ with $x_B - \Delta_1$ token$_B$ as collateral on a lending platform that uses the \ac{amm} as their sole price oracle.
    To temporarily satisfy overcollateralization,
    $\frac{x_B - \Delta_2}{x_A} < \frac{x_B - \Delta_1}{x_A + \Delta_3}$.
    \label{step:borrow}
    \STATE {\bf Repay the flash loan} with $x_A$ token$_A$.
  \end{algorithmic}
\end{algorithm}

%% file: exhibits/attack_rugpull.tex
\begin{algorithm}[tbh]
    \small
    \floatname{algorithm}{Attack Algorithm}
    \caption{\revision{Rug Pull}}
    \label{rugpull}
    \begin{algorithmic}[1]
        \STATE {\bf Mint} a new coin \coin{XYZ}.
        \STATE {\bf Create} a liquidity pool with $x_\coin{XYZ}$ \coin{XYZ} and $x_\coin{ETH}$ \coin{ETH} (or any other valuable cryptocurrency) on an \ac{amm}, and receive \ac{lp} tokens.
        \STATE {\bf Attract} unwitting traders to buy \coin{XYZ} with \coin{ETH} from the pool, effectively changing the composition of the pool.
        \STATE {\bf Withdraw} liquidity from the pool by surrendering LP tokens, and obtain $x_\coin{XYZ} - \Delta_1$ \coin{XYZ} and $x_\coin{ETH} + \Delta_2$ \coin{ETH}, where $\Delta_1, \Delta_2 > 0$.
    \end{algorithmic}
\end{algorithm}

%% file: exhibits/attack_lp.tex
\begin{algorithm}[tbhp]
    \small
    \floatname{algorithm}{Attack Algorithm}
    \caption{\revision{Sandwich LP attack}}
    \label{sandwichlp}
    \begin{algorithmic}[1]
        \STATE
        User$_A$ places a transaction order to buy $x_A$ token$_A$ with token$_B$ with a pool containing $r_A$ token$_A$ and $r_B$ token$_B$ with gas fee $g_1$.
        \STATE LP$_B$ {\bf observes} the mempool and sees the transaction.
        \STATE LP$_B$ {\bf front-runs} by withdrawing liquidity $k\,r_A$ token$_A$ and $k \, r_B$ token$_B$
        with a higher gas fee $g_2 > g_1$.
        \STATE LP$_B$ and User$_A$'s transactions are executed sequentially, resulting in a new composition of the pool with $(1-k) r_A +x_A$ token$_A$ and $(1-k) r_B - x_B$ token$_B$.
        \STATE LP$_B$ {\bf back-runs} by re-providing $k\,r_A$ token$_A$ and $k \cdot \frac{(1-k) r_B - x_B}{(1-k) r_A +x_A}$ token$_B$.
        \STATE LP$_B$ {\bf back-runs} by selling $(1-\frac{(1-k) r_B - x_B}{(1-k) r_A +x_A})$ token$_B$ for some token$_A$
    \end{algorithmic}
\end{algorithm}

%% file: exhibits/attack_frontback.tex
\begin{algorithm}[tbhp]
    \small
    \floatname{algorithm}{Attack Algorithm}
    \caption{\revision{Sandwich price attack}}
    \label{sandwichprice}
    \begin{algorithmic}[1]
        \STATE
        User$_A$ wishes to purchase $x_A$ \coin{XYZ} whose spot price is $P_1$ on an \ac{amm} with gas fee $g_1$.
        \STATE User$_B$ {\bf observes} the mempool and sees the transaction.
        \STATE User$_B$ {\bf front-runs} by buying $x_B$ \coin{XYZ} with a higher gas fee $g_2 > g_1$ on the same \ac{amm}.
        \STATE User$_B$ and User$_A$'s transactions are executed sequentially at respective average price of $P_B$ and $P_A$, pushing \coin{XYZ}'s spot price up to $P_2$, where $P_2 > P_A > P_B > P_1$ due to slippage.
        \STATE User$_B$ {\bf back-runs} by selling $x_B$ \coin{XYZ} at an average price of $P'_{B}$, with $P_{2} > P'_{B} > P_{B}$ due to slippage.
    \end{algorithmic}
\end{algorithm}

%% file: exhibits/attacks_amm.tex
\begin{table*}[tb]
  \centering
  \caption{\revision{Overview of theoretical and anecdotal attacks against \ac{amm}-based \acp{dex} as well as their mitigating solutions.}
  }
  \renewcommand{\arraystretch}{0.9}
  \vskip -8pt
  \tiny
  \begin{tabularx}{\linewidth}{lXXlrrX}
    \toprule
    \textbf{Attack layer}           & \textbf{Attacks}                  & \textbf{Literature}                                                                                                    & \textbf{Affected \acp{amm}}                   & \textbf{Estimated loss}                                                                           & \textbf{Attack time} & \textbf{\revision{Solutions}}                                                                                                                                  \\
    \midrule
    \multirow{3}{*}{Infrastructure} & Block timestamp manipulation      & \cite{CryptoMarketPool2020BlockAttack, Huang2019, Antonopoulos2018MasteringDapps, Mense2018,yaish2022blockchain}                     & ---                                           & ---                                                                                               & ---                  & \cite{Szalachowski2018,ma2020achieving}                                                                                                                        \\
                                    & Transaction sequence manipulation & \cite{obadiaa2019, Eskandari2020,Antonopoulos2018MasteringDapps}                                                       & ---                                           & ---                                                                                               & ---                  & \cite{Eskandari2020}                                                                                                                                           \\
                                    & Other infrastructure              & \cite{Saad2019, Greene2018, Perez2020d, Mirkin2020, Rembert2021, Ramdas2019, Apostolaki2017}                           & ---                                           & ---                                                                                               & ---                  & Practices in traditional cybersecurity                                                                                                                         \\
    \midrule
    \multirow{2}{*}{Middleware}     & Reentrancy attack                 &
    \cite{daoattack2016, tsankov2018securify, albert2020taming,
      uniswap-audit, huang2019smart}
                                    & Uniswap V1 \cite{ConsenSys2020}   & $\geq$ 25.00m USD                                                                                                      & 04/2020                                       & \cite{luu2016making, cecchetti2021compositional, Rodler2019a, Fatima2020,Liu2018, Alkhalifah2021}                                                                                                                                                                                         \\
                                    & Other middleware                  & \cite{sayeed2020smart, ramanan2021blockchain, Praitheeshan2019, sun2021mutation}                                       & ---                                           & ---                                                                                               & ---                  & \cite{zhang2020smartshield,lu2019neucheck,bunz2020zether} \\
    \midrule
    \multirow{12}{*}{Application}   & Oracle attack                     & \cite{harvest2020attack, peckshield2020valuedefi, pirus2020cheesbank, smartcontent2021TWAP, Oosthoek2021, gudgeon2020} & Curve \cite{Redman2020}                       & $\geq$ 30.00m USD                                                                                 & 11/2020              & \cite{smartcontent2021TWAP,wang2021promutator}                                                                                                                 \\
                                    &                                   &                                                                                                                        & QuickSwap \cite{Pancakebunny2021BSC}          & $\geq$ 2.40m USD                                                                                  & 07/2021              &                                                                                                                                                                \\
                                    &                                   &                                                                                                                        & DODO \cite{Behnke2021DODO}                    & $\geq$ 0.70m USD                                                                                  & 03/2021              &                                                                                                                                                                \\
                                    & Rug  pull                         & \cite{Xia2021a, ampleforth2021home, tmpl2021etherscan, BybitLearn, rug-pull-businessreview, rug-pull-coinmarketcap}     & UraniumFinance \cite{Malwa2021, Lyanchev2021} & $\geq$ 50.00m USD                                                                                 & 04/2021              & \cite{BybitLearn,MudraManager,Xia2021a,mazorra2022not}                                                                                                         \\
                                    &                                   &                                                                                                                        & SushiSwap \cite{Keoun2020}                    & $\geq$ 13.00m USD                                                                                 & 08/2020              &                                                                                                                                                                \\
                                    & Frontrunning                      & \cite{daian2020flash, sandwhichattack2021, Eskandari2020, Zhou2021High-Frequency,torres2021frontrunner}                & Various \cite{Mikalauskas2021}                & $\geq$ 280.00m USD                                                                                & monthly              & \cite{sandwhichattack2021,Eskandari2020,Zhou2021A2MM,Baum2021a}                                                                                                \\
                                    &                                   &                                                                                                                        & DODO \cite{Behnke2021DODO}                    & $\geq$ 0.70m USD                                                                                  & 03/2021              &                                                                                                                                                                \\
                                    & Backrunning                       & \cite{Livnev2020, Zhou2021High-Frequency, werner2020sokDeFi}                                                           & ---                                           & ---                                                                                               & ---                  & \cite{Zhou2021A2MM,feng2020application, wang2017skyshield}                                                                                                     \\
                                    & Sandwich attacks                  & \cite{Zust2021, Zhou2021High-Frequency}                                                                                & Various \cite{Dzyatkovskii2021}               & $\geq$ 0.12m USD                                                                                  & monthly              & \cite{Zust2021,Zhou2021A2MM,Heimbach2022EliminatingTheory}                                                                                                     \\
                                    & Vampire attack                    & \cite{jakub2020vampire, dale2020sushiswapvampire, Foxley2021Vampire, Lo2020UniswapExchange}                            & Uniswap V2 \cite{Wong2021}                    & $\geq$ 1000.00m USD                                                                               & 08/2020              & \cite{Foxley2021}                                                                                                                                              \\
                                    &                                   &                                                                                                                        & Uniswap V2 \cite{Ong2021}                     & $\geq$ 4300.00m USD                                                                               & 09/2020              &                                                                                                                                                                \\
                                    &                                   &                                                                                                                        & Uniswap V2 \cite{Foxley2021Vampire}           & $\geq$ 239.00m USD                                                                                & 09/2021              &                                                                                                                                                                \\
    \bottomrule
  \end{tabularx}
  \label{tab:attack}%
\end{table*}%

%% file: exhibits/literature_review.tex
\begin{table*}[t]
  \setlength{\tabcolsep}{0.5pt}
  \centering
  \caption{Overview of related \acsp{sok}, surveys, and tutorials.}
  \renewcommand{\arraystretch}{0.9}
  \vskip -8pt
  \tiny
  \begin{tabularx}{\linewidth}{lXccccccccccccc}
    \toprule
                                         &                                                                                                                                                                                   & \multicolumn{4}{c}{\textbf{Subjects covered}} & \multicolumn{5}{c}{\textbf{Methodology}} & \multicolumn{4}{c}{\textbf{Aspects}}                                                                                                                                           \\
    \cmidrule(lr){3-6}
    \cmidrule(lr){7-11}
    \cmidrule(lr){12-15}
    Reference                            & Summary                                                                                                                                                                           & \ac{dex}                                      & \ac{amm}                                 & broader                              & broader     & Literature  & Taxonomization & Modeling    & Simulation & Empirical     & Mechanical   & Economic    & Security & Privacy \\
                                         &                                                                                                                                                                                   &                                               &                                          & DeFi                                 & DLT         & review      &                &             &            & investigation & / structural &             &          &
    \\
    \midrule
    This paper                           & An \ac{sok} on \ac{amm}-based \ac{dex} protocols that uses state space modeling to abstract their mechanics and surveys their security and privacy issues                         & \CIRCLE                                       & \CIRCLE                                  & \CIRCLE                              & \CIRCLE     & \CIRCLE     & \CIRCLE        & \CIRCLE     & \CIRCLE    & \CIRCLE       & \CIRCLE      & \CIRCLE     & \CIRCLE  & \CIRCLE
    \\
    \cite{Bartoletti2021}                & A theoretical framework with parametric characterization of \ac{amm}-based \ac{dex}'s fundamental properties and behaviors, with a focus on their structural and economic aspects & \CIRCLE                                       & \CIRCLE                                  & \Circle                              & \Circle     & \Circle     & \CIRCLE        & \CIRCLE     & \Circle    & \Circle       & \CIRCLE      & \CIRCLE     & \Circle  & \Circle
    \\
    \cite{Baum2021a}                     & An \ac{sok} on front-running mitigation in \ac{defi} especially \acp{amm}                                                                                                         & \CIRCLE                                       & \CIRCLE                                  & \CIRCLE                              & \LEFTcircle & \Circle     & \Circle        & \CIRCLE     & \Circle    & \Circle       & \CIRCLE      & \LEFTcircle & \CIRCLE  & \CIRCLE
    \\
    \cite{Massacci2021}                  & An overview of security challenges and design principles of order-book-based DEX                                                                                                  & \Circle                                       & \CIRCLE                                  & \Circle                              & \Circle     & \Circle     & \CIRCLE        & \Circle     & \Circle    & \CIRCLE       & \CIRCLE      & \Circle     & \CIRCLE  & \Circle
    \\
    \cite{werner2020sokDeFi}             & An \ac{sok} on various \ac{defi} applications drawing the distinction between their technical and economic security aspects, with a focus on the latter                           & \LEFTcircle                                   & \LEFTcircle                              & \CIRCLE                              & \Circle     & \Circle     & \CIRCLE        & \Circle     & \Circle    & \CIRCLE       & \CIRCLE      & \CIRCLE     & \CIRCLE  & \CIRCLE
    \\
    \cite{Bartoletti2020sokLendingPools} & An \ac{sok} on decentralized lending protocols proposing a formalization of their archetypal implementations and properties in the context of the broader \ac{defi} ecosystem     & \LEFTcircle                                   & \LEFTcircle                              & \CIRCLE                              & \Circle     & \Circle     & \CIRCLE        & \CIRCLE     & \Circle    & \Circle       & \CIRCLE      & \CIRCLE     & \Circle  & \Circle
    \\
    \cite{Cousaert2021}                  & An \ac{sok} on DeFi yield aggregators providing a general framework for yield farming strategies with numerical simulations and an empirical evaluation on their profitability    & \Circle                                       & \Circle                                  & \CIRCLE                              & \Circle     & \Circle     & \LEFTcircle    & \CIRCLE     & \CIRCLE    & \CIRCLE       & \CIRCLE      & \CIRCLE     & \CIRCLE  & \Circle
    \\
    \cite{Eskandari2020}                 & An \ac{sok} on various front-running attacks providing a taxonomy exemplified with case studies and summarizing preventative measurements                                         & \LEFTcircle                                   & \Circle                                  & \CIRCLE                              & \CIRCLE     & \Circle     & \CIRCLE        & \Circle     & \CIRCLE    & \CIRCLE       & \CIRCLE      & \Circle     & \CIRCLE  & \CIRCLE
    \\
    \cite{Qin2021a}                      & A quantitative study that examines various forms of Blockchain Extractable Value (BEV) and estimate historical attack profit through exploitation of BEV                          & \CIRCLE                                       & \CIRCLE                                  & \CIRCLE                              & \CIRCLE     & \LEFTcircle & \CIRCLE        & \CIRCLE     & \Circle    & \CIRCLE       & \CIRCLE      & \CIRCLE     & \CIRCLE  & \CIRCLE
    \\
    \cite{Qin2021}                       & A study that systematizes lending protocols with a focus on liquidation mechanisms, and empirically assesses the liquidation risk of decentralized lending protocols              & \Circle                                       & \Circle                                  & \CIRCLE                              & \Circle     & \Circle     & \Circle        & \CIRCLE     & \Circle    & \CIRCLE       & \CIRCLE      & \CIRCLE     & \CIRCLE  & \Circle
    \\
    \cite{gudgeon2020}                   & A study that provides a generalization of the financial risk existent on \ac{defi} lending protocols and stress-tests their robustness under various scenarios                    & \LEFTcircle                                   & \LEFTcircle                              & \CIRCLE                              & \Circle     & \Circle     & \CIRCLE        & \LEFTcircle & \CIRCLE    & \CIRCLE       & \CIRCLE      & \CIRCLE     & \CIRCLE  & \Circle
    \\
    \cite{Atzei2017a}                    & A survey of Ethereum smart contract attacks with a taxonomy of common programming pitfalls and their minimum working examples                                                     & \Circle                                       & \Circle                                  & \Circle                              & \CIRCLE     & \Circle     & \CIRCLE        & \Circle     & \Circle    & \Circle       & \CIRCLE      & \Circle     & \CIRCLE  & \Circle
    \\
    \cite{Homoliak2021}                  & A survey of vulnerabilities, threats, and defenses of different security reference architecture (SRA) layers of a blockchain using a stacked model                                & \LEFTcircle                                   & \LEFTcircle                              & \CIRCLE                              & \CIRCLE     & \CIRCLE     & \CIRCLE        & \Circle     & \Circle    & \Circle       & \CIRCLE      & \Circle     & \CIRCLE  & \CIRCLE
    \\
    \cite{Chen2020a}                     & A survey of vulnerabilities, attacks, and defenses of decentralized applications (DApps) running on top of the Ethereum blockchain                                                & \Circle                                       & \Circle                                  & \LEFTcircle                          & \CIRCLE     & \CIRCLE     & \CIRCLE        & \CIRCLE     & \Circle    & \Circle       & \CIRCLE      & \Circle     & \CIRCLE  & \Circle
    \\
    \bottomrule
  \end{tabularx}
  \label{tab:litreview}%
\end{table*}%

%% file: sections/glossary.tex
\section*{Glossary}

\paragraph{call option}a financial derivative instrument giving its owner the right to buy an asset at a given price

\paragraph{hedging}to invest in offsetting positions of a security to minimize the risk of adverse price movements of an asset

\paragraph{mint-quote}the amount of tokens or currency that is being created by the protocol or more generally the monetary institution

\paragraph{numéraire}the base value for comparing values across multiple items allowing for comparison of products or financial instruments

\paragraph{out-of-the-money}a situation when an option contract is worthless as the underlying asset is under-/ overpriced accordingly compared to the strike price of the option

\paragraph{put option}a financial derivative instrument giving its owner the right to sell an asset at a given price and time

\paragraph{redeem-quote}the amount of tokens or currency that is being returned by stakeholders to the protocol

%% file: sections/acknowledgements.tex
\section*{Acknowledgments}
\label{sec:acknowledgments}

We are indebted to Nazariy Vavryk for his contribution to the early code base for the numeric illustration of various \acp{amm} and insights into security breaches on \acp{amm}.
We thank Haopeng Song and Zehua Zhang for their excellent research assistance.
This material is based upon work partially supported by Ripple under the University Blockchain Research Initiative (UBRI)~\cite{Feng2022}.

%% file: main.bbl

\begin{thebibliography}{230}


\ifx \showCODEN    \undefined \def \showCODEN     #1{\unskip}     \fi
\ifx \showDOI      \undefined \def \showDOI       #1{#1}\fi
\ifx \showISBNx    \undefined \def \showISBNx     #1{\unskip}     \fi
\ifx \showISBNxiii \undefined \def \showISBNxiii  #1{\unskip}     \fi
\ifx \showISSN     \undefined \def \showISSN      #1{\unskip}     \fi
\ifx \showLCCN     \undefined \def \showLCCN      #1{\unskip}     \fi
\ifx \shownote     \undefined \def \shownote      #1{#1}          \fi
\ifx \showarticletitle \undefined \def \showarticletitle #1{#1}   \fi
\ifx \showURL      \undefined \def \showURL       {\relax}        \fi
\providecommand\bibfield[2]{#2}
\providecommand\bibinfo[2]{#2}
\providecommand\natexlab[1]{#1}
\providecommand\showeprint[2][]{arXiv:#2}

\bibitem[202(2021a)]%
        {2021DefiPulse}
 \bibinfo{year}{2021}\natexlab{a}.
\newblock \bibinfo{title}{{Defi Pulse}}.
\newblock
\newblock
\urldef\tempurl%
\url{https://defipulse.com/}
\showURL{%
\tempurl}


\bibitem[202(2021b)]%
        {2021EthereumStandard}
 \bibinfo{year}{2021}\natexlab{b}.
\newblock \bibinfo{title}{{Ethereum Improvement Proposals - EIP-20: Token
  Standard}}.
\newblock
\newblock
\urldef\tempurl%
\url{https://eips.ethereum.org/EIPS/eip-20}
\showURL{%
\tempurl}


\bibitem[rug(2021)]%
        {rug-pull-coinmarketcap}
 \bibinfo{year}{2021}\natexlab{}.
\newblock \bibinfo{title}{{Rug Pull}}.
\newblock
\newblock
\urldef\tempurl%
\url{https://coinmarketcap.com/alexandria/glossary/rug-pull}
\showURL{%
\tempurl}


\bibitem[Adams(2018)]%
        {Adams2018}
\bibfield{author}{\bibinfo{person}{Hayden Adams}.}
  \bibinfo{year}{2018}\natexlab{}.
\newblock \bibinfo{title}{{Uniswap Whitepaper (v1)}}.
\newblock
\newblock
\urldef\tempurl%
\url{https://hackmd.io/C-DvwDSfSxuh-Gd4WKE_ig}
\showURL{%
\tempurl}


\bibitem[Adams et~al\mbox{.}(2020)]%
        {Core2020}
\bibfield{author}{\bibinfo{person}{Hayden Adams}, \bibinfo{person}{Noah
  Zinsmeister}, {and} \bibinfo{person}{Dan Robinson}.}
  \bibinfo{year}{2020}\natexlab{}.
\newblock \bibinfo{title}{{Uniswap v2 Core}}.  (\bibinfo{year}{2020}).
\newblock


\bibitem[Adams et~al\mbox{.}(2021)]%
        {adams2021v3core}
\bibfield{author}{\bibinfo{person}{Hayden Adams}, \bibinfo{person}{Noah
  Zinsmeister}, \bibinfo{person}{Moody Salem~moody},
  \bibinfo{person}{uniswaporg River~Keefer}, {and} \bibinfo{person}{Dan
  Robinson}.} \bibinfo{year}{2021}\natexlab{}.
\newblock \bibinfo{title}{{Uniswap v3 Core}}.  (\bibinfo{year}{2021}).
\newblock


\bibitem[Albert et~al\mbox{.}(2020)]%
        {albert2020taming}
\bibfield{author}{\bibinfo{person}{Elvira Albert}, \bibinfo{person}{Shelly
  Grossman}, \bibinfo{person}{Noam Rinetzky}, \bibinfo{person}{Clara
  Rodr{\'{i}}guez-N{\'{u}}{\~{n}}ez}, \bibinfo{person}{Albert Rubio}, {and}
  \bibinfo{person}{Mooly Sagiv}.} \bibinfo{year}{2020}\natexlab{}.
\newblock \showarticletitle{{Taming callbacks for smart contract modularity}}.
\newblock \bibinfo{journal}{\emph{ACM on Programming Languages}}
  \bibinfo{volume}{4}, \bibinfo{number}{OOPSLA} (\bibinfo{date}{11}
  \bibinfo{year}{2020}), \bibinfo{pages}{30}.
\newblock
\urldef\tempurl%
\url{https://doi.org/10.1145/3428277}
\showDOI{\tempurl}


\bibitem[Alkhalifah et~al\mbox{.}(2021)]%
        {Alkhalifah2021}
\bibfield{author}{\bibinfo{person}{Ayman Alkhalifah}, \bibinfo{person}{Alex
  Ng}, \bibinfo{person}{Paul~A Watters}, {and} \bibinfo{person}{A~S~M Kayes}.}
  \bibinfo{year}{2021}\natexlab{}.
\newblock \showarticletitle{{A Mechanism to Detect and Prevent Ethereum
  Blockchain Smart Contract Reentrancy Attacks}}.
\newblock \bibinfo{journal}{\emph{Frontiers in Computer Science}}
  \bibinfo{volume}{3} (\bibinfo{year}{2021}), \bibinfo{pages}{1}.
\newblock
\showISSN{2624-9898}


\bibitem[{Ampleforth}(2021)]%
        {ampleforth2021home}
\bibfield{author}{\bibinfo{person}{{Ampleforth}}.}
  \bibinfo{year}{2021}\natexlab{}.
\newblock \bibinfo{title}{{Ampleforth Home Page}}.
\newblock
\newblock
\urldef\tempurl%
\url{https://www.ampleforth.org/}
\showURL{%
\tempurl}


\bibitem[Andersson(2020)]%
        {Andersson2020}
\bibfield{author}{\bibinfo{person}{Henrik Andersson}.}
  \bibinfo{year}{2020}\natexlab{}.
\newblock \bibinfo{title}{{mStable — Introducing Constant Sum Bonding Curves
  for Tokenised Assets}}.
\newblock
\newblock
\urldef\tempurl%
\url{https://medium.com/mstable/introducing-constant-sum-bonding-curves-for-tokenised-assets-6e18879cdc5b}
\showURL{%
\tempurl}


\bibitem[Angeris and Chitra(2020)]%
        {Angeris2020}
\bibfield{author}{\bibinfo{person}{Guillermo Angeris} {and}
  \bibinfo{person}{Tarun Chitra}.} \bibinfo{year}{2020}\natexlab{}.
\newblock \showarticletitle{{Improved Price Oracles: Constant Function Market
  Makers}}. In \bibinfo{booktitle}{\emph{Advances in Financial Technologies}}.
  \bibinfo{publisher}{ACM}, \bibinfo{address}{New York, NY, USA},
  \bibinfo{pages}{80--91}.
\newblock
\showISBNx{9781450381390}
\urldef\tempurl%
\url{https://doi.org/10.1145/3419614.3423251}
\showDOI{\tempurl}


\bibitem[Angeris et~al\mbox{.}(2021a)]%
        {Angeris2021}
\bibfield{author}{\bibinfo{person}{Guillermo Angeris}, \bibinfo{person}{Alex
  Evans}, {and} \bibinfo{person}{Tarun Chitra}.}
  \bibinfo{year}{2021}\natexlab{a}.
\newblock \showarticletitle{{A Note on Bundle Profit Maximization}}.
\newblock  (\bibinfo{year}{2021}).
\newblock


\bibitem[Angeris et~al\mbox{.}(2021b)]%
        {Angeris2021AMakers}
\bibfield{author}{\bibinfo{person}{Guillermo Angeris}, \bibinfo{person}{Alex
  Evans}, {and} \bibinfo{person}{Tarun Chitra}.}
  \bibinfo{year}{2021}\natexlab{b}.
\newblock \bibinfo{title}{{A Note on Privacy in Constant Function Market
  Makers}}.  (\bibinfo{date}{3} \bibinfo{year}{2021}).
\newblock
\urldef\tempurl%
\url{http://arxiv.org/abs/2103.01193}
\showURL{%
\tempurl}


\bibitem[Angeris et~al\mbox{.}(2021c)]%
        {Angeris2021ReplicatingMakers}
\bibfield{author}{\bibinfo{person}{Guillermo Angeris}, \bibinfo{person}{Alex
  Evans}, {and} \bibinfo{person}{Tarun Chitra}.}
  \bibinfo{year}{2021}\natexlab{c}.
\newblock \bibinfo{title}{{Replicating Market Makers}}.  (\bibinfo{date}{3}
  \bibinfo{year}{2021}).
\newblock
\urldef\tempurl%
\url{http://arxiv.org/abs/2103.14769}
\showURL{%
\tempurl}


\bibitem[Antonopoulos and Wood(2018)]%
        {Antonopoulos2018MasteringDapps}
\bibfield{author}{\bibinfo{person}{Andreas~M Antonopoulos} {and}
  \bibinfo{person}{Gavin Wood}.} \bibinfo{year}{2018}\natexlab{}.
\newblock \bibinfo{booktitle}{\emph{{Mastering ethereum: building smart
  contracts and dapps}}}.
\newblock \bibinfo{publisher}{O'reilly Media}.
\newblock


\bibitem[Apostolaki et~al\mbox{.}(2017)]%
        {Apostolaki2017}
\bibfield{author}{\bibinfo{person}{Maria Apostolaki}, \bibinfo{person}{Aviv
  Zohar}, {and} \bibinfo{person}{Laurent Vanbever}.}
  \bibinfo{year}{2017}\natexlab{}.
\newblock \showarticletitle{{Hijacking bitcoin: Routing attacks on
  cryptocurrencies}}. In \bibinfo{booktitle}{\emph{2017 IEEE Symposium on
  Security and Privacy (SP)}}. \bibinfo{pages}{375--392}.
\newblock


\bibitem[{Arbitrum}(2021)]%
        {arbitrum2021}
\bibfield{author}{\bibinfo{person}{{Arbitrum}}.}
  \bibinfo{year}{2021}\natexlab{}.
\newblock \bibinfo{title}{{Arbitrum – Scaling Ethereum}}.
\newblock
\newblock
\urldef\tempurl%
\url{https://arbitrum.io/}
\showURL{%
\tempurl}


\bibitem[Atzei et~al\mbox{.}(2017)]%
        {Atzei2017a}
\bibfield{author}{\bibinfo{person}{Nicola Atzei}, \bibinfo{person}{Massimo
  Bartoletti}, {and} \bibinfo{person}{Tiziana Cimoli}.}
  \bibinfo{year}{2017}\natexlab{}.
\newblock \showarticletitle{{A survey of attacks on Ethereum smart contracts
  (SoK)}}. In \bibinfo{booktitle}{\emph{Lecture Notes in Computer Science
  (including subseries Lecture Notes in Artificial Intelligence and Lecture
  Notes in Bioinformatics)}}, Vol.~\bibinfo{volume}{10204 LNCS}.
  \bibinfo{publisher}{Springer Verlag}, \bibinfo{pages}{164--186}.
\newblock
\showISBNx{9783662544549}
\showISSN{16113349}
\urldef\tempurl%
\url{https://doi.org/10.1007/978-3-662-54455-6{\_}8}
\showDOI{\tempurl}


\bibitem[{Balancer}(2021)]%
        {balancer2021lbp}
\bibfield{author}{\bibinfo{person}{{Balancer}}.}
  \bibinfo{year}{2021}\natexlab{}.
\newblock \bibinfo{title}{{Liquidity Bootstrapping Pools (LBPs)}}.
\newblock
\newblock


\bibitem[{Balancer}(2022)]%
        {Balancer2022}
\bibfield{author}{\bibinfo{person}{{Balancer}}.}
  \bibinfo{year}{2022}\natexlab{}.
\newblock \bibinfo{title}{{Swap Fees}}.
\newblock
\newblock
\urldef\tempurl%
\url{https://docs.balancer.fi/concepts/fees#swap-fees}
\showURL{%
\tempurl}


\bibitem[{Bancor}(2020a)]%
        {bancor2020v2}
\bibfield{author}{\bibinfo{person}{{Bancor}}.}
  \bibinfo{year}{2020}\natexlab{a}.
\newblock \bibinfo{title}{{Announcing Bancor V2}}.
\newblock
\newblock
\urldef\tempurl%
\url{https://blog.bancor.network/announcing-bancor-v2-2f56b515e9d8}
\showURL{%
\tempurl}


\bibitem[{Bancor}(2020b)]%
        {Bancor2020}
\bibfield{author}{\bibinfo{person}{{Bancor}}.}
  \bibinfo{year}{2020}\natexlab{b}.
\newblock \bibinfo{title}{{Bancor V2.1 Technical Explainer}}.
  (\bibinfo{year}{2020}).
\newblock
\urldef\tempurl%
\url{https://drive.google.com/file/d/16EY7FUeS4MXnFjSf-KCgdE-Xyj4re27G/view}
\showURL{%
\tempurl}


\bibitem[{Bancor Network}(2021)]%
        {bancor2021ilprotection}
\bibfield{author}{\bibinfo{person}{{Bancor Network}}.}
  \bibinfo{year}{2021}\natexlab{}.
\newblock \bibinfo{title}{{FAQs - Bancor Network}}.
\newblock
\newblock
\urldef\tempurl%
\url{https://docs.bancor.network/faqs#how-does-impermanent-loss-insurance-work}
\showURL{%
\tempurl}


\bibitem[Bartoletti et~al\mbox{.}(2021a)]%
        {Bartoletti2020sokLendingPools}
\bibfield{author}{\bibinfo{person}{Massimo Bartoletti}, \bibinfo{person}{James
  Hsin-yu Chiang}, {and} \bibinfo{person}{Alberto~Lluch Lafuente}.}
  \bibinfo{year}{2021}\natexlab{a}.
\newblock \showarticletitle{{SoK: Lending Pools in Decentralized Finance}}. In
  \bibinfo{booktitle}{\emph{Workshop Proceedings of Financial Cryptography and
  Data Security}}. \bibinfo{publisher}{Springer, Berlin, Heidelberg},
  \bibinfo{pages}{553--578}.
\newblock
\urldef\tempurl%
\url{https://doi.org/10.1007/978-3-662-63958-0{\_}40}
\showDOI{\tempurl}


\bibitem[Bartoletti et~al\mbox{.}(2021b)]%
        {Bartoletti2021}
\bibfield{author}{\bibinfo{person}{Massimo Bartoletti}, \bibinfo{person}{James
  Hsin-Yu Chiang}, {and} \bibinfo{person}{Alberto Lluch-Lafuente}.}
  \bibinfo{year}{2021}\natexlab{b}.
\newblock \showarticletitle{{A Theory of Automated Market Makers in DeFi}}. In
  \bibinfo{booktitle}{\emph{International Conference on Coordination Languages
  and Models}}. \bibinfo{pages}{168--187}.
\newblock
\urldef\tempurl%
\url{https://doi.org/10.1007/978-3-030-78142-2{\_}11}
\showDOI{\tempurl}


\bibitem[Baum et~al\mbox{.}(2021a)]%
        {Baum2021a}
\bibfield{author}{\bibinfo{person}{Carsten Baum}, \bibinfo{person}{James
  Hsin-yu Chiang}, \bibinfo{person}{Bernardo David},
  \bibinfo{person}{Tore~Kasper Frederiksen}, {and} \bibinfo{person}{Lorenzo
  Gentile}.} \bibinfo{year}{2021}\natexlab{a}.
\newblock \bibinfo{booktitle}{\emph{{SoK: Mitigation of Front-running in
  Decentralized Finance}}}.
\newblock \bibinfo{type}{{T}echnical {R}eport}.
\newblock


\bibitem[Baum et~al\mbox{.}(2021b)]%
        {Baum2021}
\bibfield{author}{\bibinfo{person}{Carsten Baum}, \bibinfo{person}{Bernardo
  David}, {and} \bibinfo{person}{Tore~Kasper Frederiksen}.}
  \bibinfo{year}{2021}\natexlab{b}.
\newblock \showarticletitle{{P2DEX: Privacy-Preserving Decentralized
  Cryptocurrency Exchange}}.
\newblock In \bibinfo{booktitle}{\emph{Applied Cryptography and Network
  Security}}. \bibinfo{publisher}{Springer, Cham}, Chapter~7,
  \bibinfo{pages}{163--194}.
\newblock
\urldef\tempurl%
\url{https://doi.org/10.1007/978-3-030-78372-3{\_}7}
\showDOI{\tempurl}


\bibitem[Bebel and Ojha(2022)]%
        {bebel2022ferveo}
\bibfield{author}{\bibinfo{person}{Joseph Bebel} {and} \bibinfo{person}{Dev
  Ojha}.} \bibinfo{year}{2022}\natexlab{}.
\newblock \showarticletitle{Ferveo: Threshold Decryption for Mempool Privacy in
  BFT networks}.
\newblock \bibinfo{journal}{\emph{Cryptology ePrint Archive}}
  (\bibinfo{year}{2022}).
\newblock


\bibitem[Behnke(2021)]%
        {Behnke2021DODO}
\bibfield{author}{\bibinfo{person}{Rob Behnke}.}
  \bibinfo{year}{2021}\natexlab{}.
\newblock \bibinfo{title}{{Explained: The DODO DEX Hack}}.
\newblock
\newblock
\urldef\tempurl%
\url{https://halborn.com/explained-the-dodo-dex-hack-march-2021/}
\showURL{%
\tempurl}


\bibitem[Bernabe et~al\mbox{.}(2019)]%
        {Bernabe2019Privacy-preservingChallenges}
\bibfield{author}{\bibinfo{person}{Jorge~Bernal Bernabe},
  \bibinfo{person}{Jose~Luis Canovas}, \bibinfo{person}{Jose~L
  Hernandez-Ramos}, \bibinfo{person}{Rafael~Torres Moreno}, {and}
  \bibinfo{person}{Antonio Skarmeta}.} \bibinfo{year}{2019}\natexlab{}.
\newblock \showarticletitle{{Privacy-preserving solutions for blockchain:
  Review and challenges}}.
\newblock \bibinfo{journal}{\emph{IEEE Access}}  \bibinfo{volume}{7}
  (\bibinfo{year}{2019}), \bibinfo{pages}{164908--164940}.
\newblock


\bibitem[{Blank}(2021)]%
        {Blank2021}
\bibfield{author}{\bibinfo{person}{{Blank}}.} \bibinfo{year}{2021}\natexlab{}.
\newblock \bibinfo{title}{{Blank features beyond basic privacy ({\#}2):
  Protecting your IP in DeFi}}.
\newblock
\newblock
\urldef\tempurl%
\url{https://blankwallet.medium.com/blank-features-beyond-basic-privacy-2-protecting-your-ip-in-defi-11bc76f2d67b}
\showURL{%
\tempurl}


\bibitem[Bowe et~al\mbox{.}(2020)]%
        {Bowe2020}
\bibfield{author}{\bibinfo{person}{Sean Bowe}, \bibinfo{person}{Alessandro
  Chiesa}, \bibinfo{person}{Matthew Green}, \bibinfo{person}{Ian Miers},
  \bibinfo{person}{Pratyush Mishra}, {and} \bibinfo{person}{Howard Wu}.}
  \bibinfo{year}{2020}\natexlab{}.
\newblock \showarticletitle{{ZEXE: Enabling Decentralized Private
  Computation}}. In \bibinfo{booktitle}{\emph{IEEE Symposium on Security and
  Privacy}}. \bibinfo{publisher}{IEEE}, \bibinfo{pages}{947--964}.
\newblock
\showISBNx{978-1-7281-3497-0}
\urldef\tempurl%
\url{https://doi.org/10.1109/SP40000.2020.00050}
\showDOI{\tempurl}


\bibitem[Brahma et~al\mbox{.}(2012)]%
        {Brahma2012}
\bibfield{author}{\bibinfo{person}{Aseem Brahma}, \bibinfo{person}{Mithun
  Chakraborty}, \bibinfo{person}{Sanmay Das}, \bibinfo{person}{Allen Lavoie},
  {and} \bibinfo{person}{Malik Magdon-Ismail}.}
  \bibinfo{year}{2012}\natexlab{}.
\newblock \showarticletitle{{A bayesian market maker}}. In
  \bibinfo{booktitle}{\emph{ACM Conference on Electronic Commerce}}.
  \bibinfo{address}{New York}, \bibinfo{pages}{215}.
\newblock
\showISBNx{9781450314152}
\urldef\tempurl%
\url{https://doi.org/10.1145/2229012.2229031}
\showDOI{\tempurl}


\bibitem[Breidenbach et~al\mbox{.}(2018)]%
        {Breidenbach2018}
\bibfield{author}{\bibinfo{person}{Lorenz Breidenbach}, \bibinfo{person}{Philip
  Daian}, \bibinfo{person}{Florian Tram{\`{e}}r}, {and} \bibinfo{person}{Ari
  Juels}.} \bibinfo{year}{2018}\natexlab{}.
\newblock \showarticletitle{{Enter the Hydra: Towards principled bug bounties
  and exploit-resistant smart contracts}}. In \bibinfo{booktitle}{\emph{USENIX
  Security Symposium}}. \bibinfo{pages}{1335--1352}.
\newblock
\showISBNx{9781939133045}
\urldef\tempurl%
\url{https://www.usenix.org/conference/usenixsecurity18/presentation/breindenbach}
\showURL{%
\tempurl}


\bibitem[Bukov and Melnik(2020)]%
        {mooniswap2020whitepaper}
\bibfield{author}{\bibinfo{person}{Anton Bukov} {and} \bibinfo{person}{Mikhail
  Melnik}.} \bibinfo{year}{2020}\natexlab{}.
\newblock \bibinfo{title}{{Mooniswap by 1inch.exchange}}.
  (\bibinfo{year}{2020}).
\newblock


\bibitem[B{\"{u}}nz et~al\mbox{.}(2020)]%
        {bunz2020zether}
\bibfield{author}{\bibinfo{person}{Benedikt B{\"{u}}nz},
  \bibinfo{person}{Shashank Agrawal}, \bibinfo{person}{Mahdi Zamani}, {and}
  \bibinfo{person}{Dan Boneh}.} \bibinfo{year}{2020}\natexlab{}.
\newblock \showarticletitle{{Zether: Towards privacy in a smart contract
  world}}. In \bibinfo{booktitle}{\emph{International Conference on Financial
  Cryptography and Data Security}}. \bibinfo{pages}{423--443}.
\newblock


\bibitem[B{\"{u}}nz et~al\mbox{.}(2018)]%
        {Bunz2018Bulletproofs:More}
\bibfield{author}{\bibinfo{person}{Benedikt B{\"{u}}nz},
  \bibinfo{person}{Jonathan Bootle}, \bibinfo{person}{Dan Boneh},
  \bibinfo{person}{Andrew Poelstra}, \bibinfo{person}{Pieter Wuille}, {and}
  \bibinfo{person}{Greg Maxwell}.} \bibinfo{year}{2018}\natexlab{}.
\newblock \showarticletitle{{Bulletproofs: Short proofs for confidential
  transactions and more}}. In \bibinfo{booktitle}{\emph{2018 IEEE Symposium on
  Security and Privacy (SP)}}. \bibinfo{pages}{315--334}.
\newblock


\bibitem[{Burgerswap}(2020)]%
        {burgerswap2022}
\bibfield{author}{\bibinfo{person}{{Burgerswap}}.}
  \bibinfo{year}{2020}\natexlab{}.
\newblock \bibinfo{title}{{Burgerswap: Decentralized Finance Platform}}.
  (\bibinfo{year}{2020}).
\newblock
\urldef\tempurl%
\url{https://burgerswap.org/whitepaper_burgerswap.pdf}
\showURL{%
\tempurl}


\bibitem[{Bybit Learn}(2021)]%
        {BybitLearn}
\bibfield{author}{\bibinfo{person}{{Bybit Learn}}.}
  \bibinfo{year}{2021}\natexlab{}.
\newblock \bibinfo{title}{{Why Crypto Rug Pulls Happen in DeFi and How to Avoid
  It}}.
\newblock
\newblock
\urldef\tempurl%
\url{https://learn.bybit.com/investing/why-crypto-rug-pulls-happen-in-defi/}
\showURL{%
\tempurl}


\bibitem[Cao et~al\mbox{.}(2021)]%
        {cao2021flashot}
\bibfield{author}{\bibinfo{person}{Yixin Cao}, \bibinfo{person}{Chuanwei Zou},
  {and} \bibinfo{person}{Xianfeng Cheng}.} \bibinfo{year}{2021}\natexlab{}.
\newblock \bibinfo{title}{{Flashot: A Snapshot of Flash Loan Attack on DeFi
  Ecosystem}}.  (\bibinfo{date}{1} \bibinfo{year}{2021}).
\newblock
\urldef\tempurl%
\url{http://arxiv.org/abs/2102.00626}
\showURL{%
\tempurl}


\bibitem[Capponi and JIA(2021)]%
        {Capponi2021TheMaker}
\bibfield{author}{\bibinfo{person}{Agostino Capponi} {and}
  \bibinfo{person}{RUIZHE JIA}.} \bibinfo{year}{2021}\natexlab{}.
\newblock \bibinfo{title}{{The Adoption of Blockchain-based Decentralized
  Exchanges: A Market Microstructure Analysis of the Automated Market Maker}}.
  (\bibinfo{year}{2021}).
\newblock
\urldef\tempurl%
\url{https://doi.org/10.2139/ssrn.3805095}
\showDOI{\tempurl}


\bibitem[Cecchetti et~al\mbox{.}(2021)]%
        {cecchetti2021compositional}
\bibfield{author}{\bibinfo{person}{Ethan Cecchetti}, \bibinfo{person}{Siqiu
  Yao}, \bibinfo{person}{Haobin Ni}, {and} \bibinfo{person}{Andrew~C. Myers}.}
  \bibinfo{year}{2021}\natexlab{}.
\newblock \showarticletitle{{Compositional Security for Reentrant
  Applications}}. In \bibinfo{booktitle}{\emph{IEEE Symposium on Security and
  Privacy}}. \bibinfo{publisher}{IEEE}, \bibinfo{pages}{1249--1267}.
\newblock
\showISBNx{978-1-7281-8934-5}
\showISSN{1063-9578}
\urldef\tempurl%
\url{https://doi.org/10.1109/SP40001.2021.00084}
\showDOI{\tempurl}


\bibitem[Chen et~al\mbox{.}(2020b)]%
        {Chen2020a}
\bibfield{author}{\bibinfo{person}{Huashan Chen}, \bibinfo{person}{Marcus
  Pendleton}, \bibinfo{person}{Laurent Njilla}, {and} \bibinfo{person}{Shouhuai
  Xu}.} \bibinfo{year}{2020}\natexlab{b}.
\newblock \showarticletitle{{A Survey on Ethereum Systems Security}}.
\newblock \bibinfo{journal}{\emph{ACM Computing Surveys (CSUR)}}
  \bibinfo{volume}{53}, \bibinfo{number}{3} (\bibinfo{date}{6}
  \bibinfo{year}{2020}).
\newblock
\showISSN{15577341}
\urldef\tempurl%
\url{https://doi.org/10.1145/3391195}
\showDOI{\tempurl}


\bibitem[Chen et~al\mbox{.}(2020a)]%
        {Chen2020PhishingEcosystem.}
\bibfield{author}{\bibinfo{person}{Weili Chen}, \bibinfo{person}{Xiongfeng
  Guo}, \bibinfo{person}{Zhiguang Chen}, \bibinfo{person}{Zibin Zheng}, {and}
  \bibinfo{person}{Yutong Lu}.} \bibinfo{year}{2020}\natexlab{a}.
\newblock \showarticletitle{{Phishing Scam Detection on Ethereum: Towards
  Financial Security for Blockchain Ecosystem.}}. In
  \bibinfo{booktitle}{\emph{IJCAI}}. \bibinfo{pages}{4506--4512}.
\newblock


\bibitem[Cheng et~al\mbox{.}(2019)]%
        {Cheng2019Ekiden:Contracts}
\bibfield{author}{\bibinfo{person}{Raymond Cheng}, \bibinfo{person}{Fan Zhang},
  \bibinfo{person}{Jernej Kos}, \bibinfo{person}{Warren He},
  \bibinfo{person}{Nicholas Hynes}, \bibinfo{person}{Noah Johnson},
  \bibinfo{person}{Ari Juels}, \bibinfo{person}{Andrew Miller}, {and}
  \bibinfo{person}{Dawn Song}.} \bibinfo{year}{2019}\natexlab{}.
\newblock \showarticletitle{{Ekiden: A platform for confidentiality-preserving,
  trustworthy, and performant smart contracts}}. In
  \bibinfo{booktitle}{\emph{2019 IEEE European Symposium on Security and
  Privacy (EuroS{\&}P)}}. \bibinfo{pages}{185--200}.
\newblock


\bibitem[Chitra et~al\mbox{.}(2021)]%
        {Chitra2021}
\bibfield{author}{\bibinfo{person}{Tarun Chitra}, \bibinfo{person}{Guillermo
  Angeris}, {and} \bibinfo{person}{Alex Evans}.}
  \bibinfo{year}{2021}\natexlab{}.
\newblock \bibinfo{title}{{How Liveness Separates CFMMs and Order Books}}.
  (\bibinfo{year}{2021}).
\newblock


\bibitem[{ConsenSys}(2020)]%
        {ConsenSys2020}
\bibfield{author}{\bibinfo{person}{{ConsenSys}}.}
  \bibinfo{year}{2020}\natexlab{}.
\newblock \bibinfo{title}{{Thoughts on DeFi Security. A deep dive into the
  Uniswap and{\ldots} | by ConsenSys | ConsenSys Media}}.
\newblock
\newblock
\urldef\tempurl%
\url{https://media.consensys.net/thoughts-on-defi-security-640dde37bb3b}
\showURL{%
\tempurl}


\bibitem[{Consensys Diligence}(2019)]%
        {uniswap-audit}
\bibfield{author}{\bibinfo{person}{{Consensys Diligence}}.}
  \bibinfo{year}{2019}\natexlab{}.
\newblock \bibinfo{title}{{ConsenSys/Uniswap-audit-report-2018-12}}.
\newblock
\newblock
\urldef\tempurl%
\url{https://github.com/ConsenSys/Uniswap-audit-report-2018-12#31-liquidity-pool-can-be-stolen-in-some-tokens-eg-erc-777-29}
\showURL{%
\tempurl}


\bibitem[Cousaert et~al\mbox{.}(2022a)]%
        {Cousaert2022}
\bibfield{author}{\bibinfo{person}{Simon Cousaert}, \bibinfo{person}{Nikhil
  Vadgama}, {and} \bibinfo{person}{Jiahua Xu}.}
  \bibinfo{year}{2022}\natexlab{a}.
\newblock \showarticletitle{{Token-Based Insurance Solutions on Blockchain}}.
\newblock In \bibinfo{booktitle}{\emph{Blockchains and the Token Economy:
  Theory and Practice}}. \bibinfo{pages}{237--260}.
\newblock
\urldef\tempurl%
\url{https://doi.org/10.1007/978-3-030-95108-5{\_}9}
\showDOI{\tempurl}


\bibitem[Cousaert et~al\mbox{.}(2022b)]%
        {Cousaert2021}
\bibfield{author}{\bibinfo{person}{Simon Cousaert}, \bibinfo{person}{Jiahua
  Xu}, {and} \bibinfo{person}{Toshiko Matsui}.}
  \bibinfo{year}{2022}\natexlab{b}.
\newblock \showarticletitle{{SoK: Yield Aggregators in DeFi}}. In
  \bibinfo{booktitle}{\emph{IEEE International Conference on Blockchain and
  Cryptocurrency}}. \bibinfo{publisher}{IEEE}, \bibinfo{pages}{1--14}.
\newblock
\showISBNx{978-1-6654-9538-7}
\urldef\tempurl%
\url{https://doi.org/10.1109/ICBC54727.2022.9805523}
\showDOI{\tempurl}


\bibitem[Cramer et~al\mbox{.}(2015)]%
        {Cramer2015SecureComputation}
\bibfield{author}{\bibinfo{person}{Ronald Cramer}, \bibinfo{person}{Ivan~Bjerre
  Damg{\aa}rd}, {and} \bibinfo{person}{{others}}.}
  \bibinfo{year}{2015}\natexlab{}.
\newblock \bibinfo{booktitle}{\emph{{Secure multiparty computation}}}.
\newblock \bibinfo{publisher}{Cambridge University Press}.
\newblock


\bibitem[{Crypto Market Pool}(2020)]%
        {CryptoMarketPool2020BlockAttack}
\bibfield{author}{\bibinfo{person}{{Crypto Market Pool}}.}
  \bibinfo{year}{2020}\natexlab{}.
\newblock \bibinfo{title}{{Block timestamp manipulation attack}}.
\newblock
\newblock
\urldef\tempurl%
\url{https://cryptomarketpool.com/block-timestamp-manipulation-attack/}
\showURL{%
\tempurl}


\bibitem[{CryptoLocally}(2020)]%
        {CryptoLocally2020}
\bibfield{author}{\bibinfo{person}{{CryptoLocally}}.}
  \bibinfo{year}{2020}\natexlab{}.
\newblock \bibinfo{title}{{GIV Balancer Listing and Staking Rewards Updates}}.
\newblock
\newblock
\urldef\tempurl%
\url{https://cryptolocally.medium.com/giv-balancer-listing-and-staking-rewards-updates-81ebb5843e58}
\showURL{%
\tempurl}


\bibitem[Daian et~al\mbox{.}(2020)]%
        {daian2020flash}
\bibfield{author}{\bibinfo{person}{Philip Daian}, \bibinfo{person}{Steven
  Goldfeder}, \bibinfo{person}{Tyler Kell}, \bibinfo{person}{Yunqi Li},
  \bibinfo{person}{Xueyuan Zhao}, \bibinfo{person}{Iddo Bentov},
  \bibinfo{person}{Lorenz Breidenbach}, {and} \bibinfo{person}{Ari Juels}.}
  \bibinfo{year}{2020}\natexlab{}.
\newblock \showarticletitle{{Flash Boys 2.0: Frontrunning in Decentralized
  Exchanges, Miner Extractable Value, and Consensus Instability}}. In
  \bibinfo{booktitle}{\emph{2020 IEEE Symposium on Security and Privacy (SP)}},
  Vol.~\bibinfo{volume}{2020-May}. \bibinfo{publisher}{IEEE},
  \bibinfo{pages}{910--927}.
\newblock
\showISBNx{978-1-7281-3497-0}
\showISSN{10816011}
\urldef\tempurl%
\url{https://doi.org/10.1109/SP40000.2020.00040}
\showDOI{\tempurl}


\bibitem[Dale(2020)]%
        {dale2020sushiswapvampire}
\bibfield{author}{\bibinfo{person}{Brady Dale}.}
  \bibinfo{year}{2020}\natexlab{}.
\newblock \bibinfo{title}{{SushiSwap Will Withdraw Up to {\$}830M From Uniswap
  Today: Why It Matters for DeFi}}.
\newblock
\newblock
\urldef\tempurl%
\url{https://www.coindesk.com/sushiswap-uniswap-migration-defi-amm-wars}
\showURL{%
\tempurl}


\bibitem[Das et~al\mbox{.}(2021)]%
        {das2021resource}
\bibfield{author}{\bibinfo{person}{Ankush Das}, \bibinfo{person}{Stephanie
  Balzer}, \bibinfo{person}{Jan Hoffmann}, \bibinfo{person}{Frank Pfenning},
  {and} \bibinfo{person}{Ishani Santurkar}.} \bibinfo{year}{2021}\natexlab{}.
\newblock \showarticletitle{{Resource-aware session types for digital
  contracts}}. In \bibinfo{booktitle}{\emph{2021 IEEE 34th Computer Security
  Foundations Symposium (CSF)}}. \bibinfo{pages}{1--16}.
\newblock


\bibitem[De~Giglio(2021)]%
        {DeGiglio2021}
\bibfield{author}{\bibinfo{person}{Luca De~Giglio}.}
  \bibinfo{year}{2021}\natexlab{}.
\newblock \bibinfo{title}{{Geyser: Staking Rewards For Uniswap Liquidity
  Providers}}.
\newblock
\newblock
\urldef\tempurl%
\url{https://medium.com/trips-community/geyser-staking-rewards-for-uniswap-liquidity-providers-115afc6f5c07}
\showURL{%
\tempurl}


\bibitem[{DefiLlama}(2022)]%
        {DefiLlama}
\bibfield{author}{\bibinfo{person}{{DefiLlama}}.}
  \bibinfo{year}{2022}\natexlab{}.
\newblock \bibinfo{title}{{Dexes TVL Rankings}}.
\newblock
\newblock
\urldef\tempurl%
\url{https://defillama.com/protocols/Dexes}
\showURL{%
\tempurl}


\bibitem[{Degate}(2021)]%
        {sandwhichattack2021}
\bibfield{author}{\bibinfo{person}{{Degate}}.} \bibinfo{year}{2021}\natexlab{}.
\newblock \bibinfo{title}{{An Analysis of Ethereum Front-Running and its
  Defense Solutions}}.
\newblock
\newblock
\urldef\tempurl%
\url{https://globalcoinresearch.com/2021/05/04/an-analysis-of-ethereum-front-running-and-its-defense-solutions/}
\showURL{%
\tempurl}


\bibitem[{Delphi Digital}(2020)]%
        {delphi2020rollups}
\bibfield{author}{\bibinfo{person}{{Delphi Digital}}.}
  \bibinfo{year}{2020}\natexlab{}.
\newblock \bibinfo{booktitle}{\emph{{Layer 2: Rollups}}}.
\newblock \bibinfo{type}{{T}echnical {R}eport}.
\newblock


\bibitem[{demosthenes.eth}(2021)]%
        {Demosthenes.eth2021}
\bibfield{author}{\bibinfo{person}{{demosthenes.eth}}.}
  \bibinfo{year}{2021}\natexlab{}.
\newblock \bibinfo{title}{{Uniswap proposal: Managing Systemic Risk in
  Uniswap's Community Treasury using KPI Options}}.
\newblock
\newblock


\bibitem[{DODO}(2021a)]%
        {DODO2021}
\bibfield{author}{\bibinfo{person}{{DODO}}.} \bibinfo{year}{2021}\natexlab{a}.
\newblock \bibinfo{title}{{DODO Pool Incident Postmortem: With a Little Help
  from Our Friends}}.
\newblock
\newblock
\urldef\tempurl%
\url{https://medium.com/dodoex/dodo-pool-incident-postmortem-with-a-little-help-from-our-friends-327e66872d42}
\showURL{%
\tempurl}


\bibitem[{DODO}(2021b)]%
        {DODO2021a}
\bibfield{author}{\bibinfo{person}{{DODO}}.} \bibinfo{year}{2021}\natexlab{b}.
\newblock \bibinfo{title}{{How to create a pool?}}
\newblock
\newblock
\urldef\tempurl%
\url{https://dodoexhelp.zendesk.com/hc/en-us/articles/900005558243-How-to-create-a-pool-}
\showURL{%
\tempurl}


\bibitem[{DODO Team}(2020)]%
        {dodo2020whitepaper}
\bibfield{author}{\bibinfo{person}{{DODO Team}}.}
  \bibinfo{year}{2020}\natexlab{}.
\newblock \bibinfo{title}{{DODO -- A Next-Generation On-Chain Liquidity
  Provider Powered by Pro-active Market Maker Algorithm}}.
\newblock
\newblock
\urldef\tempurl%
\url{https://dodoex.github.io/docs/docs/whitepaper/}
\showURL{%
\tempurl}


\bibitem[DuPont and Squicciarini(2015)]%
        {DuPont2015TowardLocation}
\bibfield{author}{\bibinfo{person}{Jules DuPont} {and}
  \bibinfo{person}{Anna~Cinzia Squicciarini}.} \bibinfo{year}{2015}\natexlab{}.
\newblock \showarticletitle{{Toward de-anonymizing bitcoin by mapping users
  location}}. In \bibinfo{booktitle}{\emph{Proceedings of the 5th ACM
  Conference on Data and Application Security and Privacy}}.
  \bibinfo{pages}{139--141}.
\newblock


\bibitem[{dYdX}(2021)]%
        {dydx2021layer2}
\bibfield{author}{\bibinfo{person}{{dYdX}}.} \bibinfo{year}{2021}\natexlab{}.
\newblock \bibinfo{title}{{Trade now on Layer 2}}.
\newblock
\newblock
\urldef\tempurl%
\url{https://dydx.exchange/blog/public}
\showURL{%
\tempurl}


\bibitem[Dzyatkovskii(2021)]%
        {Dzyatkovskii2021}
\bibfield{author}{\bibinfo{person}{Anton Dzyatkovskii}.}
  \bibinfo{year}{2021}\natexlab{}.
\newblock \showarticletitle{{No Sandwich, Please! - Popular DeFi Attack
  Strategy Analysis | Hacker Noon}}.
\newblock \bibinfo{journal}{\emph{Hackernoon}} (\bibinfo{date}{5}
  \bibinfo{year}{2021}).
\newblock
\urldef\tempurl%
\url{https://hackernoon.com/no-sandwich-please-popular-defi-attack-strategy-analysis-jk1734rf}
\showURL{%
\tempurl}


\bibitem[Egorov(2019)]%
        {Egorov2019}
\bibfield{author}{\bibinfo{person}{Michael Egorov}.}
  \bibinfo{year}{2019}\natexlab{}.
\newblock \bibinfo{title}{{StableSwap-efficient mechanism for Stablecoin
  liquidity}}.  (\bibinfo{year}{2019}).
\newblock


\bibitem[Engelmann et~al\mbox{.}(2022)]%
        {Engelmann2022}
\bibfield{author}{\bibinfo{person}{Felix Engelmann}, \bibinfo{person}{Thomas
  Kerber}, \bibinfo{person}{Markulf Kohlweiss}, {and} \bibinfo{person}{Mikhail
  Volkhov}.} \bibinfo{year}{2022}\natexlab{}.
\newblock \showarticletitle{{Zswap: zk-SNARK Based Non-Interactive Multi-Asset
  Swaps}}.
\newblock \bibinfo{journal}{\emph{Proceedings on Privacy Enhancing Technologies
  (PoPETs)}}  \bibinfo{volume}{4} (\bibinfo{year}{2022}),
  \bibinfo{pages}{507--527}.
\newblock
\urldef\tempurl%
\url{https://doi.org/10.2478/popets-2022-0120}
\showDOI{\tempurl}


\bibitem[Eskandari et~al\mbox{.}(2020)]%
        {Eskandari2020}
\bibfield{author}{\bibinfo{person}{Shayan Eskandari},
  \bibinfo{person}{Seyedehmahsa Moosavi}, {and} \bibinfo{person}{Jeremy
  Clark}.} \bibinfo{year}{2020}\natexlab{}.
\newblock \showarticletitle{{SoK: Transparent Dishonesty: Front-Running Attacks
  on Blockchain}}. In \bibinfo{booktitle}{\emph{Lecture Notes in Computer
  Science}}, Vol.~\bibinfo{volume}{11599 LNCS}. \bibinfo{publisher}{Springer},
  \bibinfo{pages}{170--189}.
\newblock
\showISBNx{9783030437244}
\showISSN{16113349}
\urldef\tempurl%
\url{https://doi.org/10.1007/978-3-030-43725-1{\_}13}
\showDOI{\tempurl}


\bibitem[{Ethereum}(2020)]%
        {Ethereum2020}
\bibfield{author}{\bibinfo{person}{{Ethereum}}.}
  \bibinfo{year}{2020}\natexlab{}.
\newblock \bibinfo{title}{{Types}}.
\newblock
\newblock
\urldef\tempurl%
\url{https://docs.soliditylang.org/en/latest/types.html}
\showURL{%
\tempurl}


\bibitem[{ethereum.org}(2021a)]%
        {ethereum2021layertwo}
\bibfield{author}{\bibinfo{person}{{ethereum.org}}.}
  \bibinfo{year}{2021}\natexlab{a}.
\newblock \bibinfo{title}{{Layer 2 Rollups}}.
\newblock
\newblock
\urldef\tempurl%
\url{https://ethereum.org/en/developers/docs/scaling/layer-2-rollups/}
\showURL{%
\tempurl}


\bibitem[{ethereum.org}(2021b)]%
        {ethereum2021scaling}
\bibfield{author}{\bibinfo{person}{{ethereum.org}}.}
  \bibinfo{year}{2021}\natexlab{b}.
\newblock \bibinfo{title}{{Scaling}}.
\newblock
\newblock
\urldef\tempurl%
\url{https://ethereum.org/en/developers/docs/scaling/}
\showURL{%
\tempurl}


\bibitem[{Etherscan}(2021)]%
        {tmpl2021etherscan}
\bibfield{author}{\bibinfo{person}{{Etherscan}}.}
  \bibinfo{year}{2021}\natexlab{}.
\newblock \bibinfo{title}{{TruAmpl (TMPL) Token Tracker}}.
\newblock
\newblock
\urldef\tempurl%
\url{https://etherscan.io/token/0xfcb755b046ea9b9bc4586db4018b49c5a02e3d1c}
\showURL{%
\tempurl}


\bibitem[{EulerBeats}(2021)]%
        {eulerbeats2021}
\bibfield{author}{\bibinfo{person}{{EulerBeats}}.}
  \bibinfo{year}{2021}\natexlab{}.
\newblock \bibinfo{title}{{About EulerBeats}}.
\newblock
\newblock
\urldef\tempurl%
\url{https://eulerbeats.com/about}
\showURL{%
\tempurl}


\bibitem[Evans(2020)]%
        {Evans2020}
\bibfield{author}{\bibinfo{person}{Alex Evans}.}
  \bibinfo{year}{2020}\natexlab{}.
\newblock \bibinfo{title}{{Liquidity Provider Returns in Geometric Mean
  Markets}}.  (\bibinfo{date}{6} \bibinfo{year}{2020}).
\newblock
\urldef\tempurl%
\url{http://arxiv.org/abs/2006.08806}
\showURL{%
\tempurl}


\bibitem[Fatima~Samreen and Alalfi(2020)]%
        {Fatima2020}
\bibfield{author}{\bibinfo{person}{Noama Fatima~Samreen} {and}
  \bibinfo{person}{Manar~H Alalfi}.} \bibinfo{year}{2020}\natexlab{}.
\newblock \showarticletitle{{Reentrancy Vulnerability Identification in
  Ethereum Smart Contracts}}. In \bibinfo{booktitle}{\emph{2020 IEEE
  International Workshop on Blockchain Oriented Software Engineering
  (IWBOSE)}}. \bibinfo{pages}{22--29}.
\newblock
\urldef\tempurl%
\url{https://doi.org/10.1109/IWBOSE50093.2020.9050260}
\showDOI{\tempurl}


\bibitem[Feng et~al\mbox{.}(2020)]%
        {feng2020application}
\bibfield{author}{\bibinfo{person}{Yebo Feng}, \bibinfo{person}{Jun Li}, {and}
  \bibinfo{person}{Thanh Nguyen}.} \bibinfo{year}{2020}\natexlab{}.
\newblock \showarticletitle{{Application-layer DDoS defense with reinforcement
  learning}}. In \bibinfo{booktitle}{\emph{2020 IEEE/ACM 28th International
  Symposium on Quality of Service (IWQoS)}}. \bibinfo{pages}{1--10}.
\newblock


\bibitem[Feng et~al\mbox{.}(2022)]%
        {Feng2022}
\bibfield{author}{\bibinfo{person}{Yebo Feng}, \bibinfo{person}{Jiahua Xu},
  {and} \bibinfo{person}{Lauren Weymouth}.} \bibinfo{year}{2022}\natexlab{}.
\newblock \showarticletitle{{University Blockchain Research Initiative (UBRI):
  Boosting blockchain education and research}}.
\newblock \bibinfo{journal}{\emph{IEEE Potentials}} \bibinfo{volume}{41},
  \bibinfo{number}{6} (\bibinfo{date}{11} \bibinfo{year}{2022}),
  \bibinfo{pages}{19--25}.
\newblock
\showISSN{0278-6648}
\urldef\tempurl%
\url{https://doi.org/10.1109/MPOT.2022.3198929}
\showDOI{\tempurl}


\bibitem[Foxley(2021a)]%
        {Foxley2021}
\bibfield{author}{\bibinfo{person}{William Foxley}.}
  \bibinfo{year}{2021}\natexlab{a}.
\newblock \bibinfo{title}{{Uniswap V3 Introduces New License to Spoil Future
  SUSHIs}}.
\newblock
\newblock
\urldef\tempurl%
\url{https://www.coindesk.com/tech/2021/03/23/uniswap-v3-introduces-new-license-to-spoil-future-sushis/}
\showURL{%
\tempurl}


\bibitem[Foxley(2021b)]%
        {Foxley2021Vampire}
\bibfield{author}{\bibinfo{person}{William Foxley}.}
  \bibinfo{year}{2021}\natexlab{b}.
\newblock \bibinfo{title}{{‘Continuous Vampire Attack’: The AMM Wars Are
  Getting Interesting With Integral - CoinDesk}}.
\newblock
\newblock
\urldef\tempurl%
\url{https://www.coindesk.com/tech/2021/03/29/continuous-vampire-attack-the-amm-wars-are-getting-interesting-with-integral/}
\showURL{%
\tempurl}


\bibitem[Garman(1976)]%
        {Garman1976MarketMicrostructure}
\bibfield{author}{\bibinfo{person}{Mark~B. Garman}.}
  \bibinfo{year}{1976}\natexlab{}.
\newblock \showarticletitle{{Market microstructure}}.
\newblock \bibinfo{journal}{\emph{Journal of Financial Economics}}
  \bibinfo{volume}{3}, \bibinfo{number}{3} (\bibinfo{date}{6}
  \bibinfo{year}{1976}), \bibinfo{pages}{257--275}.
\newblock
\showISSN{0304405X}
\urldef\tempurl%
\url{https://doi.org/10.1016/0304-405X(76)90006-4}
\showDOI{\tempurl}


\bibitem[Garner et~al\mbox{.}(2021)]%
        {Garner2021}
\bibfield{author}{\bibinfo{person}{Ryan Garner}, \bibinfo{person}{Webb
  Mycelium}, \bibinfo{person}{Jason Potts}, \bibinfo{person}{Chris Berg}, {and}
  \bibinfo{person}{Sinclair Davidson}.} \bibinfo{year}{2021}\natexlab{}.
\newblock \showarticletitle{{Tracer: Perpetual Swaps}}.
\newblock  (\bibinfo{year}{2021}).
\newblock


\bibitem[Gentry(2009)]%
        {Gentry2009FullyLattices}
\bibfield{author}{\bibinfo{person}{Craig Gentry}.}
  \bibinfo{year}{2009}\natexlab{}.
\newblock \showarticletitle{{Fully homomorphic encryption using ideal
  lattices}}. In \bibinfo{booktitle}{\emph{Proceedings of the forty-first
  annual ACM symposium on Theory of computing}}. \bibinfo{pages}{169--178}.
\newblock


\bibitem[Gluchowski(2019)]%
        {gluchowski2019zk}
\bibfield{author}{\bibinfo{person}{Alex Gluchowski}.}
  \bibinfo{year}{2019}\natexlab{}.
\newblock \bibinfo{title}{{Zk rollup: scaling with zero-knowledge proofs}}.
\newblock
\newblock
\urldef\tempurl%
\url{https://pandax-statics.oss-cn-shenzhen.aliyuncs.com/statics/1221233526992813.pdf}
\showURL{%
\tempurl}


\bibitem[{Gnosis}(2020)]%
        {gnosiscmm}
\bibfield{author}{\bibinfo{person}{{Gnosis}}.} \bibinfo{year}{2020}\natexlab{}.
\newblock \bibinfo{title}{{Custom Market Maker {\textperiodcentered} Gnosis
  Developer Portal Gnosis Protocol}}.
\newblock
\newblock
\urldef\tempurl%
\url{https://docs.gnosis.io/protocol/docs/intro-cmm/}
\showURL{%
\tempurl}


\bibitem[Goldreich and Oren(1994)]%
        {Goldreich1994DefinitionsSystems}
\bibfield{author}{\bibinfo{person}{Oded Goldreich} {and} \bibinfo{person}{Yair
  Oren}.} \bibinfo{year}{1994}\natexlab{}.
\newblock \showarticletitle{{Definitions and properties of zero-knowledge proof
  systems}}.
\newblock \bibinfo{journal}{\emph{Journal of Cryptology}} \bibinfo{volume}{7},
  \bibinfo{number}{1} (\bibinfo{year}{1994}), \bibinfo{pages}{1--32}.
\newblock


\bibitem[Goldwasser and Rothblum(2007)]%
        {Goldwasser2007OnObfuscation}
\bibfield{author}{\bibinfo{person}{Shafi Goldwasser} {and}
  \bibinfo{person}{Guy~N Rothblum}.} \bibinfo{year}{2007}\natexlab{}.
\newblock \showarticletitle{{On best-possible obfuscation}}. In
  \bibinfo{booktitle}{\emph{Theory of Cryptography Conference}}.
  \bibinfo{pages}{194--213}.
\newblock


\bibitem[Govindarajan et~al\mbox{.}(2022)]%
        {govindarajan2022privacy}
\bibfield{author}{\bibinfo{person}{Kavya Govindarajan},
  \bibinfo{person}{Dhinakaran Vinayagamurthy}, \bibinfo{person}{Praveen
  Jayachandran}, {and} \bibinfo{person}{Chester Rebeiro}.}
  \bibinfo{year}{2022}\natexlab{}.
\newblock \showarticletitle{Privacy-preserving decentralized exchange
  marketplaces}. In \bibinfo{booktitle}{\emph{2022 IEEE International
  Conference on Blockchain and Cryptocurrency (ICBC)}}. IEEE,
  \bibinfo{pages}{1--9}.
\newblock


\bibitem[Grabundzija(2021)]%
        {Pancakebunny2021BSC}
\bibfield{author}{\bibinfo{person}{Ana Grabundzija}.}
  \bibinfo{year}{2021}\natexlab{}.
\newblock \bibinfo{title}{{BSC DeFi app 'Pancakebunny' releases post-mortem of
  {\$}2.4 million exploit}}.
\newblock
\newblock
\urldef\tempurl%
\url{https://cryptoslate.com/bsc-defi-app-pancakebunny-releases-post-mortem-of-2-4-million-exploit/}
\showURL{%
\tempurl}


\bibitem[Greene and Johnstone(2018)]%
        {Greene2018}
\bibfield{author}{\bibinfo{person}{Richard Greene} {and}
  \bibinfo{person}{Michael~N Johnstone}.} \bibinfo{year}{2018}\natexlab{}.
\newblock \bibinfo{title}{{An investigation into a denial of service attack on
  an ethereum network}}.
\newblock
\newblock


\bibitem[Gudgeon et~al\mbox{.}(2020a)]%
        {gudgeon2020layer2}
\bibfield{author}{\bibinfo{person}{Lewis Gudgeon}, \bibinfo{person}{Pedro
  Moreno-Sanchez}, \bibinfo{person}{Stefanie Roos}, \bibinfo{person}{Patrick
  McCorry}, {and} \bibinfo{person}{Arthur Gervais}.}
  \bibinfo{year}{2020}\natexlab{a}.
\newblock \showarticletitle{{SoK: Layer-Two Blockchain Protocols}}. In
  \bibinfo{booktitle}{\emph{Financial Cryptography and Data Security}},
  Vol.~\bibinfo{volume}{12059 LNCS}. \bibinfo{publisher}{Springer, Cham},
  \bibinfo{pages}{201--226}.
\newblock
\urldef\tempurl%
\url{https://doi.org/10.1007/978-3-030-51280-4{\_}12}
\showDOI{\tempurl}


\bibitem[Gudgeon et~al\mbox{.}(2020b)]%
        {gudgeon2020}
\bibfield{author}{\bibinfo{person}{Lewis Gudgeon}, \bibinfo{person}{Daniel
  Perez}, \bibinfo{person}{Dominik Harz}, \bibinfo{person}{Benjamin Livshits},
  {and} \bibinfo{person}{Arthur Gervais}.} \bibinfo{year}{2020}\natexlab{b}.
\newblock \showarticletitle{{The Decentralized Financial Crisis}}. In
  \bibinfo{booktitle}{\emph{Crypto Valley Conference on Blockchain Technology
  (CVCBT)}}. \bibinfo{publisher}{IEEE}, \bibinfo{pages}{1--15}.
\newblock
\showISBNx{978-1-7281-9390-8}
\urldef\tempurl%
\url{https://doi.org/10.1109/CVCBT50464.2020.00005}
\showDOI{\tempurl}


\bibitem[{Gyroscope Finance}(2021a)]%
        {gyroscope2021amm}
\bibfield{author}{\bibinfo{person}{{Gyroscope Finance}}.}
  \bibinfo{year}{2021}\natexlab{a}.
\newblock \bibinfo{title}{{Autonomous pricing}}.
\newblock
\newblock
\urldef\tempurl%
\url{https://docs.gyro.finance/gyroscope-protocol/stablecoin/autonomous-pricing}
\showURL{%
\tempurl}


\bibitem[{Gyroscope Finance}(2021b)]%
        {gyroscope2021}
\bibfield{author}{\bibinfo{person}{{Gyroscope Finance}}.}
  \bibinfo{year}{2021}\natexlab{b}.
\newblock \bibinfo{title}{{Gyroscope, the new all-weather stablecoin}}.
\newblock
\newblock
\urldef\tempurl%
\url{https://gyro.finance/}
\showURL{%
\tempurl}


\bibitem[Hafid et~al\mbox{.}(2020)]%
        {hafid2020scaling}
\bibfield{author}{\bibinfo{person}{Abdelatif Hafid},
  \bibinfo{person}{Abdelhakim~Senhaji Hafid}, {and} \bibinfo{person}{Mustapha
  Samih}.} \bibinfo{year}{2020}\natexlab{}.
\newblock \showarticletitle{{Scaling Blockchains: A Comprehensive Survey}}.
\newblock \bibinfo{journal}{\emph{IEEE Access}}  \bibinfo{volume}{8}
  (\bibinfo{year}{2020}), \bibinfo{pages}{125244--125262}.
\newblock
\showISSN{2169-3536}
\urldef\tempurl%
\url{https://doi.org/10.1109/ACCESS.2020.3007251}
\showDOI{\tempurl}


\bibitem[{hagaetc}(2021)]%
        {obadiaa2019}
\bibfield{author}{\bibinfo{person}{{hagaetc}}.}
  \bibinfo{year}{2021}\natexlab{}.
\newblock \bibinfo{title}{{Weekly DEX volume}}.
\newblock
\newblock
\urldef\tempurl%
\url{https://dune.xyz/queries/4323/8547 https://github.com/flashbots/pm}
\showURL{%
\tempurl}


\bibitem[Hanson(2003)]%
        {Hanson2003CombinatorialDesign}
\bibfield{author}{\bibinfo{person}{Robin Hanson}.}
  \bibinfo{year}{2003}\natexlab{}.
\newblock \showarticletitle{{Combinatorial Information Market Design}}.
\newblock \bibinfo{journal}{\emph{Information Systems Frontiers}}
  \bibinfo{volume}{5}, \bibinfo{number}{1} (\bibinfo{year}{2003}),
  \bibinfo{pages}{107--119}.
\newblock
\urldef\tempurl%
\url{http://hanson.gmu.edu}
\showURL{%
\tempurl}


\bibitem[Hanson(2012)]%
        {Hanson2012LogarithmicAggregation}
\bibfield{author}{\bibinfo{person}{Robin Hanson}.}
  \bibinfo{year}{2012}\natexlab{}.
\newblock \showarticletitle{{Logarithmic Markets Scoring Rules for Modular
  Combinatorial Information Aggregation}}.
\newblock \bibinfo{journal}{\emph{The Journal of Prediction Markets}}
  \bibinfo{volume}{1}, \bibinfo{number}{1} (\bibinfo{date}{12}
  \bibinfo{year}{2012}), \bibinfo{pages}{3--15}.
\newblock
\showISSN{1750-6751}
\urldef\tempurl%
\url{https://doi.org/10.5750/jpm.v1i1.417}
\showDOI{\tempurl}


\bibitem[{Harvest Finance}(2020)]%
        {harvest2020attack}
\bibfield{author}{\bibinfo{person}{{Harvest Finance}}.}
  \bibinfo{year}{2020}\natexlab{}.
\newblock \bibinfo{title}{{Harvest Flashloan Economic Attack Post-Mortem }}.
\newblock
\newblock
\urldef\tempurl%
\url{https://medium.com/harvest-finance/harvest-flashloan-economic-attack-post-mortem-3cf900d65217}
\showURL{%
\tempurl}


\bibitem[Heimbach and Wattenhofer(2022)]%
        {Heimbach2022EliminatingTheory}
\bibfield{author}{\bibinfo{person}{Lioba Heimbach} {and} \bibinfo{person}{Roger
  Wattenhofer}.} \bibinfo{year}{2022}\natexlab{}.
\newblock \showarticletitle{{Eliminating Sandwich Attacks with the Help of Game
  Theory}}. In \bibinfo{booktitle}{\emph{Asia Conference on Computer and
  Communications Security}}, Vol.~\bibinfo{volume}{15}.
  \bibinfo{publisher}{ACM}, \bibinfo{address}{New York, NY, USA},
  \bibinfo{pages}{153--167}.
\newblock
\showISBNx{9781450391405}
\urldef\tempurl%
\url{https://doi.org/10.1145/3488932.3517390}
\showDOI{\tempurl}


\bibitem[Hertzog et~al\mbox{.}(2018)]%
        {Hertzog2018}
\bibfield{author}{\bibinfo{person}{Eyal Hertzog}, \bibinfo{person}{Guy
  Galia~Guy Benartzi}, {and} \bibinfo{person}{Guy Galia~Guy Benartzi}.}
  \bibinfo{year}{2018}\natexlab{}.
\newblock \bibinfo{title}{{Bancor Protocol Continuous Liquidity for
  Cryptographic Tokens through their Smart Contracts}}.
  (\bibinfo{year}{2018}).
\newblock
\urldef\tempurl%
\url{https://storage.googleapis.com/website-bancor/2018/04/01ba8253-bancor_protocol_whitepaper_en.pdf}
\showURL{%
\tempurl}


\bibitem[Homoliak et~al\mbox{.}(2021)]%
        {Homoliak2021}
\bibfield{author}{\bibinfo{person}{Ivan Homoliak}, \bibinfo{person}{Sarad
  Venugopalan}, \bibinfo{person}{Daniel Reijsbergen}, \bibinfo{person}{Qingze
  Hum}, \bibinfo{person}{Richard Schumi}, {and} \bibinfo{person}{Pawel
  Szalachowski}.} \bibinfo{year}{2021}\natexlab{}.
\newblock \showarticletitle{{The Security Reference Architecture for
  Blockchains: Toward a Standardized Model for Studying Vulnerabilities,
  Threats, and Defenses}}.
\newblock \bibinfo{journal}{\emph{IEEE Communications Surveys and Tutorials}}
  \bibinfo{volume}{23}, \bibinfo{number}{1} (\bibinfo{date}{1}
  \bibinfo{year}{2021}), \bibinfo{pages}{341--390}.
\newblock
\showISSN{1553877X}
\urldef\tempurl%
\url{https://doi.org/10.1109/COMST.2020.3033665}
\showDOI{\tempurl}


\bibitem[Huang(2019)]%
        {Huang2019}
\bibfield{author}{\bibinfo{person}{Sophie Huang}.}
  \bibinfo{year}{2019}\natexlab{}.
\newblock \bibinfo{title}{{Will 2020 be The Year of DEX?}}
\newblock
\newblock
\urldef\tempurl%
\url{https://medium.com/@kidinamoto/will-2020-be-the-year-of-dex-ac7dfb6276e8}
\showURL{%
\tempurl}


\bibitem[Huang et~al\mbox{.}(2019)]%
        {huang2019smart}
\bibfield{author}{\bibinfo{person}{Yongfeng Huang}, \bibinfo{person}{Yiyang
  Bian}, \bibinfo{person}{Renpu Li}, \bibinfo{person}{J~Leon Zhao}, {and}
  \bibinfo{person}{Peizhong Shi}.} \bibinfo{year}{2019}\natexlab{}.
\newblock \showarticletitle{{Smart contract security: A software lifecycle
  perspective}}.
\newblock \bibinfo{journal}{\emph{IEEE Access}}  \bibinfo{volume}{7}
  (\bibinfo{year}{2019}), \bibinfo{pages}{150184--150202}.
\newblock


\bibitem[{HydraDX}(2021)]%
        {hydradx2021}
\bibfield{author}{\bibinfo{person}{{HydraDX}}.}
  \bibinfo{year}{2021}\natexlab{}.
\newblock \bibinfo{title}{{Intro | HydraDX Docs}}.
\newblock
\newblock
\urldef\tempurl%
\url{https://docs.hydradx.io/}
\showURL{%
\tempurl}


\bibitem[{Jakub}(2020)]%
        {jakub2020vampire}
\bibfield{author}{\bibinfo{person}{{Jakub}}.} \bibinfo{year}{2020}\natexlab{}.
\newblock \bibinfo{title}{{What is a Vampire Attack? SushiSwap Saga
  Explained}}.
\newblock
\newblock
\urldef\tempurl%
\url{https://finematics.com/vampire-attack-sushiswap-explained/}
\showURL{%
\tempurl}


\bibitem[Jourenko et~al\mbox{.}(2019)]%
        {jourenko2019layer2}
\bibfield{author}{\bibinfo{person}{Maxim Jourenko}, \bibinfo{person}{Mario
  Larangeira}, \bibinfo{person}{Kanta Kurazumi}, {and} \bibinfo{person}{Keisuke
  Tanaka}.} \bibinfo{year}{2019}\natexlab{}.
\newblock \bibinfo{title}{{SoK: A Taxonomy for Layer-2 Scalability Related
  Protocols for Cryptocurrencies}}.  (\bibinfo{year}{2019}),
  \bibinfo{numpages}{19}~pages.
\newblock
\urldef\tempurl%
\url{https://eprint.iacr.org/2019/352.pdf}
\showURL{%
\tempurl}


\bibitem[Jumadinova and Dasgupta(2012)]%
        {Jumadinova2012}
\bibfield{author}{\bibinfo{person}{Janyl Jumadinova} {and}
  \bibinfo{person}{Prithviraj Dasgupta}.} \bibinfo{year}{2012}\natexlab{}.
\newblock \showarticletitle{{A Comparison of Different Automated Market-Maker
  Strategies}}. \bibinfo{pages}{2009--2012}.
\newblock
\urldef\tempurl%
\url{http://www.cs.allegheny.edu/~jjumadinova/market-maker_AMEC.pdf}
\showURL{%
\tempurl}


\bibitem[Keoun et~al\mbox{.}(2020)]%
        {Keoun2020}
\bibfield{author}{\bibinfo{person}{Bradley Keoun}, \bibinfo{person}{Omkar
  Godbole}, {and} \bibinfo{person}{Sebastian Sinclair}.}
  \bibinfo{year}{2020}\natexlab{}.
\newblock \bibinfo{title}{{First Mover: SushiSwap's Billion-Dollar 'Rug Pull'
  Is Thriller to Crypto Geeks - CoinDesk}}.
\newblock
\newblock
\urldef\tempurl%
\url{https://www.coindesk.com/markets/2020/09/08/first-mover-sushiswaps-billion-dollar-rug-pull-is-thriller-to-crypto-geeks/}
\showURL{%
\tempurl}


\bibitem[Kisagun(2019)]%
        {Kisagun2019}
\bibfield{author}{\bibinfo{person}{Can Kisagun}.}
  \bibinfo{year}{2019}\natexlab{}.
\newblock \bibinfo{title}{{Preventing DEX Front-running with Enigma}}.
\newblock
\newblock
\urldef\tempurl%
\url{https://blog.enigma.co/preventing-dex-front-running-with-enigma-df3f0b5b9e78}
\showURL{%
\tempurl}


\bibitem[Konstantopoulos(2021)]%
        {konstantopoulos2021rollups}
\bibfield{author}{\bibinfo{person}{Georgios Konstantopoulos}.}
  \bibinfo{year}{2021}\natexlab{}.
\newblock \bibinfo{title}{{(Almost) Everything you need to know about
  Optimistic Rollup}}.
\newblock
\newblock
\urldef\tempurl%
\url{https://research.paradigm.xyz/rollups}
\showURL{%
\tempurl}


\bibitem[Kosba et~al\mbox{.}(2016)]%
        {Kosba2016Hawk:Contracts}
\bibfield{author}{\bibinfo{person}{Ahmed Kosba}, \bibinfo{person}{Andrew
  Miller}, \bibinfo{person}{Elaine Shi}, \bibinfo{person}{Zikai Wen}, {and}
  \bibinfo{person}{Charalampos Papamanthou}.} \bibinfo{year}{2016}\natexlab{}.
\newblock \showarticletitle{{Hawk: The blockchain model of cryptography and
  privacy-preserving smart contracts}}. In \bibinfo{booktitle}{\emph{2016 IEEE
  symposium on security and privacy (SP)}}. \bibinfo{pages}{839--858}.
\newblock


\bibitem[Krishnamachari et~al\mbox{.}(2021)]%
        {nguyen2021damm}
\bibfield{author}{\bibinfo{person}{Bhaskar Krishnamachari}, \bibinfo{person}{Qi
  Feng}, {and} \bibinfo{person}{Eugenio Grippo}.}
  \bibinfo{year}{2021}\natexlab{}.
\newblock \showarticletitle{{Dynamic Automated Market Makers for Decentralized
  Cryptocurrency Exchange}}. In \bibinfo{booktitle}{\emph{IEEE International
  Conference on Blockchain and Cryptocurrency (ICBC)}}.
  \bibinfo{publisher}{IEEE}, \bibinfo{pages}{1--2}.
\newblock
\showISBNx{9781665435789}
\urldef\tempurl%
\url{https://doi.org/10.1109/icbc51069.2021.9461100}
\showDOI{\tempurl}


\bibitem[{Kyber Network}(2021)]%
        {Kyber2021whitepaper}
\bibfield{author}{\bibinfo{person}{{Kyber Network}}.}
  \bibinfo{year}{2021}\natexlab{}.
\newblock \bibinfo{title}{{Kyber 3.0: Architecture Revamp, Dynamic MM, and KNC
  Migration Proposal}}.
\newblock
\newblock
\urldef\tempurl%
\url{https://blog.kyber.network/kyber-3-0-architecture-revamp-dynamic-mm-and-knc-migration-proposal-acae41046513}
\showURL{%
\tempurl}


\bibitem[Li et~al\mbox{.}(2020)]%
        {Li2020ASystems}
\bibfield{author}{\bibinfo{person}{Xiaoqi Li}, \bibinfo{person}{Peng Jiang},
  \bibinfo{person}{Ting Chen}, \bibinfo{person}{Xiapu Luo}, {and}
  \bibinfo{person}{Qiaoyan Wen}.} \bibinfo{year}{2020}\natexlab{}.
\newblock \showarticletitle{{A survey on the security of blockchain systems}}.
\newblock \bibinfo{journal}{\emph{Future Generation Computer Systems}}
  \bibinfo{volume}{107} (\bibinfo{year}{2020}), \bibinfo{pages}{841--853}.
\newblock


\bibitem[Lin and Liao(2017)]%
        {Lin2017AChallenges.}
\bibfield{author}{\bibinfo{person}{Iuon-Chang Lin} {and}
  \bibinfo{person}{Tzu-Chun Liao}.} \bibinfo{year}{2017}\natexlab{}.
\newblock \showarticletitle{{A survey of blockchain security issues and
  challenges.}}
\newblock \bibinfo{journal}{\emph{Int. J. Netw. Secur.}} \bibinfo{volume}{19},
  \bibinfo{number}{5} (\bibinfo{year}{2017}), \bibinfo{pages}{653--659}.
\newblock


\bibitem[Liu et~al\mbox{.}(2018)]%
        {Liu2018}
\bibfield{author}{\bibinfo{person}{Chao Liu}, \bibinfo{person}{Han Liu},
  \bibinfo{person}{Zhao Cao}, \bibinfo{person}{Zhong Chen},
  \bibinfo{person}{Bangdao Chen}, {and} \bibinfo{person}{Bill Roscoe}.}
  \bibinfo{year}{2018}\natexlab{}.
\newblock \showarticletitle{{Reguard: finding reentrancy bugs in smart
  contracts}}. In \bibinfo{booktitle}{\emph{2018 IEEE/ACM 40th International
  Conference on Software Engineering: Companion (ICSE-Companion)}}.
  \bibinfo{publisher}{IEEE}, \bibinfo{pages}{65--68}.
\newblock
\showISBNx{145035663X}


\bibitem[{livnev}(2020)]%
        {Livnev2020}
\bibfield{author}{\bibinfo{person}{{livnev}}.} \bibinfo{year}{2020}\natexlab{}.
\newblock \bibinfo{title}{{Random ordering of equally-priced transactions
  incentivises competitive spam}}.
\newblock
\newblock
\urldef\tempurl%
\url{https://github.com/ethereum/go-ethereum/issues/21350}
\showURL{%
\tempurl}


\bibitem[Lo and Medda(2020)]%
        {Lo2020UniswapExchange}
\bibfield{author}{\bibinfo{person}{Yuen Lo} {and} \bibinfo{person}{Francesca
  Medda}.} \bibinfo{year}{2020}\natexlab{}.
\newblock \bibinfo{title}{{Uniswap and the rise of the decentralized
  exchange}}.  (\bibinfo{date}{11} \bibinfo{year}{2020}).
\newblock


\bibitem[Lu et~al\mbox{.}(2019)]%
        {lu2019neucheck}
\bibfield{author}{\bibinfo{person}{Ning Lu}, \bibinfo{person}{Bin Wang},
  \bibinfo{person}{Yongxin Zhang}, \bibinfo{person}{Wenbo Shi}, {and}
  \bibinfo{person}{Christian Esposito}.} \bibinfo{year}{2019}\natexlab{}.
\newblock \showarticletitle{{NeuCheck: A more practical Ethereum smart contract
  security analysis tool}}.
\newblock \bibinfo{journal}{\emph{Software: Practice and Experience}}
  (\bibinfo{year}{2019}).
\newblock


\bibitem[Luu et~al\mbox{.}(2016)]%
        {luu2016making}
\bibfield{author}{\bibinfo{person}{Loi Luu}, \bibinfo{person}{Duc-Hiep Chu},
  \bibinfo{person}{Hrishi Olickel}, \bibinfo{person}{Prateek Saxena}, {and}
  \bibinfo{person}{Aquinas Hobor}.} \bibinfo{year}{2016}\natexlab{}.
\newblock \showarticletitle{{Making Smart Contracts Smarter}}. In
  \bibinfo{booktitle}{\emph{ACM SIGSAC Conference on Computer and
  Communications Security}}. \bibinfo{publisher}{ACM}, \bibinfo{address}{New
  York, NY, USA}, \bibinfo{pages}{254--269}.
\newblock
\showISBNx{9781450341394}
\showISSN{15437221}
\urldef\tempurl%
\url{https://doi.org/10.1145/2976749.2978309}
\showDOI{\tempurl}


\bibitem[Lyanchev(2021)]%
        {Lyanchev2021}
\bibfield{author}{\bibinfo{person}{Jordan Lyanchev}.}
  \bibinfo{year}{2021}\natexlab{}.
\newblock \showarticletitle{{{\$}50M Drained From Uranium Finance: Hack or Rug
  Pull?}}
\newblock  (\bibinfo{year}{2021}).
\newblock
\urldef\tempurl%
\url{https://cryptopotato.com/50m-drained-from-uranium-finance-hack-or-rug-pull/}
\showURL{%
\tempurl}


\bibitem[Ma et~al\mbox{.}(2020)]%
        {ma2020achieving}
\bibfield{author}{\bibinfo{person}{Guangkai Ma}, \bibinfo{person}{Chunpeng Ge},
  {and} \bibinfo{person}{Lu Zhou}.} \bibinfo{year}{2020}\natexlab{}.
\newblock \showarticletitle{Achieving reliable timestamp in the bitcoin
  platform}.
\newblock \bibinfo{journal}{\emph{Peer-to-Peer Networking and Applications}}
  \bibinfo{volume}{13}, \bibinfo{number}{6} (\bibinfo{year}{2020}),
  \bibinfo{pages}{2251--2259}.
\newblock


\bibitem[Makarov and Schoar(2020)]%
        {makarov2020trading}
\bibfield{author}{\bibinfo{person}{Igor Makarov} {and}
  \bibinfo{person}{Antoinette Schoar}.} \bibinfo{year}{2020}\natexlab{}.
\newblock \showarticletitle{Trading and arbitrage in cryptocurrency markets}.
\newblock \bibinfo{journal}{\emph{Journal of Financial Economics}}
  \bibinfo{volume}{135}, \bibinfo{number}{2} (\bibinfo{year}{2020}),
  \bibinfo{pages}{293--319}.
\newblock


\bibitem[Malamud and Rostek(2017)]%
        {Malamud2017}
\bibfield{author}{\bibinfo{person}{Semyon Malamud} {and}
  \bibinfo{person}{Marzena Rostek}.} \bibinfo{year}{2017}\natexlab{}.
\newblock \showarticletitle{{Decentralized Exchange}}.
\newblock \bibinfo{journal}{\emph{American Economic Review}}
  \bibinfo{volume}{107}, \bibinfo{number}{11} (\bibinfo{date}{11}
  \bibinfo{year}{2017}), \bibinfo{pages}{3320--3362}.
\newblock
\showISSN{0002-8282}
\urldef\tempurl%
\url{https://doi.org/10.1257/aer.20140759}
\showDOI{\tempurl}


\bibitem[Malwa(2021)]%
        {Malwa2021}
\bibfield{author}{\bibinfo{person}{Shaurya Malwa}.}
  \bibinfo{year}{2021}\natexlab{}.
\newblock \showarticletitle{{DeFi ‘Rug Pull’ Scams Pulled In {\$}2.8B This
  Year: Chainalysis}}.
\newblock  (\bibinfo{date}{12} \bibinfo{year}{2021}).
\newblock
\urldef\tempurl%
\url{https://www.coindesk.com/markets/2021/12/17/defi-rug-pull-scams-pulled-in-28b-this-year-chainalysis/}
\showURL{%
\tempurl}


\bibitem[Martinelli(2021)]%
        {martinelli2021balancerv2}
\bibfield{author}{\bibinfo{person}{Fernando Martinelli}.}
  \bibinfo{year}{2021}\natexlab{}.
\newblock \bibinfo{title}{{Introducing Balancer V2: Generalized AMMs}}.
\newblock
\newblock
\urldef\tempurl%
\url{https://medium.com/balancer-protocol/balancer-v2-generalizing-amms-16343c4563ff}
\showURL{%
\tempurl}


\bibitem[Martinelli and Mushegian(2019)]%
        {martinelli2019whitepaper}
\bibfield{author}{\bibinfo{person}{Fernando Martinelli} {and}
  \bibinfo{person}{Nikolai Mushegian}.} \bibinfo{year}{2019}\natexlab{}.
\newblock \bibinfo{title}{{Balancer: A non-custodial portfolio manager,
  liquidity provider, and price sensor}}.  (\bibinfo{year}{2019}).
\newblock
\urldef\tempurl%
\url{https://balancer.finance/whitepaper/}
\showURL{%
\tempurl}


\bibitem[Massacci and Ngo(2021)]%
        {Massacci2021}
\bibfield{author}{\bibinfo{person}{Fabio Massacci} {and}
  \bibinfo{person}{Chan~Nam Ngo}.} \bibinfo{year}{2021}\natexlab{}.
\newblock \showarticletitle{{Distributed Financial Exchanges: Security
  Challenges and Design Principles}}.
\newblock \bibinfo{journal}{\emph{IEEE Security {\&} Privacy}}
  \bibinfo{volume}{19}, \bibinfo{number}{1} (\bibinfo{date}{1}
  \bibinfo{year}{2021}), \bibinfo{pages}{54--64}.
\newblock
\showISSN{1540-7993}
\urldef\tempurl%
\url{https://doi.org/10.1109/MSEC.2020.2994826}
\showDOI{\tempurl}


\bibitem[Mazorra et~al\mbox{.}(2022)]%
        {mazorra2022not}
\bibfield{author}{\bibinfo{person}{Bruno Mazorra}, \bibinfo{person}{Victor
  Adan}, {and} \bibinfo{person}{Vanesa Daza}.} \bibinfo{year}{2022}\natexlab{}.
\newblock \showarticletitle{Do not rug on me: Zero-dimensional Scam Detection}.
\newblock \bibinfo{journal}{\emph{arXiv preprint arXiv:2201.07220}}
  (\bibinfo{year}{2022}).
\newblock


\bibitem[Mense and Flatscher(2018)]%
        {Mense2018}
\bibfield{author}{\bibinfo{person}{Alexander Mense} {and}
  \bibinfo{person}{Markus Flatscher}.} \bibinfo{year}{2018}\natexlab{}.
\newblock \bibinfo{title}{{Security vulnerabilities in ethereum smart
  contracts}}.
\newblock , \bibinfo{numpages}{375--380}~pages.
\newblock


\bibitem[Miers et~al\mbox{.}(2013)]%
        {Miers2013Zerocoin:Bitcoin}
\bibfield{author}{\bibinfo{person}{Ian Miers}, \bibinfo{person}{Christina
  Garman}, \bibinfo{person}{Matthew Green}, {and} \bibinfo{person}{Aviel~D
  Rubin}.} \bibinfo{year}{2013}\natexlab{}.
\newblock \showarticletitle{{Zerocoin: Anonymous distributed e-cash from
  bitcoin}}. In \bibinfo{booktitle}{\emph{2013 IEEE Symposium on Security and
  Privacy}}. \bibinfo{pages}{397--411}.
\newblock


\bibitem[Mikalauskas(2021)]%
        {Mikalauskas2021}
\bibfield{author}{\bibinfo{person}{Edvardas Mikalauskas}.}
  \bibinfo{year}{2021}\natexlab{}.
\newblock \showarticletitle{{{\$}280 million stolen per month from crypto
  transactions}}.
\newblock \bibinfo{journal}{\emph{Cybernews}} (\bibinfo{year}{2021}).
\newblock
\urldef\tempurl%
\url{https://cybernews.com/crypto/flash-boys-2-0-front-runners-draining-280-million-per-month-from-crypto-transactions/}
\showURL{%
\tempurl}


\bibitem[Mirkin et~al\mbox{.}(2020)]%
        {Mirkin2020}
\bibfield{author}{\bibinfo{person}{Michael Mirkin}, \bibinfo{person}{Yan Ji},
  \bibinfo{person}{Jonathan Pang}, \bibinfo{person}{Ariah Klages-Mundt},
  \bibinfo{person}{Ittay Eyal}, {and} \bibinfo{person}{Ari Juels}.}
  \bibinfo{year}{2020}\natexlab{}.
\newblock \showarticletitle{{BDoS: Blockchain Denial-of-Service}}. In
  \bibinfo{booktitle}{\emph{ACM SIGSAC Conference on Computer and
  Communications Security}}. \bibinfo{publisher}{ACM}, \bibinfo{address}{New
  York, NY, USA}, \bibinfo{pages}{601--619}.
\newblock
\showISBNx{9781450370899}
\urldef\tempurl%
\url{https://doi.org/10.1145/3372297.3417247}
\showDOI{\tempurl}


\bibitem[{Mudra Manager}(2021)]%
        {MudraManager}
\bibfield{author}{\bibinfo{person}{{Mudra Manager}}.}
  \bibinfo{year}{2021}\natexlab{}.
\newblock \bibinfo{title}{{Why Locking Liquidity is Important for
  Cryptocurrency}}.
\newblock
\newblock
\urldef\tempurl%
\url{https://hackernoon.com/why-locking-liquidity-is-important-for-cryptocurrency-qv4d37hd}
\showURL{%
\tempurl}


\bibitem[Narayanan et~al\mbox{.}(2016)]%
        {Narayanan2016BitcoinIntroduction}
\bibfield{author}{\bibinfo{person}{Arvind Narayanan}, \bibinfo{person}{Joseph
  Bonneau}, \bibinfo{person}{Edward Felten}, \bibinfo{person}{Andrew Miller},
  {and} \bibinfo{person}{Steven Goldfeder}.} \bibinfo{year}{2016}\natexlab{}.
\newblock \bibinfo{booktitle}{\emph{{Bitcoin and cryptocurrency technologies: a
  comprehensive introduction}}}.
\newblock \bibinfo{publisher}{Princeton University Press}.
\newblock


\bibitem[{Nasdaq}(2021)]%
        {nasdaq2021layer2}
\bibfield{author}{\bibinfo{person}{{Nasdaq}}.} \bibinfo{year}{2021}\natexlab{}.
\newblock \bibinfo{title}{{How Ethereum Layer 2's Are Leveling Up DeFi }}.
\newblock
\newblock
\urldef\tempurl%
\url{https://www.nasdaq.com/articles/how-ethereum-layer-2s-are-leveling-up-defi-2021-06-08}
\showURL{%
\tempurl}


\bibitem[Nava(2015)]%
        {Nava2015EfficiencyMarkets}
\bibfield{author}{\bibinfo{person}{Francesco Nava}.}
  \bibinfo{year}{2015}\natexlab{}.
\newblock \showarticletitle{{Efficiency in decentralized oligopolistic
  markets}}.
\newblock \bibinfo{journal}{\emph{Journal of Economic Theory}}
  \bibinfo{volume}{157} (\bibinfo{date}{5} \bibinfo{year}{2015}),
  \bibinfo{pages}{315--348}.
\newblock
\showISSN{10957235}
\urldef\tempurl%
\url{https://doi.org/10.1016/j.jet.2015.01.009}
\showDOI{\tempurl}


\bibitem[Niemerg et~al\mbox{.}(2020)]%
        {Niemerg2020}
\bibfield{author}{\bibinfo{person}{Allan Niemerg}, \bibinfo{person}{Dan
  Robinson}, {and} \bibinfo{person}{Lev Livnev}.}
  \bibinfo{year}{2020}\natexlab{}.
\newblock \bibinfo{title}{{YieldSpace: An Automated Liquidity Provider for
  Fixed Yield Tokens}}.  (\bibinfo{year}{2020}).
\newblock
\urldef\tempurl%
\url{https://yield.is/Yield.pdf}
\showURL{%
\tempurl}


\bibitem[Noether(2015)]%
        {Noether2015RingMonero.}
\bibfield{author}{\bibinfo{person}{Shen Noether}.}
  \bibinfo{year}{2015}\natexlab{}.
\newblock \showarticletitle{{Ring SIgnature Confidential Transactions for
  Monero.}}
\newblock \bibinfo{journal}{\emph{IACR Cryptol. ePrint Arch.}}
  \bibinfo{volume}{2015} (\bibinfo{year}{2015}), \bibinfo{pages}{1098}.
\newblock


\bibitem[{Notional Finance}(2020)]%
        {notional2020amm}
\bibfield{author}{\bibinfo{person}{{Notional Finance}}.}
  \bibinfo{year}{2020}\natexlab{}.
\newblock \bibinfo{title}{{Notional AMM}}.
\newblock
\newblock
\urldef\tempurl%
\url{https://docs.notional.finance/traders/technical-topics/notional-amm}
\showURL{%
\tempurl}


\bibitem[{Notional Finance}(2021)]%
        {notionalfinance2021}
\bibfield{author}{\bibinfo{person}{{Notional Finance}}.}
  \bibinfo{year}{2021}\natexlab{}.
\newblock \bibinfo{title}{{Notional Finance}}.
\newblock
\newblock
\urldef\tempurl%
\url{https://notional.finance/}
\showURL{%
\tempurl}


\bibitem[Ong(2021)]%
        {Ong2021}
\bibfield{author}{\bibinfo{person}{Jeremy Ong}.}
  \bibinfo{year}{2021}\natexlab{}.
\newblock \bibinfo{title}{{PancakeSwap : A Perpetual Vampire? - Delphi
  Digital}}.
\newblock
\newblock
\urldef\tempurl%
\url{https://members.delphidigital.io/reports/pancakeswap-a-perpetual-vampire/}
\showURL{%
\tempurl}


\bibitem[Oosthoek(2021)]%
        {Oosthoek2021}
\bibfield{author}{\bibinfo{person}{Kris Oosthoek}.}
  \bibinfo{year}{2021}\natexlab{}.
\newblock \bibinfo{title}{{Flash Crash for Cash: Cyber Threats in Decentralized
  Finance}}.  (\bibinfo{date}{6} \bibinfo{year}{2021}).
\newblock
\urldef\tempurl%
\url{https://arxiv.org/abs/2106.10740v1}
\showURL{%
\tempurl}


\bibitem[{Optimism}(2021)]%
        {optimism2021}
\bibfield{author}{\bibinfo{person}{{Optimism}}.}
  \bibinfo{year}{2021}\natexlab{}.
\newblock \bibinfo{title}{{Optimism home page}}.
\newblock
\newblock
\urldef\tempurl%
\url{https://optimism.io/}
\showURL{%
\tempurl}


\bibitem[Othman et~al\mbox{.}(2013)]%
        {Othman2013}
\bibfield{author}{\bibinfo{person}{Abraham Othman}, \bibinfo{person}{David~M.
  Pennock}, \bibinfo{person}{Daniel~M. Reeves}, {and} \bibinfo{person}{Tuomas
  Sandholm}.} \bibinfo{year}{2013}\natexlab{}.
\newblock \showarticletitle{{A Practical Liquidity-Sensitive Automated Market
  Maker}}.
\newblock \bibinfo{journal}{\emph{ACM Transactions on Economics and
  Computation}} \bibinfo{volume}{1}, \bibinfo{number}{3} (\bibinfo{date}{9}
  \bibinfo{year}{2013}), \bibinfo{pages}{1--25}.
\newblock
\showISSN{2167-8375}
\urldef\tempurl%
\url{https://doi.org/10.1145/2509413.2509414}
\showDOI{\tempurl}


\bibitem[Othman and Sandholm(2011)]%
        {Othman2011}
\bibfield{author}{\bibinfo{person}{Abraham Othman} {and}
  \bibinfo{person}{Tuomas Sandholm}.} \bibinfo{year}{2011}\natexlab{}.
\newblock \showarticletitle{{Liquidity-Sensitive Automated Market Makers via
  Homogeneous Risk Measures}}.
\newblock In \bibinfo{booktitle}{\emph{Lecture Notes in Computer Science
  (including subseries Lecture Notes in Artificial Intelligence and Lecture
  Notes in Bioinformatics)}}. Vol.~\bibinfo{volume}{7090 LNCS}.
  \bibinfo{publisher}{Springer, Berlin, Heidelberg}, \bibinfo{pages}{314--325}.
\newblock
\showISBNx{9783642255090}
\showISSN{16113349}
\urldef\tempurl%
\url{https://doi.org/10.1007/978-3-642-25510-6{\_}27}
\showDOI{\tempurl}


\bibitem[{PeckShield}(2020a)]%
        {uniswap2020april}
\bibfield{author}{\bibinfo{person}{{PeckShield}}.}
  \bibinfo{year}{2020}\natexlab{a}.
\newblock \bibinfo{title}{{Uniswap/Lendf.Me Hacks: Root Cause and Loss
  Analysis}}.
\newblock
\newblock
\urldef\tempurl%
\url{https://peckshield.medium.com/uniswap-lendf-me-hacks-root-cause-and-loss-analysis-50f3263dcc09}
\showURL{%
\tempurl}


\bibitem[{PeckShield}(2020b)]%
        {peckshield2020valuedefi}
\bibfield{author}{\bibinfo{person}{{PeckShield}}.}
  \bibinfo{year}{2020}\natexlab{b}.
\newblock \bibinfo{title}{{Value DeFi Incident: Root Cause Analysis}}.
\newblock
\newblock
\urldef\tempurl%
\url{https://peckshield.medium.com/value-defi-incident-root-cause-analysis-fbab71faf373}
\showURL{%
\tempurl}


\bibitem[Perez et~al\mbox{.}(2021)]%
        {Perez}
\bibfield{author}{\bibinfo{person}{Daniel Perez}, \bibinfo{person}{Sam~M.
  Werner}, \bibinfo{person}{Jiahua Xu}, {and} \bibinfo{person}{Benjamin
  Livshits}.} \bibinfo{year}{2021}\natexlab{}.
\newblock \showarticletitle{{Liquidations: DeFi on a Knife-Edge}}. In
  \bibinfo{booktitle}{\emph{Financial Cryptography and Data Security}},
  Vol.~\bibinfo{volume}{12675 LNCS}. \bibinfo{pages}{457--476}.
\newblock
\showISBNx{9783662643303}
\showISSN{16113349}
\urldef\tempurl%
\url{https://doi.org/10.1007/978-3-662-64331-0{\_}24}
\showDOI{\tempurl}


\bibitem[Perez et~al\mbox{.}(2020)]%
        {Perez2020d}
\bibfield{author}{\bibinfo{person}{Daniel Perez}, \bibinfo{person}{Jiahua Xu},
  {and} \bibinfo{person}{Benjamin Livshits}.} \bibinfo{year}{2020}\natexlab{}.
\newblock \showarticletitle{{Revisiting Transactional Statistics of
  High-scalability Blockchains}}. In \bibinfo{booktitle}{\emph{Proceedings of
  the ACM Internet Measurement Conference}}, Vol.~\bibinfo{volume}{16}.
  \bibinfo{publisher}{ACM}, \bibinfo{address}{New York, NY, USA},
  \bibinfo{pages}{535--550}.
\newblock
\showISBNx{9781450381383}
\urldef\tempurl%
\url{https://doi.org/10.1145/3419394.3423628}
\showDOI{\tempurl}


\bibitem[{Perpetual Protocol}(2021)]%
        {Protocol2021}
\bibfield{author}{\bibinfo{person}{{Perpetual Protocol}}.}
  \bibinfo{year}{2021}\natexlab{}.
\newblock \bibinfo{title}{{vAMM}}.
\newblock
\newblock
\urldef\tempurl%
\url{https://docs.perp.fi/getting-started/how-it-works/vamm}
\showURL{%
\tempurl}


\bibitem[Perraudin and Vitale(1996)]%
        {Perraudin1996}
\bibfield{author}{\bibinfo{person}{William Perraudin} {and}
  \bibinfo{person}{Paolo Vitale}.} \bibinfo{year}{1996}\natexlab{}.
\newblock \showarticletitle{{Interdealer Trade and Information Flows in a
  Decentralized Foreign Exchange Market}}.
\newblock In \bibinfo{booktitle}{\emph{The Microstructure of Foreign Exchange
  Markets}}. \bibinfo{publisher}{University of Chicago Press},
  \bibinfo{pages}{73--106}.
\newblock
\showISBNx{0226260003}


\bibitem[Peterson and Krug(2015)]%
        {augur}
\bibfield{author}{\bibinfo{person}{Jack Peterson} {and} \bibinfo{person}{Joseph
  Krug}.} \bibinfo{year}{2015}\natexlab{}.
\newblock \bibinfo{title}{{Augur: a decentralized, open-source platform for
  prediction markets}}.  (\bibinfo{year}{2015}), \bibinfo{numpages}{13}~pages.
\newblock


\bibitem[Phillips and Wilder(2020)]%
        {Phillips2020TracingWebsites}
\bibfield{author}{\bibinfo{person}{Ross Phillips} {and} \bibinfo{person}{Heidi
  Wilder}.} \bibinfo{year}{2020}\natexlab{}.
\newblock \showarticletitle{{Tracing cryptocurrency scams: Clustering
  replicated advance-fee and phishing websites}}. In
  \bibinfo{booktitle}{\emph{2020 IEEE International Conference on Blockchain
  and Cryptocurrency (ICBC)}}. \bibinfo{pages}{1--8}.
\newblock


\bibitem[Pirus(2020)]%
        {pirus2020cheesbank}
\bibfield{author}{\bibinfo{person}{Benjamin Pirus}.}
  \bibinfo{year}{2020}\natexlab{}.
\newblock \bibinfo{title}{{Cheese Bank’s multi-million-dollar hack explained
  by security firm}}.
\newblock
\newblock
\urldef\tempurl%
\url{https://cointelegraph.com/news/cheese-bank-s-multi-million-dollar-hack-explained-by-security-firm}
\showURL{%
\tempurl}


\bibitem[{Pods Finance}(2021)]%
        {pods2021}
\bibfield{author}{\bibinfo{person}{{Pods Finance}}.}
  \bibinfo{year}{2021}\natexlab{}.
\newblock \bibinfo{title}{{The easiest way to hedge crypto}}.
\newblock
\newblock
\urldef\tempurl%
\url{https://www.pods.finance/}
\showURL{%
\tempurl}


\bibitem[{Polygon}(2021)]%
        {polygon2021}
\bibfield{author}{\bibinfo{person}{{Polygon}}.}
  \bibinfo{year}{2021}\natexlab{}.
\newblock \bibinfo{title}{{Ethereum's Internet of Blockchains}}.
\newblock
\newblock
\urldef\tempurl%
\url{https://polygon.technology/}
\showURL{%
\tempurl}


\bibitem[Popper and {Nathaniel Popper}(2016)]%
        {daoattack2016}
\bibfield{author}{\bibinfo{person}{Nathaniel Popper} {and}
  \bibinfo{person}{{Nathaniel Popper}}.} \bibinfo{year}{2016}\natexlab{}.
\newblock \bibinfo{title}{{A Hacking of More Than {\$}50 Million Dashes Hopes
  in the World of Virtual Currency}}.
\newblock
\newblock
\urldef\tempurl%
\url{https://www.nytimes.com/2016/06/18/business/dealbook/hacker-may-have-removed-more-than-50-million-from-experimental-cybercurrency-project.html}
\showURL{%
\tempurl}


\bibitem[Praitheeshan et~al\mbox{.}(2019)]%
        {Praitheeshan2019}
\bibfield{author}{\bibinfo{person}{et~al Praitheeshan, Purathani},
  \bibinfo{person}{Lei Pan}, \bibinfo{person}{Jiangshan Yu},
  \bibinfo{person}{Joseph Liu}, {and} \bibinfo{person}{Robin Doss}.}
  \bibinfo{year}{2019}\natexlab{}.
\newblock \showarticletitle{{Security analysis methods on ethereum smart
  contract vulnerabilities: a survey}}.
\newblock \bibinfo{journal}{\emph{arXiv preprint arXiv:1908.08605}}
  (\bibinfo{year}{2019}).
\newblock


\bibitem[Qin et~al\mbox{.}(2021)]%
        {Qin2021}
\bibfield{author}{\bibinfo{person}{Kaihua Qin}, \bibinfo{person}{Liyi Zhou},
  \bibinfo{person}{Pablo Gamito}, \bibinfo{person}{Philipp Jovanovic}, {and}
  \bibinfo{person}{Arthur Gervais}.} \bibinfo{year}{2021}\natexlab{}.
\newblock \showarticletitle{{An empirical study of DeFi liquidations}}. In
  \bibinfo{booktitle}{\emph{Proceedings of the 21st ACM Internet Measurement
  Conference}}. \bibinfo{publisher}{ACM}, \bibinfo{address}{New York, NY, USA},
  \bibinfo{pages}{336--350}.
\newblock
\showISBNx{9781450391290}
\urldef\tempurl%
\url{https://doi.org/10.1145/3487552.3487811}
\showDOI{\tempurl}


\bibitem[Qin et~al\mbox{.}(2022)]%
        {Qin2021a}
\bibfield{author}{\bibinfo{person}{Kaihua Qin}, \bibinfo{person}{Liyi Zhou},
  {and} \bibinfo{person}{Arthur Gervais}.} \bibinfo{year}{2022}\natexlab{}.
\newblock \showarticletitle{{Quantifying Blockchain Extractable Value: How dark
  is the forest?}}. In \bibinfo{booktitle}{\emph{IEEE Symposium on Security and
  Privacy}}.
\newblock


\bibitem[Qin et~al\mbox{.}(2020)]%
        {Qin2020a}
\bibfield{author}{\bibinfo{person}{Kaihua Qin}, \bibinfo{person}{Liyi Zhou},
  \bibinfo{person}{Benjamin Livshits}, {and} \bibinfo{person}{Arthur Gervais}.}
  \bibinfo{year}{2020}\natexlab{}.
\newblock \bibinfo{booktitle}{\emph{{Attacking the DeFi Ecosystem with Flash
  Loans for Fun and Profit}}}.
\newblock \bibinfo{type}{{T}echnical {R}eport}.
\newblock
\urldef\tempurl%
\url{http://arxiv.org/abs/2003.03810}
\showURL{%
\tempurl}


\bibitem[{QuickSwap Official}(2020)]%
        {QuickSwapOfficial2020}
\bibfield{author}{\bibinfo{person}{{QuickSwap Official}}.}
  \bibinfo{year}{2020}\natexlab{}.
\newblock \bibinfo{title}{{QuickSwap FAQ}}.
\newblock
\newblock
\urldef\tempurl%
\url{https://quickswap-layer2.medium.com/welcome-to-quickswap-exchange-93d47e057633}
\showURL{%
\tempurl}


\bibitem[Raikwar and Gligoroski(2021)]%
        {Raikwar2021}
\bibfield{author}{\bibinfo{person}{Mayank Raikwar} {and}
  \bibinfo{person}{Danilo Gligoroski}.} \bibinfo{year}{2021}\natexlab{}.
\newblock \showarticletitle{{Aggregation in Blockchain Ecosystem}}. In
  \bibinfo{booktitle}{\emph{International Conference on Software Defined
  Systems (SDS)}}. \bibinfo{publisher}{IEEE}, \bibinfo{pages}{1--6}.
\newblock
\showISBNx{978-1-6654-5820-7}
\urldef\tempurl%
\url{https://doi.org/10.1109/SDS54264.2021.9732100}
\showDOI{\tempurl}


\bibitem[Ramanan et~al\mbox{.}(2021)]%
        {ramanan2021blockchain}
\bibfield{author}{\bibinfo{person}{Paritosh Ramanan}, \bibinfo{person}{Dan Li},
  {and} \bibinfo{person}{Nagi Gebraeel}.} \bibinfo{year}{2021}\natexlab{}.
\newblock \showarticletitle{{Blockchain-Based Decentralized Replay Attack
  Detection for Large-Scale Power Systems}}.
\newblock \bibinfo{journal}{\emph{IEEE Transactions on Systems, Man, and
  Cybernetics: Systems}} (\bibinfo{year}{2021}).
\newblock


\bibitem[Ramdas and Muthukrishnan(2019)]%
        {Ramdas2019}
\bibfield{author}{\bibinfo{person}{Anju Ramdas} {and}
  \bibinfo{person}{Ramakrishnan Muthukrishnan}.}
  \bibinfo{year}{2019}\natexlab{}.
\newblock \showarticletitle{{A survey on dns security issues and mitigation
  techniques}}. In \bibinfo{booktitle}{\emph{2019 International Conference on
  Intelligent Computing and Control Systems (ICCS)}}.
  \bibinfo{publisher}{IEEE}, \bibinfo{pages}{781--784}.
\newblock
\showISBNx{1538681137}


\bibitem[{Rango}(2022)]%
        {Rango2022}
\bibfield{author}{\bibinfo{person}{{Rango}}.} \bibinfo{year}{2022}\natexlab{}.
\newblock \bibinfo{title}{{Rango Docs}}.
\newblock
\newblock
\urldef\tempurl%
\url{https://docs.rango.exchange/}
\showURL{%
\tempurl}


\bibitem[Redman(2020)]%
        {Redman2020}
\bibfield{author}{\bibinfo{person}{Jamie Redman}.}
  \bibinfo{year}{2020}\natexlab{}.
\newblock \bibinfo{title}{{Report: Blockchain Price Oracle Manipulation
  Produces Millions in Losses, Shows No Signs of Slowing – Altcoins Bitcoin
  News}}.
\newblock
\newblock
\urldef\tempurl%
\url{https://news.bitcoin.com/report-blockchain-price-oracle-manipulation-produces-millions-in-losses-shows-no-signs-of-slowing/}
\showURL{%
\tempurl}


\bibitem[Rembert(2021)]%
        {Rembert2021}
\bibfield{author}{\bibinfo{person}{Ludovic Rembert}.}
  \bibinfo{year}{2021}\natexlab{}.
\newblock \bibinfo{title}{{The 51{\%} Attack}}.
\newblock
\newblock
\urldef\tempurl%
\url{https://privacycanada.net/cryptocurrency/51-attack/}
\showURL{%
\tempurl}


\bibitem[Richardson and Xu(2020)]%
        {Richardson2020c}
\bibfield{author}{\bibinfo{person}{Andreas Richardson} {and}
  \bibinfo{person}{Jiahua Xu}.} \bibinfo{year}{2020}\natexlab{}.
\newblock \showarticletitle{{Carbon Trading with Blockchain}}.
\newblock In \bibinfo{booktitle}{\emph{Mathematical Research for Blockchain
  Economy}}, \bibfield{editor}{\bibinfo{person}{Panos Pardalos},
  \bibinfo{person}{Ilias Kotsireas}, \bibinfo{person}{Yike Guo}, {and}
  \bibinfo{person}{William Knottenbelt}} (Eds.). Chapter~7,
  \bibinfo{pages}{105--124}.
\newblock
\urldef\tempurl%
\url{https://doi.org/10.1007/978-3-030-53356-4{\_}7}
\showDOI{\tempurl}


\bibitem[Robinson and Niemerg(2020)]%
        {robinson2020yield}
\bibfield{author}{\bibinfo{person}{Dan Robinson} {and} \bibinfo{person}{Allan
  Niemerg}.} \bibinfo{year}{2020}\natexlab{}.
\newblock \bibinfo{booktitle}{\emph{{The Yield Protocol: On-Chain Lending With
  Interest Rate Discovery}}}.
\newblock \bibinfo{type}{{T}echnical {R}eport}.
\newblock


\bibitem[Rodler et~al\mbox{.}(2019)]%
        {Rodler2019a}
\bibfield{author}{\bibinfo{person}{Michael Rodler}, \bibinfo{person}{Wenting
  Li}, \bibinfo{person}{Ghassan~O Karame}, {and} \bibinfo{person}{Lucas Davi}.}
  \bibinfo{year}{2019}\natexlab{}.
\newblock \showarticletitle{{Sereum: Protecting Existing Smart Contracts
  Against Re-Entrancy Attacks}}. In \bibinfo{booktitle}{\emph{Network and
  Distributed System Security Symposium}}. \bibinfo{publisher}{Internet
  Society}, \bibinfo{address}{Reston, VA}.
\newblock
\showISBNx{1-891562-55-X}
\urldef\tempurl%
\url{https://doi.org/10.14722/ndss.2019.23413}
\showDOI{\tempurl}


\bibitem[Saad et~al\mbox{.}(2019)]%
        {Saad2019}
\bibfield{author}{\bibinfo{person}{Muhammad Saad}, \bibinfo{person}{Jeffrey
  Spaulding}, \bibinfo{person}{Laurent Njilla}, \bibinfo{person}{Charles
  Kamhoua}, \bibinfo{person}{Sachin Shetty}, \bibinfo{person}{DaeHun Nyang},
  {and} \bibinfo{person}{Aziz Mohaisen}.} \bibinfo{year}{2019}\natexlab{}.
\newblock \showarticletitle{{Exploring the attack surface of blockchain: A
  systematic overview}}.
\newblock \bibinfo{journal}{\emph{arXiv preprint arXiv:1904.03487}}
  (\bibinfo{year}{2019}).
\newblock


\bibitem[{Saber}(2021)]%
        {saber2021}
\bibfield{author}{\bibinfo{person}{{Saber}}.} \bibinfo{year}{2021}\natexlab{}.
\newblock \bibinfo{title}{{Saber | Solana AMM and DEX}}.
\newblock
\newblock
\urldef\tempurl%
\url{https://saber.so/}
\showURL{%
\tempurl}


\bibitem[{Sam M. Werner} et~al\mbox{.}(2022)]%
        {werner2020sokDeFi}
\bibfield{author}{\bibinfo{person}{{Sam M. Werner}}, \bibinfo{person}{{Daniel
  Perez}}, \bibinfo{person}{{Lewis Gudgeon}}, \bibinfo{person}{{Ariah
  Klages-Mundt}}, \bibinfo{person}{{Dominik Harz}}, {and}
  \bibinfo{person}{{William J. Knottenbelt}}.} \bibinfo{year}{2022}\natexlab{}.
\newblock \showarticletitle{{SoK: Decentralized Finance (DeFi)}}.
\newblock


\bibitem[Sasson et~al\mbox{.}(2014)]%
        {Sasson2014Zerocash:Bitcoin}
\bibfield{author}{\bibinfo{person}{Eli~Ben Sasson}, \bibinfo{person}{Alessandro
  Chiesa}, \bibinfo{person}{Christina Garman}, \bibinfo{person}{Matthew Green},
  \bibinfo{person}{Ian Miers}, \bibinfo{person}{Eran Tromer}, {and}
  \bibinfo{person}{Madars Virza}.} \bibinfo{year}{2014}\natexlab{}.
\newblock \showarticletitle{{Zerocash: Decentralized anonymous payments from
  bitcoin}}. In \bibinfo{booktitle}{\emph{2014 IEEE Symposium on Security and
  Privacy}}. \bibinfo{pages}{459--474}.
\newblock


\bibitem[Sayeed et~al\mbox{.}(2020)]%
        {sayeed2020smart}
\bibfield{author}{\bibinfo{person}{Sarwar Sayeed}, \bibinfo{person}{Hector
  Marco-Gisbert}, {and} \bibinfo{person}{Tom Caira}.}
  \bibinfo{year}{2020}\natexlab{}.
\newblock \showarticletitle{{Smart contract: Attacks and protections}}.
\newblock \bibinfo{journal}{\emph{IEEE Access}}  \bibinfo{volume}{8}
  (\bibinfo{year}{2020}), \bibinfo{pages}{24416--24427}.
\newblock


\bibitem[Sch{\"{a}}r(2021)]%
        {Schar2021Defi}
\bibfield{author}{\bibinfo{person}{Fabian Sch{\"{a}}r}.}
  \bibinfo{year}{2021}\natexlab{}.
\newblock \showarticletitle{{Decentralized finance: on blockchain-and smart
  contract-based financial markets}}.
\newblock \bibinfo{journal}{\emph{Federal Reserve Bank of St. Louis Review}}
  \bibinfo{volume}{103}, \bibinfo{number}{2} (\bibinfo{year}{2021}),
  \bibinfo{pages}{153--174}.
\newblock
\showISSN{00149187}
\urldef\tempurl%
\url{https://doi.org/10.20955/r.103.153-74}
\showDOI{\tempurl}


\bibitem[Senchenko(2020)]%
        {Senchenko2020}
\bibfield{author}{\bibinfo{person}{Dmitri Senchenko}.}
  \bibinfo{year}{2020}\natexlab{}.
\newblock \bibinfo{title}{{Impermanent Losses in Uniswap-Like Markets}}.
\newblock
\newblock
\urldef\tempurl%
\url{https://dsenchenko.medium.com/impermanent-losses-in-uniswap-like-markets-4315359ea9b1}
\showURL{%
\tempurl}


\bibitem[Sguanci et~al\mbox{.}(2021)]%
        {sguanci2021layer}
\bibfield{author}{\bibinfo{person}{Cosimo Sguanci}, \bibinfo{person}{Roberto
  Spatafora}, {and} \bibinfo{person}{Andrea~Mario Vergani}.}
  \bibinfo{year}{2021}\natexlab{}.
\newblock \showarticletitle{Layer 2 blockchain scaling: A survey}.
\newblock \bibinfo{journal}{\emph{arXiv preprint arXiv:2107.10881}}
  (\bibinfo{year}{2021}).
\newblock


\bibitem[Shorish(2018)]%
        {Shorish2018}
\bibfield{author}{\bibinfo{person}{Jamsheed Shorish}.}
  \bibinfo{year}{2018}\natexlab{}.
\newblock \bibinfo{title}{{Blockchain State Machine Representation}}.
  (\bibinfo{year}{2018}).
\newblock
\urldef\tempurl%
\url{https://doi.org/10.31235/osf.io/eusxg}
\showDOI{\tempurl}


\bibitem[{Siren}(2021)]%
        {Siren2021}
\bibfield{author}{\bibinfo{person}{{Siren}}.} \bibinfo{year}{2021}\natexlab{}.
\newblock \bibinfo{title}{{SIREN Markets Summary}}.  (\bibinfo{year}{2021}).
\newblock
\urldef\tempurl%
\url{https://siren.xyz/whitepaper}
\showURL{%
\tempurl}


\bibitem[Slamka et~al\mbox{.}(2013)]%
        {Slamka2013}
\bibfield{author}{\bibinfo{person}{Christian Slamka}, \bibinfo{person}{Bernd
  Skiera}, {and} \bibinfo{person}{Martin Spann}.}
  \bibinfo{year}{2013}\natexlab{}.
\newblock \showarticletitle{{Prediction Market Performance and Market
  Liquidity: A Comparison of Automated Market Makers}}.
\newblock \bibinfo{journal}{\emph{IEEE Transactions on Engineering Management}}
  \bibinfo{volume}{60}, \bibinfo{number}{1} (\bibinfo{date}{2}
  \bibinfo{year}{2013}), \bibinfo{pages}{169--185}.
\newblock
\showISSN{0018-9391}
\urldef\tempurl%
\url{https://doi.org/10.1109/TEM.2012.2191618}
\showDOI{\tempurl}


\bibitem[{SmartContent}(2021)]%
        {smartcontent2021TWAP}
\bibfield{author}{\bibinfo{person}{{SmartContent}}.}
  \bibinfo{year}{2021}\natexlab{}.
\newblock \bibinfo{title}{{TWAP Oracles vs. Chainlink Price Feeds: A
  Comparative Analysis}}.
\newblock
\newblock
\urldef\tempurl%
\url{https://smartcontentpublication.medium.com/twap-oracles-vs-chainlink-price-feeds-a-comparative-analysis-8155a3483cbd}
\showURL{%
\tempurl}


\bibitem[{StarkWare Industries Ltd.}(2021)]%
        {starknet2021}
\bibfield{author}{\bibinfo{person}{{StarkWare Industries Ltd.}}}
  \bibinfo{year}{2021}\natexlab{}.
\newblock \bibinfo{title}{{StarkNet}}.
\newblock
\newblock
\urldef\tempurl%
\url{https://starkware.co/product/starknet/}
\showURL{%
\tempurl}


\bibitem[Stone(2021)]%
        {Stone2021TrustlessBridges}
\bibfield{author}{\bibinfo{person}{Drew Stone}.}
  \bibinfo{year}{2021}\natexlab{}.
\newblock \bibinfo{title}{{Trustless, privacy-preserving blockchain bridges}}.
  (\bibinfo{year}{2021}).
\newblock
\urldef\tempurl%
\url{http://arxiv.org/abs/2102.04660}
\showURL{%
\tempurl}


\bibitem[Sun et~al\mbox{.}(2021)]%
        {sun2021mutation}
\bibfield{author}{\bibinfo{person}{Jinlei Sun}, \bibinfo{person}{Song Huang},
  \bibinfo{person}{Changyou Zheng}, \bibinfo{person}{Tingyong Wang},
  \bibinfo{person}{Cheng Zong}, {and} \bibinfo{person}{Zhanwei Hui}.}
  \bibinfo{year}{2021}\natexlab{}.
\newblock \showarticletitle{{Mutation testing for integer overflow in ethereum
  smart contracts}}.
\newblock \bibinfo{journal}{\emph{Tsinghua Science and Technology}}
  \bibinfo{volume}{27}, \bibinfo{number}{1} (\bibinfo{year}{2021}),
  \bibinfo{pages}{27--40}.
\newblock


\bibitem[{Sushiswap}(2020)]%
        {sushi2020}
\bibfield{author}{\bibinfo{person}{{Sushiswap}}.}
  \bibinfo{year}{2020}\natexlab{}.
\newblock \bibinfo{title}{{The SushiSwap Project}}.
\newblock
\newblock
\urldef\tempurl%
\url{https://sushiswapchef.medium.com/the-sushiswap-project-dd6eb80c6ba2}
\showURL{%
\tempurl}


\bibitem[Szalachowski(2018)]%
        {Szalachowski2018}
\bibfield{author}{\bibinfo{person}{Pawel Szalachowski}.}
  \bibinfo{year}{2018}\natexlab{}.
\newblock \bibinfo{title}{{(Short paper) towards more reliable bitcoin
  timestamps}}.
\newblock , \bibinfo{numpages}{101--104}~pages.
\newblock


\bibitem[Taylor(2021)]%
        {frontrunningnews}
\bibfield{author}{\bibinfo{person}{Dan Taylor}.}
  \bibinfo{year}{2021}\natexlab{}.
\newblock \bibinfo{title}{{Privacy first DeFi Sienna Network raises {\$}11.2
  million, takes front-running head on}}.
\newblock
\newblock
\urldef\tempurl%
\url{https://tech.eu/brief/privacy-first-defi-sienna-network-raises-11-2-million-takes-front-running-head-on/}
\showURL{%
\tempurl}


\bibitem[{The European Business Review}(2021)]%
        {rug-pull-businessreview}
\bibfield{author}{\bibinfo{person}{{The European Business Review}}.}
  \bibinfo{year}{2021}\natexlab{}.
\newblock \bibinfo{title}{{What Is A ``Rug Pull'' In Crypto? DeFi Exploits
  Explained}}.
\newblock
\newblock
\urldef\tempurl%
\url{https://www.europeanbusinessreview.com/what-is-a-rug-pull-in-crypto-defi-exploits-explained/}
\showURL{%
\tempurl}


\bibitem[Torres et~al\mbox{.}(2021)]%
        {torres2021frontrunner}
\bibfield{author}{\bibinfo{person}{Christof~Ferreira Torres},
  \bibinfo{person}{Ramiro Camino}, {et~al\mbox{.}}}
  \bibinfo{year}{2021}\natexlab{}.
\newblock \showarticletitle{Frontrunner jones and the raiders of the dark
  forest: An empirical study of frontrunning on the ethereum blockchain}. In
  \bibinfo{booktitle}{\emph{30th USENIX Security Symposium (USENIX Security
  21)}}. \bibinfo{pages}{1343--1359}.
\newblock


\bibitem[Tsankov et~al\mbox{.}(2018)]%
        {tsankov2018securify}
\bibfield{author}{\bibinfo{person}{Petar Tsankov}, \bibinfo{person}{Andrei
  Dan}, \bibinfo{person}{Dana Drachsler-Cohen}, \bibinfo{person}{Arthur
  Gervais}, \bibinfo{person}{Florian B{\"{u}}nzli}, {and}
  \bibinfo{person}{Martin Vechev}.} \bibinfo{year}{2018}\natexlab{}.
\newblock \showarticletitle{{Securify: Practical security analysis of smart
  contracts}}.
\newblock \bibinfo{journal}{\emph{ACM Conference on Computer and Communications
  Security}} (\bibinfo{date}{10} \bibinfo{year}{2018}),
  \bibinfo{pages}{67--82}.
\newblock
\showISBNx{9781450356930}
\showISSN{15437221}
\urldef\tempurl%
\url{https://doi.org/10.1145/3243734.3243780}
\showDOI{\tempurl}


\bibitem[{Uniswap}(2020)]%
        {Uniswap2020}
\bibfield{author}{\bibinfo{person}{{Uniswap}}.}
  \bibinfo{year}{2020}\natexlab{}.
\newblock \bibinfo{title}{{Flash Swaps}}.
\newblock
\newblock
\urldef\tempurl%
\url{https://uniswap.org/docs/v2/core-concepts/flash-swaps/}
\showURL{%
\tempurl}


\bibitem[{Uniswap}(2022)]%
        {Uniswap2022}
\bibfield{author}{\bibinfo{person}{{Uniswap}}.}
  \bibinfo{year}{2022}\natexlab{}.
\newblock \bibinfo{title}{{Liquidity provider fees}}.
\newblock
\newblock
\urldef\tempurl%
\url{https://docs.uniswap.org/protocol/V2/concepts/advanced-topics/fees#liquidity-provider-fees}
\showURL{%
\tempurl}


\bibitem[{Uniswap Governance}(2021)]%
        {UniswapGovernance}
\bibfield{author}{\bibinfo{person}{{Uniswap Governance}}.}
  \bibinfo{year}{2021}\natexlab{}.
\newblock \bibinfo{title}{{Temperature Check - [Fee Switch V2 should be turned
  on]}}.
\newblock
\newblock
\urldef\tempurl%
\url{https://gov.uniswap.org/t/temperature-check-fee-switch-v2-should-be-turned-on/13537}
\showURL{%
\tempurl}


\bibitem[{Uranium.finance}(2021)]%
        {Uranium.finance}
\bibfield{author}{\bibinfo{person}{{Uranium.finance}}.}
  \bibinfo{year}{2021}\natexlab{}.
\newblock \bibinfo{booktitle}{\emph{{How It Works - OUSD}}}.
\newblock \bibinfo{type}{{T}echnical {R}eport}.
\newblock
\urldef\tempurl%
\url{https://docs.ousd.com/how-it-works}
\showURL{%
\tempurl}


\bibitem[Ushida and Angel(2021)]%
        {Ushida2021}
\bibfield{author}{\bibinfo{person}{Ryosuke Ushida} {and} \bibinfo{person}{James
  Angel}.} \bibinfo{year}{2021}\natexlab{}.
\newblock \showarticletitle{{Regulatory Considerations on Centralized Aspects
  of DeFi Managed by DAOs}}.
\newblock In \bibinfo{booktitle}{\emph{FC International Workshops}}.
  Vol.~\bibinfo{volume}{12676 LNCS}. \bibinfo{publisher}{Springer},
  \bibinfo{pages}{21--36}.
\newblock
\showISBNx{9783662639573}
\showISSN{16113349}
\urldef\tempurl%
\url{https://doi.org/10.1007/978-3-662-63958-0{\_}2}
\showDOI{\tempurl}


\bibitem[vbuterin(2022)]%
        {pbs}
\bibfield{author}{\bibinfo{person}{vbuterin}.} \bibinfo{year}{2022}\natexlab{}.
\newblock \bibinfo{title}{State of research: increasing censorship resistance
  of transactions under proposer/builder separation ({PBS})}.
\newblock
\newblock
\urldef\tempurl%
\url{https://notes.ethereum.org/@vbuterin/pbs_censorship_resistance}
\showURL{%
\tempurl}


\bibitem[Victor and Weintraud(2021)]%
        {Victor2021DetectingExchanges}
\bibfield{author}{\bibinfo{person}{Friedhelm Victor} {and}
  \bibinfo{person}{Andrea~Marie Weintraud}.} \bibinfo{year}{2021}\natexlab{}.
\newblock \showarticletitle{{Detecting and Quantifying Wash Trading on
  Decentralized Cryptocurrency Exchanges}}.
\newblock  (\bibinfo{date}{2} \bibinfo{year}{2021}), \bibinfo{pages}{10}.
\newblock
\urldef\tempurl%
\url{https://doi.org/10.1145/3442381.3449824}
\showDOI{\tempurl}


\bibitem[Wang et~al\mbox{.}(2017)]%
        {wang2017skyshield}
\bibfield{author}{\bibinfo{person}{Chenxu Wang}, \bibinfo{person}{Tony T~N
  Miu}, \bibinfo{person}{Xiapu Luo}, {and} \bibinfo{person}{Jinhe Wang}.}
  \bibinfo{year}{2017}\natexlab{}.
\newblock \showarticletitle{{SkyShield: A sketch-based defense system against
  application layer DDoS attacks}}.
\newblock \bibinfo{journal}{\emph{IEEE Transactions on Information Forensics
  and Security}} \bibinfo{volume}{13}, \bibinfo{number}{3}
  (\bibinfo{year}{2017}), \bibinfo{pages}{559--573}.
\newblock


\bibitem[Wang et~al\mbox{.}(2020)]%
        {Wang2020FlashLoan}
\bibfield{author}{\bibinfo{person}{Dabao Wang}, \bibinfo{person}{Siwei Wu},
  \bibinfo{person}{Ziling Lin}, \bibinfo{person}{Lei Wu},
  \bibinfo{person}{Xingliang Yuan}, \bibinfo{person}{Yajin Zhou},
  \bibinfo{person}{Haoyu Wang}, {and} \bibinfo{person}{Kui Ren}.}
  \bibinfo{year}{2020}\natexlab{}.
\newblock \bibinfo{title}{{Towards understanding flash loan and its
  applications in defi ecosystem}}.  (\bibinfo{date}{10} \bibinfo{year}{2020}).
\newblock
\urldef\tempurl%
\url{http://arxiv.org/abs/2010.12252}
\showURL{%
\tempurl}


\bibitem[Wang et~al\mbox{.}(2021)]%
        {wang2021promutator}
\bibfield{author}{\bibinfo{person}{Shih-Hung Wang}, \bibinfo{person}{Chia-Chien
  Wu}, \bibinfo{person}{Yu-Chuan Liang}, \bibinfo{person}{Li-Hsun Hsieh}, {and}
  \bibinfo{person}{Hsu-Chun Hsiao}.} \bibinfo{year}{2021}\natexlab{}.
\newblock \showarticletitle{ProMutator: Detecting Vulnerable Price Oracles in
  DeFi by Mutated Transactions}. In \bibinfo{booktitle}{\emph{2021 IEEE
  European Symposium on Security and Privacy Workshops (EuroS\&PW)}}. IEEE,
  \bibinfo{pages}{380--385}.
\newblock


\bibitem[Wang(2020)]%
        {Wang2020AutomatedDeFi}
\bibfield{author}{\bibinfo{person}{Yongge Wang}.}
  \bibinfo{year}{2020}\natexlab{}.
\newblock \bibinfo{title}{{Automated market makers for decentralized finance
  (DeFi)}}.  (\bibinfo{date}{9} \bibinfo{year}{2020}).
\newblock
\showISSN{23318422}
\urldef\tempurl%
\url{http://arxiv.org/abs/2009.01676}
\showURL{%
\tempurl}


\bibitem[Warren and Bandeali(2017)]%
        {0xProject}
\bibfield{author}{\bibinfo{person}{Will Warren} {and} \bibinfo{person}{Amir
  Bandeali}.} \bibinfo{year}{2017}\natexlab{}.
\newblock \bibinfo{title}{{0x: An open protocol for decentralized exchange on
  the Ethereum blockchain}}.
\newblock
\newblock
\urldef\tempurl%
\url{https://github.com/0xProject/whitepaper/blob/master/0x_white_paper.pdf}
\showURL{%
\tempurl}


\bibitem[Wintermute(2020)]%
        {Wintermute2020}
\bibfield{author}{\bibinfo{person}{Molly Wintermute}.}
  \bibinfo{year}{2020}\natexlab{}.
\newblock \bibinfo{booktitle}{\emph{{Hegic: On-chain Options Trading Protocol
  on Ethereum Powered by Hedge Contracts and Liquidity Pools}}}.
\newblock \bibinfo{type}{{T}echnical {R}eport}.
\newblock
\urldef\tempurl%
\url{https://github.com/hegic/whitepaper/blob/master/Hegic Protocol
  Whitepaper.pdf}
\showURL{%
\tempurl}


\bibitem[Wong(2021)]%
        {Wong2021}
\bibfield{author}{\bibinfo{person}{Joon~Ian Wong}.}
  \bibinfo{year}{2021}\natexlab{}.
\newblock \bibinfo{title}{{SushiSwap drained UniSwap of {\$}1 billion in
  liquidity and no one knows who was behind it to this day}}.
\newblock
\newblock
\urldef\tempurl%
\url{https://www.businessofbusiness.com/articles/satoshi-30-billion-bitcoin-sushiswap-uniswap-defi-summer-crypto-anonymity-sybil-attacks/}
\showURL{%
\tempurl}


\bibitem[Wu et~al\mbox{.}(2020)]%
        {Wu2020WhoEmbedding}
\bibfield{author}{\bibinfo{person}{Jiajing Wu}, \bibinfo{person}{Qi Yuan},
  \bibinfo{person}{Dan Lin}, \bibinfo{person}{Wei You}, \bibinfo{person}{Weili
  Chen}, \bibinfo{person}{Chuan Chen}, {and} \bibinfo{person}{Zibin Zheng}.}
  \bibinfo{year}{2020}\natexlab{}.
\newblock \showarticletitle{{Who are the phishers? phishing scam detection on
  ethereum via network embedding}}.
\newblock \bibinfo{journal}{\emph{IEEE Transactions on Systems, Man, and
  Cybernetics: Systems}} (\bibinfo{year}{2020}).
\newblock


\bibitem[Xia et~al\mbox{.}(2021)]%
        {Xia2021a}
\bibfield{author}{\bibinfo{person}{Pengcheng Xia}, \bibinfo{person}{Haoyu
  Wang}, \bibinfo{person}{Bingyu Gao}, \bibinfo{person}{Weihang Su},
  \bibinfo{person}{Zhou Yu}, \bibinfo{person}{Xiapu Luo}, \bibinfo{person}{Chao
  Zhang}, \bibinfo{person}{Xusheng Xiao}, {and} \bibinfo{person}{Guoai Xu}.}
  \bibinfo{year}{2021}\natexlab{}.
\newblock \showarticletitle{{Trade or Trick? Detecting and Characterizing Scam
  Tokens on Uniswap Decentralized Exchange}}.
\newblock \bibinfo{journal}{\emph{Proceedings of the ACM on Measurement and
  Analysis of Computing Systems}} \bibinfo{volume}{5}, \bibinfo{number}{3}
  (\bibinfo{date}{12} \bibinfo{year}{2021}), \bibinfo{pages}{1--26}.
\newblock
\showISSN{2476-1249}
\urldef\tempurl%
\url{https://doi.org/10.1145/3491051}
\showDOI{\tempurl}


\bibitem[Xie and Yu(2008)]%
        {xie2008monitoring}
\bibfield{author}{\bibinfo{person}{Yi Xie} {and} \bibinfo{person}{Shun-Zheng
  Yu}.} \bibinfo{year}{2008}\natexlab{}.
\newblock \showarticletitle{{Monitoring the application-layer DDoS attacks for
  popular websites}}.
\newblock \bibinfo{journal}{\emph{IEEE/ACM Transactions on networking}}
  \bibinfo{volume}{17}, \bibinfo{number}{1} (\bibinfo{year}{2008}),
  \bibinfo{pages}{15--25}.
\newblock


\bibitem[Xu and Feng(2022)]%
        {Xu2022g}
\bibfield{author}{\bibinfo{person}{Jiahua Xu} {and} \bibinfo{person}{Yebo
  Feng}.} \bibinfo{year}{2022}\natexlab{}.
\newblock \showarticletitle{{Reap the Harvest on Blockchain: A Survey of Yield
  Farming Protocols}}.
\newblock \bibinfo{journal}{\emph{IEEE Transactions on Network and Service
  Management}} (\bibinfo{date}{10} \bibinfo{year}{2022}),
  \bibinfo{pages}{1--1}.
\newblock
\showISSN{1932-4537}
\urldef\tempurl%
\url{https://doi.org/10.1109/TNSM.2022.3222815}
\showDOI{\tempurl}


\bibitem[Xu and Vadgama(2022)]%
        {Xu2022d}
\bibfield{author}{\bibinfo{person}{Jiahua Xu} {and} \bibinfo{person}{Nikhil
  Vadgama}.} \bibinfo{year}{2022}\natexlab{}.
\newblock \showarticletitle{{From Banks to DeFi: the Evolution of the Lending
  Market}}.
\newblock In \bibinfo{booktitle}{\emph{Enabling the Internet of Value}},
  \bibfield{editor}{\bibinfo{person}{Nikhil Vadgama}, \bibinfo{person}{Jiahua
  Xu}, {and} \bibinfo{person}{Paolo Tasca}} (Eds.).
  \bibinfo{publisher}{Springer}, Chapter~6, \bibinfo{pages}{53--66}.
\newblock
\urldef\tempurl%
\url{https://doi.org/10.1007/978-3-030-78184-2{\_}6}
\showDOI{\tempurl}


\bibitem[Xu and Xu(2022)]%
        {Xu2022a}
\bibfield{author}{\bibinfo{person}{Teng~Andrea Xu} {and}
  \bibinfo{person}{Jiahua Xu}.} \bibinfo{year}{2022}\natexlab{}.
\newblock \showarticletitle{{A Short Survey on Business Models of Decentralized
  Finance (DeFi) Protocols}}. In \bibinfo{booktitle}{\emph{Workshop Proceedings
  of Financial Cryptography and Data Security}}.
\newblock
\urldef\tempurl%
\url{https://doi.org/10.48550/arxiv.2202.07742}
\showDOI{\tempurl}


\bibitem[Xu et~al\mbox{.}(2022)]%
        {Xu2022DeFiFlow}
\bibfield{author}{\bibinfo{person}{Teng~Andrea Xu}, \bibinfo{person}{Jiahua
  Xu}, {and} \bibinfo{person}{Kristof Lommers}.}
  \bibinfo{year}{2022}\natexlab{}.
\newblock \showarticletitle{{DeFi vs TradFi: Valuation Using Multiples and
  Discounted Cash Flow}}.
\newblock  (\bibinfo{date}{10} \bibinfo{year}{2022}).
\newblock
\urldef\tempurl%
\url{https://doi.org/10.48550/arxiv.2210.16846}
\showDOI{\tempurl}


\bibitem[Yaish et~al\mbox{.}(2022)]%
        {yaish2022blockchain}
\bibfield{author}{\bibinfo{person}{Aviv Yaish}, \bibinfo{person}{Saar Tochner},
  {and} \bibinfo{person}{Aviv Zohar}.} \bibinfo{year}{2022}\natexlab{}.
\newblock \showarticletitle{Blockchain Stretching \& Squeezing: Manipulating
  Time for Your Best Interest}. In \bibinfo{booktitle}{\emph{Proceedings of the
  23rd ACM Conference on Economics and Computation}}. \bibinfo{pages}{65--88}.
\newblock


\bibitem[{YCharts}(2021)]%
        {ethereumgasprice2021}
\bibfield{author}{\bibinfo{person}{{YCharts}}.}
  \bibinfo{year}{2021}\natexlab{}.
\newblock \bibinfo{title}{{Ethereum Average Gas Price}}.
\newblock
\newblock
\urldef\tempurl%
\url{https://ycharts.com/indicators/ethereum_average_gas_price}
\showURL{%
\tempurl}


\bibitem[Y{\"{u}}ksel et~al\mbox{.}(2021)]%
        {Yuksel2021}
\bibfield{author}{\bibinfo{person}{Akif Y{\"{u}}ksel}, \bibinfo{person}{Oguzhan
  Ersoy}, {and} \bibinfo{person}{Zekeriya Erkin}.}
  \bibinfo{year}{2021}\natexlab{}.
\newblock \bibinfo{title}{{Mitigating sandwich attacks in Kyber DMM}}.
  (\bibinfo{year}{2021}).
\newblock
\urldef\tempurl%
\url{https://repository.tudelft.nl/islandora/object/uuid%3A58ac3b00-10fb-44cd-b1eb-1e1139c39fd7}
\showURL{%
\tempurl}


\bibitem[Zargham et~al\mbox{.}(2020a)]%
        {Zargham2020b}
\bibfield{author}{\bibinfo{person}{Michael Zargham}, \bibinfo{person}{Krzysztof
  Paruch}, {and} \bibinfo{person}{Jamsheed Shorish}.}
  \bibinfo{year}{2020}\natexlab{a}.
\newblock \showarticletitle{{Economic Games as Estimators}}. In
  \bibinfo{booktitle}{\emph{Mathematical Research for Blockchain Economy}}.
  \bibinfo{publisher}{Springer, Cham}, \bibinfo{pages}{125--142}.
\newblock
\urldef\tempurl%
\url{https://doi.org/10.1007/978-3-030-53356-4{\_}8}
\showDOI{\tempurl}


\bibitem[Zargham et~al\mbox{.}(2020b)]%
        {Zargham2020a}
\bibfield{author}{\bibinfo{person}{Michael Zargham}, \bibinfo{person}{Jamsheed
  Shorish}, {and} \bibinfo{person}{Krzysztof Paruch}.}
  \bibinfo{year}{2020}\natexlab{b}.
\newblock \showarticletitle{{From Curved Bonding to Configuration Spaces}}. In
  \bibinfo{booktitle}{\emph{IEEE International Conference on Blockchain and
  Cryptocurrency (ICBC)}}. \bibinfo{publisher}{IEEE}, \bibinfo{pages}{1--3}.
\newblock
\showISBNx{978-1-7281-6680-3}
\urldef\tempurl%
\url{https://doi.org/10.1109/ICBC48266.2020.9169474}
\showDOI{\tempurl}


\bibitem[Zargham et~al\mbox{.}(2018)]%
        {Zargham2018}
\bibfield{author}{\bibinfo{person}{Michael Zargham}, \bibinfo{person}{Zixuan
  Zhang}, {and} \bibinfo{person}{Victor Preciado}.}
  \bibinfo{year}{2018}\natexlab{}.
\newblock \bibinfo{title}{{A State-Space Modeling Framework for Engineering
  Blockchain-Enabled Economic Systems}}.  (\bibinfo{date}{7}
  \bibinfo{year}{2018}).
\newblock
\urldef\tempurl%
\url{http://arxiv.org/abs/1807.00955}
\showURL{%
\tempurl}


\bibitem[Zhang et~al\mbox{.}(2019)]%
        {Zhang2019SecurityBlockchain}
\bibfield{author}{\bibinfo{person}{Rui Zhang}, \bibinfo{person}{Rui Xue}, {and}
  \bibinfo{person}{Ling Liu}.} \bibinfo{year}{2019}\natexlab{}.
\newblock \showarticletitle{{Security and privacy on blockchain}}.
\newblock \bibinfo{journal}{\emph{ACM Computing Surveys (CSUR)}}
  \bibinfo{volume}{52}, \bibinfo{number}{3} (\bibinfo{year}{2019}),
  \bibinfo{pages}{1--34}.
\newblock


\bibitem[Zhang et~al\mbox{.}(2020a)]%
        {zhang2020smartshield}
\bibfield{author}{\bibinfo{person}{Yuyao Zhang}, \bibinfo{person}{Siqi Ma},
  \bibinfo{person}{Juanru Li}, \bibinfo{person}{Kailai Li},
  \bibinfo{person}{Surya Nepal}, {and} \bibinfo{person}{Dawu Gu}.}
  \bibinfo{year}{2020}\natexlab{a}.
\newblock \showarticletitle{{Smartshield: Automatic smart contract protection
  made easy}}. In \bibinfo{booktitle}{\emph{2020 IEEE 27th International
  Conference on Software Analysis, Evolution and Reengineering (SANER)}}.
  \bibinfo{pages}{23--34}.
\newblock


\bibitem[Zhang et~al\mbox{.}(2020b)]%
        {Zhang2020}
\bibfield{author}{\bibinfo{person}{Zixuan Zhang}, \bibinfo{person}{Michael
  Zargham}, {and} \bibinfo{person}{Victor~M. Preciado}.}
  \bibinfo{year}{2020}\natexlab{b}.
\newblock \showarticletitle{{On modeling blockchain-enabled economic networks
  as stochastic dynamical systems}}.
\newblock \bibinfo{journal}{\emph{Applied Network Science}}
  \bibinfo{volume}{5}, \bibinfo{number}{1} (\bibinfo{date}{12}
  \bibinfo{year}{2020}), \bibinfo{pages}{19}.
\newblock
\showISSN{2364-8228}
\urldef\tempurl%
\url{https://doi.org/10.1007/s41109-020-0254-9}
\showDOI{\tempurl}


\bibitem[Zhou et~al\mbox{.}(2021b)]%
        {zhou2021just}
\bibfield{author}{\bibinfo{person}{Liyi Zhou}, \bibinfo{person}{Kaihua Qin},
  \bibinfo{person}{Antoine Cully}, \bibinfo{person}{Benjamin Livshits}, {and}
  \bibinfo{person}{Arthur Gervais}.} \bibinfo{year}{2021}\natexlab{b}.
\newblock \showarticletitle{On the just-in-time discovery of profit-generating
  transactions in defi protocols}. In \bibinfo{booktitle}{\emph{2021 IEEE
  Symposium on Security and Privacy (SP)}}. IEEE, \bibinfo{pages}{919--936}.
\newblock


\bibitem[Zhou et~al\mbox{.}(2021a)]%
        {Zhou2021A2MM}
\bibfield{author}{\bibinfo{person}{Liyi Zhou}, \bibinfo{person}{Kaihua Qin},
  {and} \bibinfo{person}{Arthur Gervais}.} \bibinfo{year}{2021}\natexlab{a}.
\newblock \bibinfo{title}{{A2MM: Mitigating Frontrunning, Transaction
  Reordering and Consensus Instability in Decentralized Exchanges}}.
  (\bibinfo{date}{6} \bibinfo{year}{2021}).
\newblock
\urldef\tempurl%
\url{http://arxiv.org/abs/2106.07371}
\showURL{%
\tempurl}


\bibitem[Zhou et~al\mbox{.}(2021c)]%
        {Zhou2021High-Frequency}
\bibfield{author}{\bibinfo{person}{Liyi Zhou}, \bibinfo{person}{Kaihua Qin},
  \bibinfo{person}{Christof~Ferreira Torres}, \bibinfo{person}{Duc~V. Le},
  {and} \bibinfo{person}{Arthur Gervais}.} \bibinfo{year}{2021}\natexlab{c}.
\newblock \showarticletitle{{High-frequency trading on decentralized on-chain
  exchanges}}. In \bibinfo{booktitle}{\emph{IEEE Symposium on Security and
  Privacy}}. \bibinfo{pages}{428--445}.
\newblock
\showISBNx{9781728189345}
\showISSN{10816011}
\urldef\tempurl%
\url{https://doi.org/10.1109/SP40001.2021.00027}
\showDOI{\tempurl}


\bibitem[{ZKSwap}(2021)]%
        {zkswap2021}
\bibfield{author}{\bibinfo{person}{{ZKSwap}}.} \bibinfo{year}{2021}\natexlab{}.
\newblock \bibinfo{title}{{ZKSwap home page}}.
\newblock
\newblock
\urldef\tempurl%
\url{https://zks.org/en}
\showURL{%
\tempurl}


\bibitem[Z{\"{u}}st et~al\mbox{.}(2021)]%
        {Zust2021}
\bibfield{author}{\bibinfo{person}{Patrick Z{\"{u}}st},
  \bibinfo{person}{Tejaswi Nadahalli}, {and} \bibinfo{person}{Ye Wang
  Roger~Wattenhofer}.} \bibinfo{year}{2021}\natexlab{}.
\newblock \showarticletitle{{Analyzing and Preventing Sandwich Attacks in
  Ethereum}}.
\newblock  (\bibinfo{year}{2021}).
\newblock
\urldef\tempurl%
\url{www.DeFi-Sandwi.ch.}
\showURL{%
\tempurl}


\end{thebibliography}
